\documentclass[sn-mathphys-num]{sn-jnl}

\usepackage{graphicx}%
\usepackage{multirow}%
\usepackage{amsmath,amssymb,amsfonts,mathabx}%
\usepackage{amsthm}%
\usepackage{mathrsfs}%
\usepackage[title]{appendix}%
\usepackage{xcolor}%
\usepackage{textcomp}%
\usepackage{manyfoot}%
\usepackage{booktabs}%
\usepackage{algorithm}%
\usepackage{algorithmicx}%
\usepackage{algpseudocode}%
\usepackage{listings}%
\usepackage{marginnote}
\usepackage{mathtools}
\usepackage{scalerel}
\usepackage{comment}
\usepackage{dsfont}
\usepackage{tikz}
\usepackage{tikz-cd}
\usetikzlibrary[patterns]
\usepackage{mdframed}
\usepackage{subfigure}

\theoremstyle{definition}%
\newtheorem{example}{Example}%
\newtheorem{theorem}{Theorem}

\newtheorem{cor}{Corollary}
\newtheorem{lemma}{Lemma}
\newtheorem{definition}{Definition}
\newtheorem{proposition}[theorem]{Proposition}%

\theoremstyle{thmstyletwo}%

\def\CA{{\mathcal A}}

\def\CH {{\mathcal H}}
\def\CQ{{\mathcal Q}}

\def\IC{\mathbb{C}}

\def\IR{{\mathbb{R}}}

\def\IZ{{\mathbb{Z}}}

\newcommand{\ket}[1]{|#1\rangle}

\def\[#1\]{%
  \begin{equation}\begin{gathered}#1\end{gathered}\end{equation}%
}

\def\1{\mathds{1}}

\begin{document}

\title[Article Title]{Anomalies on the Lattice, Homotopy of Quantum Cellular Automata, and a Spectrum of Invertible States}

\author{\fnm{Alexander M.} \sur{Czajka}, \fnm{Roman} \sur{Geiko}, \fnm{Ryan} \sur{Thorngren}

\affil{\orgdiv{Mani L. Bhaumik Institute for Theoretical Physics

Department of Physics and Astronomy}, \orgname{University of California},

\city{Los Angeles}, \postcode{90095}, \state{CA}, \country{USA}}}

\abstract{We develop a rigorous topological theory of anomalies on the lattice, which are obstructions to gauging global symmetries and the existence of trivial symmetric states. We also construct $\Omega$-spectra of a class of invertible states and quantum cellular automata, which allows us to classify both anomalies and symmetry protected topological phases up to blend equivalence.}


\maketitle

\tableofcontents

\section{Introduction}

In the past 20 years, there have been significant advances in understanding the interplay between topological phases and symmetry. In particular, the discovery of topological insulators \cite{hasan2010colloquium,qi2011topological} and their subsequent generalizations to symmetry-protected topological (SPT) phases \cite{fidkowski2011topological,schuch2011classifying,pollmann2012symmetry,Chen_2013,Gu_2014} has highlighted a class of topological phases with rich structure and deep connections between condensed matter, high energy theory, and quantum information. Moreover, this is a class which we can hope to completely understand.

Perhaps the most salient feature of SPT phases is the bulk-boundary correspondence, which connects the symmetry-protected entanglement in the bulk of the SPT state to symmetry-protected anomalous modes localized on the boundary, such as the famous Dirac cone on the surface of a 3d topological insulator. It is conjectured and widely believed that the continuum limit of an SPT phase should be describable in terms of a topological quantum field theory (TQFT) coupled to background gauge fields for the global symmetries \cite{dijkgraaf1990topological,kapustin2014symmetryprotectedtopologicalphases,Freed_2021}.

In this continuum formalism, the quantum field theory (QFT) of the boundary has a 't Hooft anomaly, which must be matched by non-trivial low energy degrees of freedom, such as gapless surface states. The connection to 't Hooft anomalies makes the classification and characterization of SPTs very important. Indeed, 't Hooft anomaly matching provides one of the few universally applicable non-perturbative methods in strongly-interacting QFT.

Many works have also applied anomaly-matching reasoning to make predictions of lattice models. These predictions seem in concert with known rigorous theorems such as the Lieb-Schultz-Mattis theorem \cite{cho2017anomaly,else2020topological,else2025t}, and so we believe there should be a rigorous theory of anomalies on the lattice.

In this paper, we hope to lay some of the foundations of this theory, by developing topological methods to study lattice anomalies and the classification of SPT phases, without appeal to the continuum.

In particular, we will outline a construction of a ``space''\footnote{We outline the construction of this space as an $\infty$-groupoid, which defines the space up to weak homotopy equivalence. We will mainly focus on applications of the construction, and postpone demonstrating all the $\infty$-groupoid axioms to future work \cite{followup}. Although we do not prove our structure is equivalent to an $\infty$-groupoid, it is enough to compute homotopy groups we need for applications in this paper.} of quantum cellular automata (QCA), which are a class of locality preserving unitary transformations of the algebra of local operators on a lattice. This space will allow us to define and study lattice anomalies of symmetries using homotopy theory. We find that there are actually two distinct natural notions of anomaly on the lattice, as anticipated in recent works \cite{tu2025anomaliesglobalsymmetrieslattice,shirley2025anomalyfreesymmetriesobstructionsgauging}: one is an obstruction to on-siteability/gauging, and the other is an obstruction to there being a symmetric trivial state.

This space also allows us to define an $\Omega$-spectrum of FDQC-invertible\footnote{This is a subclass of all invertible states, which can be disentangled with their inverse using a finite depth quantum circuit (FDQC). Precise definitions for all terms are given in the main text.} states. This $\Omega$-spectrum is a very structured type of topological space, whose homotopy groups classify FDQC-invertible states up to an equivalence we call blend equivalence. The conjectural existence of this $\Omega$-spectrum is another expectation from the continuum, and lends many very nice features to the theory of anomalies, such as making anomalies form an abelian group \cite{kitaev2013sre,kitaevIPAMtalk,Xiong_2018,Gaiotto_2019,Beaudry_2024,kubota2025stablehomotopytheoryinvertible}. In general, the $\Omega$-spectrum means that anomalies and FDQC-invertible states are classified by a generalized cohomology theory.

Let us see how the a space of QCA helps us analyze anomalies on the lattice with homotopy theory. Let $\CQ^d$ be the classifying space of QCA in $d$ spatial dimensions, which we construct\footnote{We call this a classifying space because it is a connected space whose loops correspond to QCA, i.e. its loop space can be considered the space of QCA themselves.}. A global $G$ symmetry $\alpha$ which acts by QCA satisfying the group law defines a map from the classifying space $BG$ of the group to $\CQ^d$, which we also call $\alpha$:
\[\alpha:BG \to \CQ^d.\]
We consider the homotopy class of this map and prove it is an obstruction to ``disentangling'' $\alpha$, meaning to find a QCA change of basis that makes $\alpha$ on-site. It can thus be viewed as an obstruction to making the symmetry on-site (``on-sitability''). It is also an obstruction to a weaker condition that there exists a truncation of the $G$-representation to half the system in a way that still satisfies locality and the group law. This is called a blend, so we thus call the homotopy class of $\alpha$ the ``blend anomaly''. Note that the existence of a blend is a weaker condition than gaugability, which requires localizing the symmetry to finite regions, so the blend anomaly is also an obstruction to gauging.

We can study blend anomalies both in the case of a fixed algebra on which the QCA acts, or in a ``stable setting'' where we allow addition of ancillas. Once we do this appropriately, $\CQ^d$ also becomes an $\Omega$-spectrum and the blend anomalies for a fixed $G$ form an abelian group $\CQ^d(BG)$ with the group law given by stacking. Moreover, we obtain a generalized cohomology theory, allowing one to use tools like spectral sequences to compute this group. We use these tools to compute the homotopy type of $\CQ^1$, classifying stable blend anomalies in $d = 1$ for both bosons and fermions.

This QCA spectrum turns out to be closely related to the spectrum of FDQC-invertible states. In particular, we can apply QCA to product states to obtain FDQC-invertible states. We use this to construct the spectrum $\CQ^d_{inv}$ of FDQC-invertible states by considering a ``cofiber spectrum'' of the QCA spectrum. One can view this as taking a quotient of all QCA by those which fix a chosen reference product state. We show that the homotopy groups of this cofiber spectrum are in one-to-one correspondence with blend equivalence classes of FDQC-invertible states. It seems almost clear that our construction will yield an $\Omega$-spectrum for all invertible states once we generalize our QCA space to a more general class of locality preserving automorphisms.

From our construction of the spectrum $\CQ^d_{inv}$ of FDQC-invertible states we also obtain a map
\[\CQ^d \to B\CQ^d_{inv}\]
and thus for a global $G$ symmetry we obtain a map
\[\alpha:BG \to B\CQ^d_{inv}.\]
The homotopy class of this map is the ``SRE anomaly''. We prove it is an obstruction to the existence of a $G$-symmetric short-range-entangled state. This is the other kind of anomaly. We obtain a long exact sequence relating this to the blend anomaly.

In the above, we considered $G$ acting by QCA, as occurs naturally at the boundary of a (FDQC-entanglable) $G$-SPT. There is another set of constructions corresponding to the bulk of the SPT which we can obtain by letting $G$ act as an on-site symmetry. In particular, let us fix a unitary representation of $G$ acting on the local Hilbert space. We then construct a (naive) $G$-$\Omega$-spectrum $\CQ^d_G$ of QCA commuting with the $G$ action. By the cofiber construction we obtain also a $G$-$\Omega$-spectrum $\CQ^d_{G,inv}$ of $G$-SPT phases.

There does not seem to be any obvious bulk-boundary correspondence for SRE anomalies and SPT phases. This is supported by our calculations of $\IZ_2$ blend and SRE anomalies in 1d fermion chains. However, we are able to prove a bulk-boundary correspondence between blend anomalies and SPT entanglers.

The paper is organized as follows. In \hyperref[secmainsection]{Section \ref{secmainsection}}, we describe our theory of blend anomalies. Basic definitions are given in \hyperref[subseconsitability]{Section \ref{subseconsitability}}. In \hyperref[subsecspacefirstpass]{Section \ref{subsecspacefirstpass}}, we give a short intro to the construction of the space of QCA and define the blend anomaly, which is an obstruction to on-siteability. In \hyperref[subsecstabilization]{Section \ref{subsecstabilization}} we discuss ``stabilization'' and the introduction of ancillas, and prove in this case we obtain an $\Omega$-spectrum of QCA. In \hyperref[subseccomptheanom]{Section \ref{subseccomptheanom}} we give an abstract definition of the ``anomaly indices'', which generalize the constructions of Else and Nayak \cite{Else_2014}.

In \hyperref[secspectrasreanom]{Section \ref{secspectrasreanom}}, we describe the $\Omega$-spectrum of FDQC-invertible states, SRE anomalies, and bulk-boundary correspondence. In particular in \hyperref[subsecrelationtoinvertiblestates]{Section \ref{subsecrelationtoinvertiblestates}} we construct the $\Omega$-spectrum of FDQC-invertible states as a certain cofiber spectrum, and prove its main properties. In \hyperref[subsecsreanom]{Section \ref{subsecsreanom}} we use this $\Omega$-spectrum to define the ``SRE anomaly'', which is the obstruction to the existence of a symmetric short-range-entangled (SRE) state. In \hyperref[secsptandbulkboundary]{Section \ref{secsptandbulkboundary}} we construct an equivariant version of the $\Omega$-spectrum of FDQC-invertible states for classifying SPTs, and discuss the bulk-boundary correspondence (or lack thereof, as it turns out).

In \hyperref[secconstruction]{Section \ref{secconstruction}}, we give more details of the construction of the space of QCA and its applications. In particular, in \hyperref[subsecglobular]{Section \ref{subsecglobular}} we define it as an $\infty$-groupoid, which is a globular set with certain composition rules (although we postpone checking the coherence axioms to future work \cite{followup}). We use these composition rules to compute its homotopy groups. In \hyperref[subsecelsenayak]{Section \ref{subsecelsenayak}} we use these composition rules to give formulas for the anomaly indices. This is one of the main applications of the $\infty$-groupoid formalism, since it allows us to express the anomaly indices, which are ``nonabelian cocycles''.

In \hyperref[secanomalousferms1d]{Section \ref{secanomalousferms1d}} we use homotopy theoretic methods to study the homotopy types of the spaces of 1d QCA for bosons and fermions, and the spaces of 1d invertible states for bosons and fermions. We compute the blend anomaly group for 1d $\IZ_2$ symmetries with fermions and find $\CQ^1_f(B\IZ_2) = \IZ_4$, demonstrating a non-trivial Postnikov invariant, which we compute. We also compute the SRE anomaly group for 1d $\IZ_2$ symmetries with fermions and find $\CQ^1_{f,inv}(B\IZ_2) = \IZ_2 \times \IZ_4$, which is different from the continuum expectation $\Omega^3_{\rm Spin}(B\IZ_2) = \IZ_8$. We show our answer is physically sensible and describe a resolution to this paradox as a subtle matching of anomalies between the lattice and the continuum.

In the appendices, we compute the Else-Nayak index for a QCA $G$-representation based on an arbitrary group cohomology class in $H^d(BG,U(1))$, and separately we discuss the dependence of anomaly indices (including the Else-Nayak index) on choices made in its construction.

\section{Blend Anomalies and Obstructions to On-Siteability}\label{secmainsection}

\subsection{On-site and disentanglable symmetries}\label{subseconsitability}

Consider a lattice $\Lambda \subset \mathbb{R}^d$ with Hilbert spaces $\mathcal{H}_x$ associated to each lattice point $x \in \Lambda$. The simple internal symmetries to consider on such a system are on-site symmetries:
\vspace{3mm}
\begin{definition}
    An \textbf{on-site unitary $G$-representation} on $\{\mathcal{H}_x\}_{x \in \Lambda}$ is a collection of unitary $G$-representations $\alpha_x$ on each $\mathcal{H}_x$.
\end{definition}
\vspace{3mm}
\noindent On-site symmetries may be promoted to local symmetries with the addition of gauge fields to the Hilbert space \cite{kogut1979introduction,haegeman2015gauging,chen2022exactly}. They also admit symmetric short-range-entangled (SRE) states under mild assumptions\footnote{For example, if only finitely many types of representations appear among the $\CH_x$, we can locally block sites until the tensor product representation contains a singlet. Then we may consider the tensor product of those singlet states between blocks as a symmetric SRE state.}. They may thus be considered ``anomaly-free'' by analogy with anomaly-free symmetries of quantum field theories, which can be gauged and are thought to admit symmetric deformations to a trivial state (after adding massive fields) \cite{hooft1980naturalness,kapustin2014symmetryprotectedtopologicalphases,kapustin2014anomalous}.

More generally, one can study symmetries which are not-on-site but still send local operators to local (or quasilocal) operators. There are different ways to formulate this. We will focus on quantum cellular-automata (QCA)  \cite{schumacher2004reversiblequantumcellularautomata} which send (strictly) local operators to local operators, with a uniform upper bound on how the support of these operators can grow. The precise definitions are as follows.
\vspace{3mm}
\begin{definition}
    The \textbf{algebra $\CA$ of local operators on $\{\mathcal{H}_x\}_{x \in \Lambda}$} is the algebra of finite sums of finite products of \textbf{single-site operators}, which are of the form
    \[a_x \otimes \left(\bigotimes_{y \neq x} 1_y \right),\]
    where $a_x$ is an operator on $\mathcal{H}_x$, and $1_y$ is the identity on $\mathcal{H}_y$\footnote{We will typically suppress these identity factors from the expressions of operators and just write $a_x$.}. An algebra of this sort will be called a \textbf{local algebra}\footnote{Fermions may be considered in this definition by taking $\CH_x$ to be $\IZ_2$-graded (a supervector space) and taking supertensor products of operators. More generally one can use objects in a suitably braided category in $n$ dimensions.}.
    \end{definition}
    \begin{definition}
    The \textbf{support of a local operator} $a \in \CA$ is the set ${\rm supp}(a)$ of lattice points $x$ such that there exists a single-site operator $b_x$ at $x$ which does not commute with $a$.\footnote{Note the support of a scalar operator is the empty set. For fermionic operators, we phrase this definition in terms of supercommutation.} For a subset $S \subseteq \Lambda$ we denote $\CA(S)$ as the subalgebra of operators whose support is contained in $S$.\footnote{These subalgebras have unital inclusions $\CA(S) \hookrightarrow \CA(S')$ whenever $S \subseteq S'$. This is often abstracted as a ``net of algebras'' \cite{jones2025localtopologicalorderboundary}.}
    \end{definition}
    \begin{definition}
    A \textbf{quantum cellular automaton (QCA)} \cite{schumacher2004reversiblequantumcellularautomata} $\alpha$ on $\CA$ is an $\star$-algebra automorphism, i.e. an invertible map $\alpha:\CA \to \CA$ satisfying
    \begin{enumerate}
        \item $\alpha(x+y) = \alpha(x) + \alpha(y)$
        \item $\alpha(0) = 0$
        \item $\alpha(cx) = c \alpha(x)$
        \item $\alpha(xy) = \alpha(x)\alpha(y)$
        \item $\alpha(1) = 1$
        \item $\alpha(x^\dagger) = \alpha(x)^\dagger$
    \end{enumerate}
    where $x,y \in \CA$ and $c \in \IC$, which further has a \textbf{spread} $r$ such that for all local operators $x \in \CA$,
    \[{\rm supp}(\alpha(x)) \subseteq {\rm supp}(x)^{+r} := \{v \in \Lambda\ |\ {
    \rm dist}(v,{\rm supp}(x)) \le r\},\]
    where dist is the Euclidean distance\footnote{We do not expect that any of our results depend on this choice of metric.}.
\end{definition}
\vspace{3mm}

 \begin{figure}[t]
     \centering
     \includegraphics[width=0.8\linewidth]{figs/circuit-action-2.png}
     \caption{This figure shows a 1-dimensional depth 2 circuit $C = \prod_i V_i\prod_i U_i$ composed of two-site unitaries $U_i$ and $V_i$, acting on a local two-site operator $a$ (green). Only a finite segment of an infinite circuit is shown. However, all the blue circuit elements combine with their inverses in the formal product $C a C^{-1}$, leaving a finite product involving only $a$ and the orange circuit elements (the ``light cone''). Evaluated this way, $C a C^{-1}$ is a well-defined local operator. We see that the map $\alpha_C(a) = C a C^{-1}$ is a $\star$-algebra isomorphism with spread 2, and is a QCA.}
     \label{figFDQCgivesQCA}
 \end{figure}

One can think of a QCA as a generalization of a finite-depth quantum circuit (FDQC). Indeed, as shown in Fig. \ref{figFDQCgivesQCA}, every FDQC defines a QCA. However, QCA are more general than FDQC. For example, lattice translations are a QCA but cannot be expressed as an FDQC\footnote{Note that rotations are not QCA by our definition.}. See \cite{Haah_2022} for more examples.

\begin{figure}
    \centering
    \includegraphics[width=6cm]{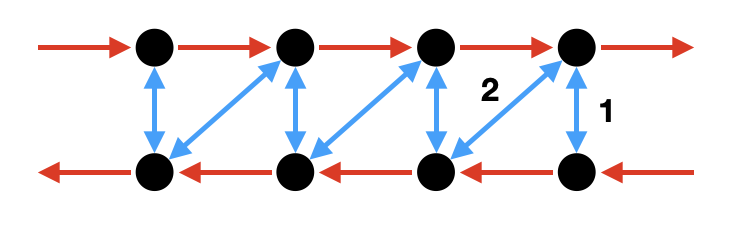}
    \caption{A depth 2 circuit acting on sites (black), consisting of swap gates (blue) performed in the order shown (first vertical, then diagonal), and realizing counter-propagating translations (red). By combining many of these circuits together in the vertical direction, adding more sites, we may obtain a depth 4 circuit which acts as the identity in the interior of a half-space, but realizes a translation on the boundary. Thus, QCA arise naturally at the boundaries of FDQC. With a little more work, one can show every QCA arises this way (see \hyperref[propinvertibilityqcaentangler]{Proposition \ref{propinvertibilityqcaentangler}}).}
    \label{figdoubletranslation}
\end{figure}

Even if one only wishes to study FDQC, QCA arise naturally. Indeed, an FDQC which acts as the identity in the bulk of a half-space defines a QCA acting only along its boundary, and but that QCA may not be expressible as an FDQC (see \hyperref[figdoubletranslation]{Figure \ref{figdoubletranslation}}). We will often be interested in situations like this where we need to extract a boundary action of an FDQC, so it is necessary to enlarge the study of FDQC to QCA for the purpose of studying anomalies. QCA can be composed to yield new QCA, and the inverse of a QCA is also a QCA \cite{Freedman_2020}, making them suitable to consider as a class of unitary symmetries.

Thus, we will study the following generalization of an on-site unitary\footnote{These symmetries are assumed to be complex-linear in the definition of QCA. However, we could also consider anti-unitary QCA which satisfy $\alpha(cx) = c^*\alpha(x)$ with the other axioms unmodified. Our constructions may be easily modified to include such anti-unitary symmetries, although we will not comment on it further.} symmetry:
\vspace{3mm}
\begin{definition}
    A \textbf{QCA $G$-representation on $\CA$} is a collection of QCA $\{\alpha(g)\}_{g \in G}$ on $\CA$ satisfying the $G$ group law under composition:
    \[\alpha(g) \alpha(h) = \alpha(gh).\]
\end{definition}
\vspace{3mm}
\noindent More generally, one might want to consider symmetries $\alpha(g)$ which have a Lieb-Robinson bound, meaning they have a light cone, but only up to some exponentially small tail. Such transformations do not preserve the algebra of local operators, but instead act on a suitable completion of $\CA$ such as quasi-local \cite{bratteli2012operator} or almost-local operators \cite{fredenhagen1982locality}. We expect similar results to ours to hold for such symmetries, but, there are some technical advantages of working with QCA which we use, and recently a lot of theory has been developed for them \cite{Freedman_2020,haah2023invertible}.

Given a QCA-$G$-representation $\alpha(g)$, a natural question is whether we can make a ``change of basis'' by a QCA such that $\alpha(g)$ becomes on-site.
\vspace{3mm}
\begin{definition}
    A QCA $G$-representation $\alpha(g)$ is \textbf{disentanglable} if there exists another QCA $\varepsilon$ such that
\[\varepsilon \circ \alpha(g) \circ \varepsilon^{-1}\]
is an on-site $G$-representation ($\varepsilon$ is independent of $g$). It is \textbf{stably disentanglable} if there exists an on-site unitary $G$-representation $\alpha_0$ on some ancilla Hilbert spaces $\mathcal{H}_x'$, such that
\[\alpha(g) \otimes \alpha_0(g)\]
is disentanglable as  QCA $G$-representation on the larger local algebra built on $\mathcal{H}_x \otimes \mathcal{H}_x'$ (ancillas may be added at every site).
\end{definition}
\vspace{3mm}
\noindent Disentanglable and stably disentanglable QCA $G$-representations behave very much like on-site ones. Others may be considered ``anomalous''.

Given a QCA $G$-representation $\alpha$, we will define an obstruction to (stably) disentangling $\alpha$ which we call the blend anomaly (to be defined in the next section). In particular, we will show
\vspace{3mm}
  \begin{mdframed}[backgroundcolor=blue!5]
\begin{theorem}\label{thmanomalyobstructiontodisent}
    If a QCA $G$-representation is (stably) disentanglable, then the (stable) blend anomaly must be trivial.
\end{theorem}
\end{mdframed}
\vspace{3mm}
\noindent This theorem is proved below in \hyperref[propconstructionofhomotopymap]{Proposition \ref{propconstructionofhomotopymap}} and \hyperref[propstabledisentimpliesanomfree]{Proposition \ref{propstabledisentimpliesanomfree}}. 

We will also see below that this anomaly is the obstruction to finding a ``blend of $G$-representations'' from $\alpha$ to the trivial representation, meaning to finding QCA $G$-representation $\beta$ such that $\beta(g)$ equals $\alpha(g)$ on the interior of a half-space, and equals the identity on the interior of the complementary half-space. This is where the term ``blend anomaly'' comes from. Note that to gauge the symmetry $G$, we would like to express it as a commuting product of local $G$-representations, so the blend anomaly is also an obstruction to gauging.

\subsection{A space of QCA and the homotopy theory of lattice anomalies}\label{subsecspacefirstpass}

To define the blend anomaly, for each local algebra $\CA$ we will define a ``classifying space of QCA'' $\CQ(\CA)$ whose loops correspond to QCA. To see how such a space helps us to understand anomalies, recall that to any finite group $G$ we can associate another space $BG$, the classifying space of $G$ \cite{hatcher2005algebraic,brown2012cohomology} (to see this space in a physics context see \cite{dijkgraaf1990topological}). For discrete $G$, which will be our focus, this space is constructed so that it is connected, $\pi_1 BG = G$, and $\pi_{\ge 2} BG = 0$. Our space $\CQ(A)$ is quite analogous to $BG$.

In fact, a simple corollary of the construction of $\CQ(\CA)$ is that any QCA $G$-representation $\alpha$ on $\CA$ defines a (based\footnote{$BG$ and $\CQ(\CA)$ both have a canonical basepoint and this map and all homotopies we construct are based.}) map
\[\alpha:BG \to \mathcal{Q}(\CA).\]
The idea is that $\alpha$ sends the loop in $BG$ corresponding to $g$ to the loop in $\CQ(\CA)$ corresponding to the QCA $\alpha(g)$. See \hyperref[propconstructionofhomotopymap]{Proposition \ref{propconstructionofhomotopymap}} below.

We can consider the homotopy class $[\alpha]$ of this map. It turns out that all disentanglable representations are homotopy equivalent to the trivial representation (see below), which corresponds to a constant map. This motivates the following definition: 
\vspace{3mm}
\begin{mdframed}[backgroundcolor=blue!5]
\begin{definition}\label{deflatanom}
    The \textbf{blend anomaly} of a QCA $G$-representation $\alpha$ is the (based) homotopy class $[\alpha]$ of the induced map $\alpha:BG \to \mathcal{Q}(\CA)$.
\end{definition}
\end{mdframed}
\vspace{3mm}
\hyperref[thmanomalyobstructiontodisent]{Theorem 1} in the unstable case will then follow from \hyperref[propconstructionofhomotopymap]{Proposition \ref{propconstructionofhomotopymap}}. The stable case of \hyperref[thmanomalyobstructiontodisent]{Theorem 1} will be similar once we define a stable blend anomaly in terms of a stable version of $\CQ(\CA)$ in \hyperref[subsecstabilization]{Section \ref{subsecstabilization}}. In particular it will follow from \hyperref[propstabledisentimpliesanomfree]{Proposition \ref{propstabledisentimpliesanomfree}}. In \hyperref[subseccomptheanom]{Section \ref{subseccomptheanom}} and \hyperref[subsecelsenayak]{Section \ref{subsecelsenayak}} we will discuss methods to compute this anomaly.

To appreciate this definition, we will need to delve slightly deeper into the definition of $\mathcal{Q}(\CA)$. The main construction of this space will be outlined in \hyperref[secconstruction]{Section \ref{secconstruction}}. Here we just give enough details to prove the results above and those that will follow in this section.

Our aim is to construct $\CQ(\CA)$ as an $\infty$-groupoid, which is an $\infty$-category with all morphisms invertible. One can think of an $\infty$-groupoid as a model of a topological space, given by specifying points, paths, paths of paths, and so on, as well as their composition rules. In particular, any topological space defines an $\infty$-groupoid, called its ``fundamental $\infty$-groupoid'', which determines the space up to weak homotopy equivalence\footnote{See \cite{Ara_2013} for a nice introduction to the globular framework we use. This is natural from the point of view of QCA. More commonly one will encounter a simplicial picture of $\infty$-groupoids known as Kan complexes, see \cite{friedman2023elementaryillustratedintroductionsimplicial} for an introduction which is a bit simpler than Ara's and which discusses the fundamental $\infty$-groupoid. In \hyperref[subsecelsenayak]{Section \ref{subsecelsenayak}} we perform calculations which demonstrate some aspects of the equivalence of the simplicial and globular pictures.}. Here we just construct an $\infty$-groupoid directly, without constructing a topological space. We leave constructing a topology on the set of QCAs directly with this same homotopy type to future work.

The $\infty$-groupoid approach has the advantage of being definable purely algebraically in terms of the composition laws of QCA, and we suspect that analogous definitions hold for more general locality-preserving automorphisms. Furthermore, we will see in \hyperref[subsecelsenayak]{Section \ref{subsecelsenayak}} the utility of the $\infty$-groupoid approach for calculating the anomalies on the lattice.

\begin{figure}
    \centering
    \includegraphics[width=0.8\linewidth]{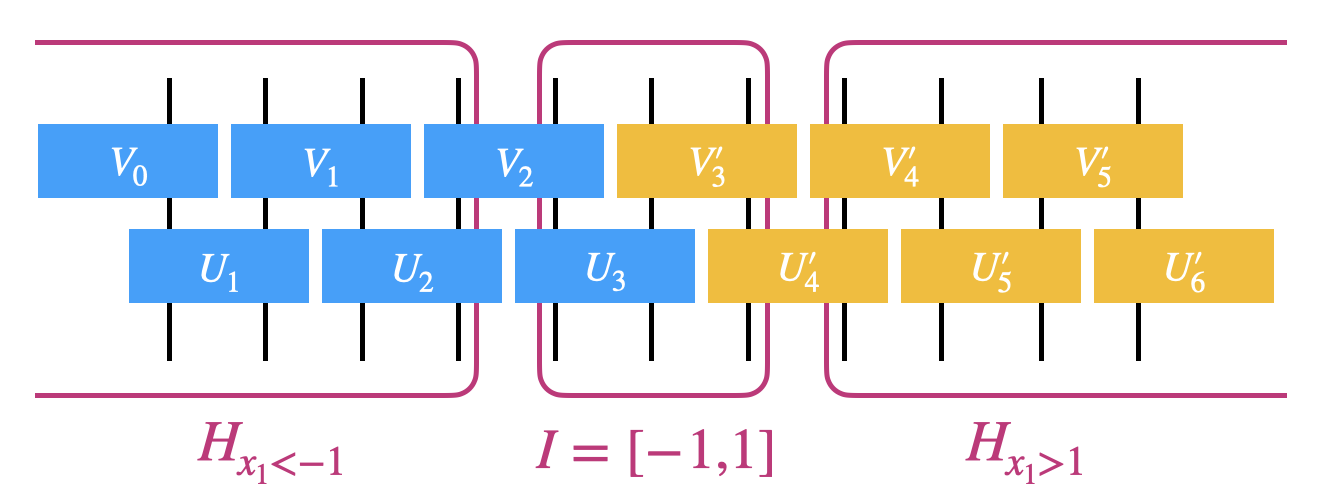}
    \caption{This figure shows a blend between two depth two circuits $C = \prod_i V_i \prod_i U_i$ (blue) and $\prod_i V_i' \prod U_i'$ (orange), induces a blend of the corresponding QCA $\alpha_C$ and $\alpha_{C'}$, with blending interval $I = [-1,1]$ (magenta). For local operators supported in the region $H_{x_1<-1}$, only the blue circuit elements of $C$ will act (compare Fig. \ref{figFDQCgivesQCA}), while for $H_{x_1 > 1}$, only elements of $C'$ act. Any two finite depth circuits admit blends, including to the identity, by generalizing this construction.}
    \label{figcircuitblend}
\end{figure}

The notion of path for QCA we will use is that of a blend \cite{Gross_2012}:
\vspace{3mm}
\begin{definition}
    Given QCA $\alpha$ and $\beta$ on a local algebra $\CA$ defined over the lattice $\mathbb{Z}^d$, a \textbf{blend from $\alpha$ to $\beta$ along the $i$th axis} is a QCA $\gamma$ on $\CA$, denoted\footnote{Note that in this notation $\alpha \equiv_i \beta$ is not the same as $\beta \equiv_i \alpha$. The later is a blend that goes the other way.}
    \[\gamma:\alpha \equiv_i \beta\]
    such that there is a finite interval $I = [y,z] \subset \mathbb{Z}$ (the ``blending interval''), such that for operators for operators $a$ supported in the left region
    \[H_{x_i < y} = \{(x_1,\ldots,x_d) \in \mathbb{Z}^d\ |\ x_i < y \},\]
    \[\gamma(a) = \alpha(a),\]
    while for operators $b$ supported in the right region
    \[H_{x_i > z} = \{(x_1,\ldots,x_d) \in \mathbb{Z}^d\ |\ x_i > z \},\]
    \[\gamma(b) = \beta(b).\]
    Note that the action of $\gamma$ on operators whose support intersects the ``blending region''
    \[\{(x_1,\ldots,x_d) \in \mathbb{Z}^d\ |\ x_i \in I\}\]
    is unconstrained, except that $\gamma$ must be a QCA. Two QCA which admit a blend along the $i$th axis are \textbf{blend equivalent (along the $i$th axes)}.
\end{definition}
\vspace{3mm}
\noindent For example, a QCA given by a finite depth quantum circuit admits blends to the identity QCA along any axis by appropriately discarding circuit elements (see Fig. \ref{figcircuitblend}). Translations however do not admit blends to the identity along the translation axis. In one dimension with finite-dimensional bosonic Hilbert spaces, it is known that all QCAs are blend equivalent to a translation of Hilbert spaces of dimension $n$-bits to the right times a translation of Hilbert spaces of dimension $m$ to the left, where $gcd(n,m) = 1$, with the ratio $n/m \in \mathbb{Q}$ known as the GNVW invariant \cite{Gross_2012}. In three dimensions and above, non-translation blend equivalence classes are thought to exist \cite{Haah_2022}.

The space $\mathcal{Q}(\CA)$ is defined by starting with a single point $\star$, and then taking ``paths'' from $\star$ to $\star$, ie. 1-morphisms ${\rm Hom}(\star,\star)$, to be labeled by QCA. We may represent such a path as a diagram
\[\star \xrightarrow{\alpha} \star.\]
Composition of these paths corresponds to composition of QCA:
\[\star \xrightarrow{\alpha} \star \xrightarrow{\beta} \star = \star \xrightarrow{\beta \alpha} \star.\]
So far, this defines a 1-groupoid with a single object, which is a model for the classifying space $BQCA(\CA)$, where $QCA(\CA)$ is the group of QCA on $\CA$ with composition. This space is too coarse for us, since it does not take into account blend equivalence of QCA. Therefore, we must continue the construction to higher groupoids.

In the ``globular'' presentation of $\infty$-groupoids \cite{Ara_2013}, ``Paths of paths'' are represented as ``2-globes'', which as a diagram look like
\begin{equation}\label{eqn2globe}
\begin{tikzcd}[column sep=huge]
\star \arrow[r, bend left=40, "\alpha"{name=f}] 
  \arrow[r, bend right=40, "\beta"'{name=g}] 
  & \star
  \arrow[Rightarrow, from=f, to=g, "\gamma"]
\end{tikzcd}
\end{equation}
and are labeled by blends $\gamma:\alpha \equiv_1 \beta$ from $\alpha$ to $\beta$ along the 1st axis (note that this notation keeps track of the orientation of the axis). One can think of 2-globes as a special class of 2-morphisms of the $\infty$-groupoid, ${\rm Hom}(\alpha,\beta)$. The idea of the globular formalism is that all 2-morphisms are expressible in terms of 2-globes \cite{Ara_2013,baez2004higherdimensionalalgebrav2groups} (see \hyperref[subsecelsenayak]{Section \ref{subsecelsenayak}} for some examples in the calculation of anomalies).

Globes may be composed in two ways. If we compose them ``horizontally'', it corresponds to ordinary QCA composition.
\begin{equation}\label{eqnhorizontal2globecomposition}
\begin{tikzcd}[column sep=huge]
\star \arrow[r, bend left=40, "\alpha_1"{name=f}] 
  \arrow[r, bend right=40, "\beta_2"'{name=g}] 
  & \star
  \arrow[Rightarrow, from=f, to=g, "\gamma_1"] \arrow[r, bend left=40, "\alpha_2"{name=f2}] 
  \arrow[r, bend right=40, "\beta_2"'{name=g2}] &  \arrow[Rightarrow, from=f2, to=g2, "\gamma_2"]\star
\end{tikzcd} = \begin{tikzcd}[column sep=huge]
\star \arrow[r, bend left=40, "\alpha_2\alpha_1"{name=f}] 
  \arrow[r, bend right=40, "\beta_2\beta_1"'{name=g}] 
  & \star
  \arrow[Rightarrow, from=f, to=g, "\gamma_2\gamma_1"]
\end{tikzcd}
\end{equation}
Note that the source and target make sense, ie. $\gamma_2 \gamma_1:\alpha_1 \alpha_2 \equiv_1 \beta_1 \beta_2$.

To compose ``vertically'', we need a
\vspace{3mm}
\begin{definition}\label{defblendcomposition}
    Suppose $\beta_{12}:\alpha_1 \equiv_i \alpha_2$ is a blend from $\alpha_1$ to $\alpha_2$ along the $i$th axis, and $\beta_{23}:\alpha_2 \equiv_i \alpha_3$ is a blend from $\alpha_2$ to $\alpha_3$ along the $i$th axis. We define the \textbf{1-blend composition} to be the QCA
    \[\label{eqnblendcomposition}\beta_{23} \circ_1 \beta_{12} = \beta_{23}\alpha_2^{-1} \beta_{12}.\]
\end{definition}
\vspace{3mm}
\noindent It is easy to check that $\beta_{23} \circ_1\beta_{12}$ is a blend $\alpha_1 \equiv_i \alpha_3$ whose blending region is the convex hull of the blending regions of $\beta_{12}$ and $\beta_{23}$. Thus the ``vertical composition'' is defined by
\begin{equation}\label{eqnvertical2globecomposition}
\begin{tikzcd}[column sep=huge]
\star \arrow[r, bend left=40, "\alpha_1"{name=f}] 
  \arrow[r, bend right=40, "\alpha_2"'{name=g}] 
  & \star
  \arrow[Rightarrow, from=f, to=g, "\gamma_1"] \\
  & \\
  \star \arrow[r, bend left=40, "\alpha_2"{name=f2}] 
  \arrow[r, bend right=40, "\alpha_3"'{name=g2}]  & \star \arrow[Rightarrow, from=f2, to=g2, "\gamma_2"]
\end{tikzcd} = \begin{tikzcd}[column sep=3cm]
\star \arrow[r, bend left=50, "\alpha_1"{name=f}] 
  \arrow[r, bend right=50, "\alpha_3"'{name=g}] 
  & \star
  \arrow[Rightarrow, from=f, to=g, "\gamma_2\circ_1 \gamma_1"]
\end{tikzcd}
\end{equation}
Note to be composable, the edges which are glued must match.

We build higher dimensional globes by taking blends of blends along successive axes (see \hyperref[defglobularset]{Definition \ref{defglobularset}}). When all the axes of $\mathbb{Z}^d$ are used up, the highest level of globes are defined using local unitary operators. To define an $\infty$-groupoid, we need to define compositions of all higher globes along each direction and prove a number of coherence relations. In \hyperref[secconstruction]{Section \ref{secconstruction}} we give a general definition of the composition relations, but postpone proving the coherence relations to future work \cite{followup}. Homotopy groups of $\CQ(\CA)$ correspond to blend equivalence classes of QCA, see \hyperref[prophomotopycalc]{Proposition \ref{prophomotopycalc}}.

Taking this construction, we can show how to get the map $\alpha:BG \to \CQ(\CA)$ from a QCA $G$-representation, and show that in the disentanglable case it is null-homotopic.
\vspace{3mm}
\begin{mdframed}[backgroundcolor=blue!5]
\begin{proposition}\label{propconstructionofhomotopymap}
    Let $\alpha(g)$ be a QCA $G$-representation on the local algebra $\CA$. Recall $BG$ is the $\infty$-groupoid with a single object $\star$, and 1-morphisms ${\rm Hom}(\star,\star) = G$, with composition given by the group law, and all higher morphisms are the identity. There is a map of $\infty$-groupoids
    \[\alpha:BG \to \CQ(\CA)\]
    mapping the unique object $\star \in BG$ to the unique object $\star \in \CQ(\CA)$, mapping 1-morphisms as
    \[\alpha(\star \xrightarrow{g} \star) = (\star \xrightarrow{\alpha(g)} \star),\]
    and mapping all higher morphisms to the identity.
    
    Moreover, if $\alpha(g)$ is disentanglable, then $\alpha:BG \to \CQ(\CA)$ is homotopy equivalent to the identity.
\end{proposition}
\end{mdframed}
\begin{proof}
    For the existence of the map $\alpha$, we just need to check that the rules so defined respect the composition laws. For instance, at the 2-morphism level, we want to use the identity blend $id:\alpha(g) \alpha(h) \equiv_1 \alpha(gh)$, which exists as long as the source and target are the same. This is ensured by $\alpha(g)$ being a $G$-representation. All higher composition laws in $BG$ correspond to identities in $G$, and these are automatically satisfied as well. Note that since none of the non-identity higher morphisms are activated, $\alpha$ factors as
    \[BG \to B\text{QCA}(\CA) \xrightarrow{i} \CQ(\CA)\]
    where $\text{QCA}(\CA)$ is the group of all QCA on $\CA$ with composition, and $i$ extends by the identity on all higher morphisms.

    Now suppose that $\alpha(g)$ is disentanglable, so there exists $\varepsilon$ such that $\varepsilon \alpha(g) \varepsilon^{-1} = \sigma(g)$ is an on-site representation. Then we may choose a truncation $\tilde \sigma(g)$ to half of the system along the first axis to obtain a set of blends $\tilde\sigma(g):\sigma(g) \equiv_1 1$ satisfying the group law. Applying $\varepsilon$, we obtain as well
    \[\tilde \alpha(g) = \varepsilon^{-1} \tilde \sigma(g) \varepsilon:\alpha(g) \equiv_1 1\]
    satisfying the group law. We then apply \hyperref[lemmaGblend]{Lemma \ref{lemmaGblend}} to obtain the result.
\end{proof}

\begin{mdframed}[backgroundcolor=blue!5]
\begin{lemma}\label{lemmaGblend}
    Suppose $\alpha$ and $\beta$ are QCA $G$-representations which admit a blend along the first axis as $G$-representations, i.e. a set of blends $\gamma(g):\alpha(g)\equiv_1\beta(g)$ such that
    \[\gamma(g) \gamma(h) = \gamma(gh).\]
    Then $\alpha:BG \to \CQ(\CA)$ and $\beta:BG \to \CQ(\CA)$ are homotopic.
\end{lemma}
\end{mdframed}
\begin{proof}
    A homotopy can be thought of as a map from $\gamma:BG \times [0,1] \to \CQ(\CA)$ such that restricting to $BG \times 0$ we get $\alpha$, and restricting to $BG \times 1$ we get $\beta$. We can treat this homotopy as well as a map of $\infty$-groupoids.
    
    The globes of $BG \times [0,1]$ contain the globes of $BG \times 0$ and the globes of $BG \times 1$, as well as for each $g$ a special 2-globe, which we can map to our blend $ \gamma(g)$:
    \begin{equation}
    \begin{tikzcd}[column sep=huge]
    \star \arrow[r, bend left=40, "g \times 0"{name=f}] 
      \arrow[r, bend right=40, "g\times 1"'{name=g}] 
      & \star
      \arrow[Rightarrow, from=f, to=g, "!"]
    \end{tikzcd} \mapsto \begin{tikzcd}[column sep=huge]
    \star \arrow[r, bend left=40, "\alpha(g)"{name=f}] 
      \arrow[r, bend right=40, "\beta(g)"'{name=g}] 
      & \star
      \arrow[Rightarrow, from=f, to=g, "\gamma(g)"]
    \end{tikzcd}
    \end{equation}
    The source globes satisfy horizontal composition laws given by the group law. Since $\gamma(g)$ satisfy the group law as well, this defines a map of $\infty$-groupoids by extending to the identity on all higher globes, yielding the required homotopy.
\end{proof}

\subsection{Stabilization and an $\Omega$-spectrum}\label{subsecstabilization}

In this section we extend the definition of QCA representation and blend anomaly to study obstructions to stable disentangling. In doing so, we construct an $\Omega$-spectrum associated to QCA with a fixed algebra of single-site operators. As well as an interesting result in its own right, it allows many technical simplifications in the question of the homotopy class of a QCA representation. For example it implies that these homotopy classes, ie. the ``stable'' blend anomalies, form an abelian group.

To motivate the definition, suppose we consider a blend from the identity to the identity along the 1st axis, $\beta: 1 \equiv_1 1$. Such a blend is a QCA $\beta$ which acts as the identity on operators supported outside of the blending region
\[I \times \IZ^{d-1} \subset \IZ^d.\]
If the spread of $\beta$ is $s$, then it is simple to show that $\beta$ preserves the subalgebra $\CA(I^{+s} \times \IZ^{d-1})$ of operators supported in
\[I^{+s} \times \IZ^{d-1},\]
where $I^{+s}$ is $\{x \in \IZ\ |\ \text{dist}(x,I) \le s\}$. Furthermore, $\beta$ is determined by its action on $\CA(I^{+s} \times \IZ^{d-1})$. This is a local algebra on $\IZ^{d-1}$ since the finite factor $I^{+s}$ can be compressed into a single site of $\IZ^{d-1}$, growing its dimension by a finite amount\footnote{It is dangerous to consider infinite compressions of this sort, as the site Hilbert spaces would become uncountably infinite dimensional and are in fact \emph{not} Hilbert spaces at all, which is a classic difficulty in many-body quantum mechanics in infinite volume \cite{bratteli2012operator}. Countably infinite dimensional site Hilbert spaces are admissible in our definitions, however.}. Thus, $\beta$ is equivalent to a QCA on $\CA(I^{+s} \times \IZ^{d-1})$.

\begin{figure}
    \centering
    \includegraphics[width=8cm]{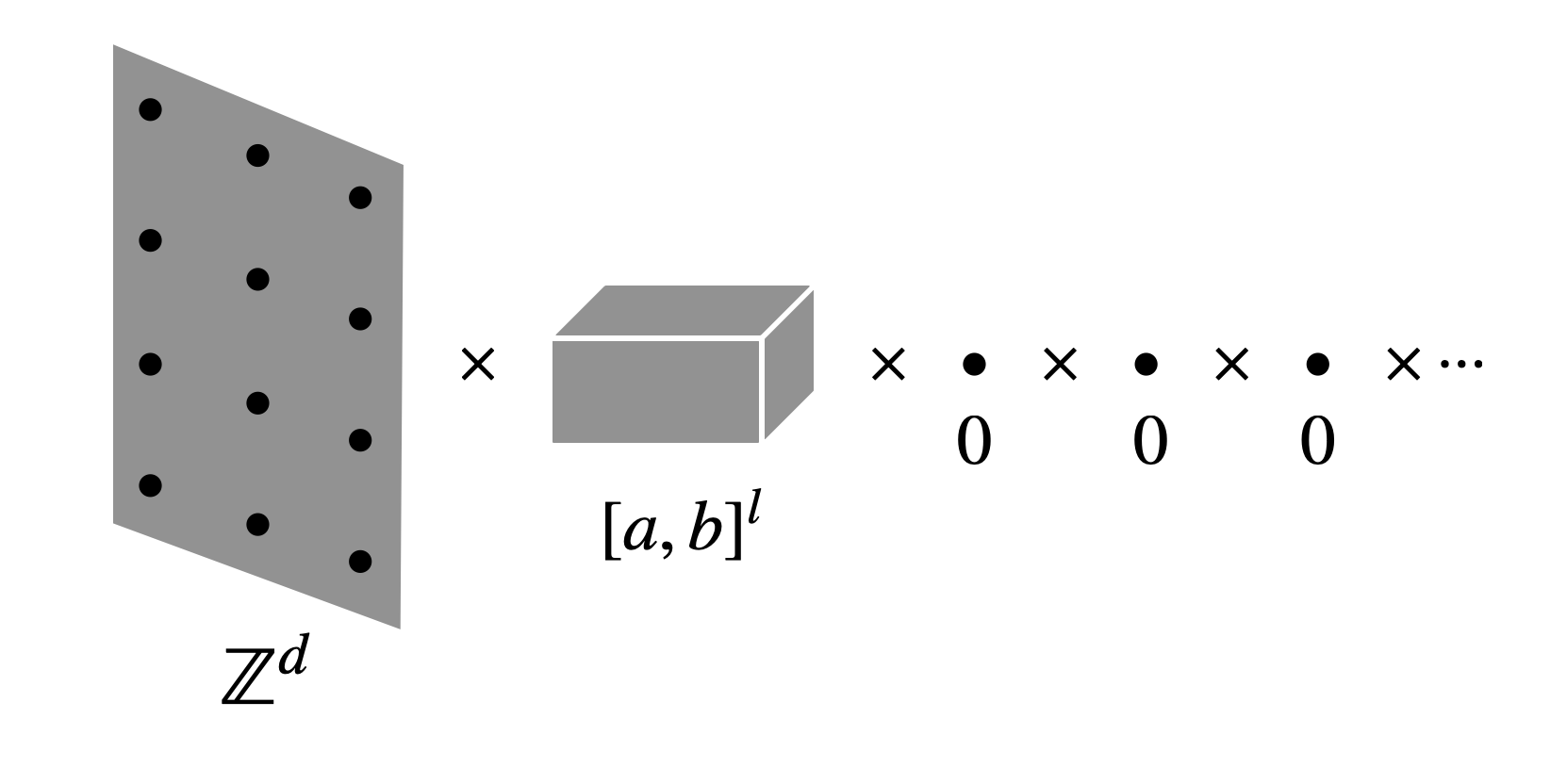}
    \caption{The domain $D(\alpha)$ of a stable $d$-QCA $\alpha$ is illustrated. $\alpha$ is required to preserve the subalgebra of operators supported in $D(\alpha)$ and act by the identity on all operators supported outside $D(\alpha)$. It is thus determined by a QCA on the ``thickened'' $d$-dimensional space $D(\alpha)$. This allows us to freely utilize ancillas and is crucial for obtaining an $\Omega$ spectrum of QCA.}
    \label{figstableqca}
\end{figure}

When we compose blends from the identity to itself, they behave almost as $d-1$-dimensional QCA, except that the blend interval $I$ in general will grow with each composition. This motivates the following definition.
\vspace{3mm}
\begin{mdframed}[backgroundcolor=blue!5]
\begin{definition}\label{defstableqca}
    Let $\CH$ be a Hilbert space which will be the Hilbert space associated with each site of our lattice. Call this the \textbf{site Hilbert space}. We will assume this Hilbert space is finite-dimensional, although this and many definitions extend to general Hilbert spaces. For fermions we use a super Hilbert space.
    
    We define the local algebra $\CA^\omega_\CH$ on the infinite-dimensional lattice $\mathbb{Z}^\omega$ with axes ordered by positive integers $1,2,3,\dots$ in the usual way: for each region $R \subseteq \IZ^\omega$, $\CA^\omega_\CH(R) = \bigotimes_{\vec x \in R} M(\CH)$ is the algebra of operators whose support is contained in $R$, where $M(\CH)$ the $\star$-algebra of endomorphisms of $\CH$ (typically finite matrices). We embed $\IZ^\omega$ in $\mathbb{R}^\omega$ equipped with any product metric, and we consider QCA on this algebra.

    A \textbf{stable $d$-QCA} on $\CA^\omega_\CH$ is a QCA on $\CA^\omega_\CH$ such that there exist an integer $l$ and and interval $[a,b]$ which define the \textbf{domain of $\alpha$}:
    \[D(\alpha) = \left\{(x_1,\ldots) \in \IZ^\omega\ \Bigg|\ \begin{cases}
        x_i \text{ unconstrained} & 1 \le i \le d \\
        a \le x_i \le b & d < i \le d+l \\
        x_i = 0 & i > d+l
    \end{cases}\right\} \\
    = \IZ^d \times [a,b]^l \times 0 \times 0 \times \cdots\]
    such that
    \begin{enumerate}
        \item $\alpha$ preserves the subalgebra $\CA^\omega_\CH(D(\alpha))$ of operators supported in $D(\alpha)$.
        \item $\alpha$ acts by the identity on operators supported outside $D(\alpha)$.
    \end{enumerate}
     $\alpha$ is thus determined by a QCA on $\CA^\omega_\CH(D(\alpha))$. This is illustrated in Fig. \ref{figstableqca}.
     
     A \textbf{stable local unitary} is an equivalence class of an integer $l$ and an interval $[a,b]$ and a local unitary operator acting on the (finite dimensional) Hilbert space
     \[\bigotimes_{\vec x \in [a,b]^l} \CH,\]
     which is equivalent to its extension by the identity under inclusions $[a,b]^l \subset [a',b']^{l'}$ for $l' \ge l$, $a' \le a$, and $b' \ge b$. In other words, the group of stable local unitaries is the direct limit of the unitary groups $U([a,b]^l)$ under these inclusions. Stable local unitaries also define stable 0-QCA on $\IZ^\omega$ with domain $[a,b]^l$ by the adjoint action of these unitaries.

    Given two stable $d$-QCA, with $l_1,a_1,b_1$ and $l_2,a_2,b_2$, we can always consider them on their ``common domain'' taking $l = \max(l_1,l_2)$, $a = \min(a_1,a_2)$, $b = \max(b_1,b_2)$, by extending by the identity. They can then be composed on their common domain to give a stable $d$-QCA. Composition is associative and invertible, with the inverse of a stable $d$-QCA given by the QCA $\alpha^{-1}$ on the same domain.
    
    A \textbf{blend $\gamma:\alpha \equiv_i \beta$ between stable $d$-QCA} is a stable $d$-QCA whose domain contains the common domain of $\alpha$ and $\beta$, and equals $\alpha$ (resp. $\beta$) to the left (right) of a blending interval along the $i$th axis.
    
    All of this gives enough structure to define an \textbf{$\infty$-groupoid of stable $d$-QCA} $\mathcal{Q}^d_\CH$. This $\infty$-groupoid is defined by taking $k+2$-globes to be blends along the $d-k$th axis for $0 \le k \le d-1$ and $d+2$-globes to be stable local unitaries. See \hyperref[secconstruction]{Section \ref{secconstruction}} for more details.
\end{definition}
\end{mdframed}
\vspace{3mm}

Consider the local algebra $\CA_\CH^d$ on $\IZ^d$ with site Hilbert space $\CH$. We can consider any QCA $\alpha$ on $\CA_\CH^d$ to be a stable $d$-QCA with the domain ($l = 0$)
\[D(\alpha) = \IZ^d \times 0 \times 0 \times \cdots.\]
This defines a map of $\infty$-groupoids
\[\CQ(\CA_\CH^d) \to \CQ^d,\]
which can be thought of as a ``stabilization'', allowing the addition of ancillas to the study of QCA. 
In particular we have the following easy corollary of the definition (see \hyperref[prophomotopycalc]{Proposition \ref{prophomotopycalc}}):
\vspace{3mm}
\begin{mdframed}[backgroundcolor=blue!5]
\begin{theorem}\label{thmspaceofstableqca}
    The $\infty$-groupoid $\CQ^d_\CH$ of stable $d$-QCA with site Hilbert space $\CH$ has fundamental group (for $d \ge 1$)
    \[\pi_1(\CQ^d_\CH) = \{ (\alpha,n)\ |\ n\in \IZ_{\ge 1},\ \alpha \text{ is a QCA on }\CA^d_{\CH^{\otimes n}}\}/\sim\]
    where $\sim$ is the equivalence relation generated by the two basic equivalences
    \begin{enumerate}
        \item $(\alpha,n) \sim (\beta,n)$ if there is a blend along the $d$th axis from $\alpha$ to $\beta$ as QCA on $\CA^d_{\CH^{\otimes n}}$
        \item $(\alpha,n) \sim (\alpha \otimes 1,n+m)$ where $\alpha \otimes 1$ is the QCA on $\CA^d_{\CH^{\otimes(n+m)}} = \CA^d_{\CH^{\otimes n}} \otimes \CA^d_{\CH^{\otimes m}}$ which acts as $\alpha$ on the first factor and the identity on the second.
    \end{enumerate}
    Let $[\alpha,n]$ denote the equivalence classes. The (abelian) group structure on $\pi_1(\CQ^d_\CH)$ is given by
    \[ [\alpha,n] + [\beta,m] = [\alpha \otimes \beta,n+m] \]
    which is also equal to
    \[ [\alpha,n] + [\beta,n] = [\alpha \circ \beta,n] .\]
    We have
    \[ [\alpha^\text{rev},n] = [\alpha^{-1},n] = -[\alpha,n],\]
    where $\alpha^\text{rev}$ is the reversal of $\alpha$, obtained by $\alpha^\text{rev} = R \circ \alpha \circ R^{-1}$, where $R$ is any permutation of sites, acting on $\CA^d_{\CH^{\otimes n}}$ by some reflection over the $d$th axis ($R$ is not a QCA but $R \circ \alpha \circ R^{-1}$ is).
\end{theorem}
\end{mdframed}
\vspace{3mm}

We likewise define a notion of stable QCA symmetry.
\vspace{3mm}
\begin{mdframed}[backgroundcolor=blue!5]
\begin{definition}
    A \textbf{stable $d$-QCA $G$-representation (with site Hilbert space $\mathcal{H}$)} is a collection of stable $d$-QCA $\{\alpha(g)\}_{g \in G}$ with site Hilbert space $\mathcal{H}$ satisfying the $G$ group law under composition:
    \[\alpha(g) \alpha(h) = \alpha(gh).\]
\end{definition}
\end{mdframed}
\vspace{3mm}

These definitions allow the easy inclusion of ancillas using the extra directions, allowing us also to characterize stably disentanglable $G$-representations. Indeed, a QCA $G$-representation $\alpha$ on $\mathbb{Z}^d$ with site Hilbert space $\mathcal{H}$ defines a stable $d$-QCA $G$-representation, and therefore a map
\[\alpha:BG \to \mathcal{Q}^d_\CH.\]
As before, we find that if $\alpha$ is stably-disentanglable, then $\alpha$ is homotopic to the constant map (see \hyperref[propstabledisentimpliesanomfree]{Proposition \ref{propstabledisentimpliesanomfree}} below). This motivates the following definition.
\vspace{3mm}
\begin{mdframed}[backgroundcolor=blue!5]
\begin{definition}\label{deflatanomstab}
    The \textbf{stable blend anomaly} of a QCA $G$-representation $\alpha$ on $\mathbb{Z}^d$ with site Hilbert space $\mathcal{H}$ is the (based\footnote{Below we will show $\CQ^d_\CH$ has an $H$-space structure, being a component of a loop space. Therefore, it does not matter if we consider homotopies of based or unbased maps. See Proposition 1.4.3 of \cite{may2012more}.}) homotopy class $[\alpha]$ of the induced map $\alpha:BG \to \mathcal{Q}^d_\CH$.
\end{definition}
\end{mdframed}
\vspace{3mm}
\hyperref[thmanomalyobstructiontodisent]{Theorem 1} now follows easily from the following proposition.

\vspace{3mm}
\begin{mdframed}[backgroundcolor=blue!5]
\begin{proposition}\label{propstabledisentimpliesanomfree}
    Let $\CH$ be a Hilbert space. Suppose $\alpha$ is a QCA $G$-representation on the local algebra $\CA^d_\CH$ on $\IZ^d$ with site Hilbert space $\CH$. Suppose further that $\alpha$ is stably disentanglable upon introducing ancillas in $\CH$ (a finite number per site). Then the associated map
    \[\alpha:BG \to \CQ^d_\CH\]
    is null-homotopic, i.e. the stable blend anomaly $[\alpha]$ vanishes and \hyperref[thmanomalyobstructiontodisent]{Theorem \ref{thmanomalyobstructiontodisent}} holds.
\end{proposition}
\end{mdframed}
\begin{proof}
    The introduction of $n$ ancillas per site amounts to considering the enlargement
    \[\alpha'(g) = \alpha(g) \otimes 1\]
    acting in $\CA^d_{\CH^{\otimes (n+1)}} = \CA^d_\CH \otimes (\CA^d_{\CH})^{\otimes n}$. Considered as stable $d$-QCA, $\alpha(g)$, acting in the domain
    \[\IZ^d \times \{0\} \times \cdots\]
    and $\alpha'(g)$, acting in the domain
    \[\IZ^d \times \{0,1,\ldots,n\} \times \{0\} \times \cdots,\]
    which extends $\alpha(g)$ by the identity, these stable $d$-QCA are equal. Thus, if $\alpha'$ is disentanglable, the same argument as in \hyperref[propconstructionofhomotopymap]{Proposition \ref{propconstructionofhomotopymap}} constructs a set of stable blends $\alpha(g) \equiv_1 1$ satisfying the group law, and this yields a nullhomotopy of $\alpha:BG \to \CQ^d_\CH$ by \hyperref[lemmaGblend]{Lemma \ref{lemmaGblend}} (this lemma has the same proof in the stable case).
\end{proof}

There is another fortuitous consequence of this definition. In particular, we find that a stable blend of $d$-QCA along the $d$th axis from the identity to itself is the same thing as a stable $d-1$-QCA. Consider the space $\Omega_\star \CQ^d_\CH$ of loops beginning and ending at the distinguished point $\star$ (the unique object of $\CQ^d_\CH$ as we have constructed it). This space itself is an $\infty$-groupoid whose objects are endomorphisms ${\rm Hom}(\star,\star)$, ie. stable $d$-QCAs, and whose 1-morphisms are 2-morphisms, and so on. Thus $\Omega_\star \CQ^d$ is a space of QCA. If we consider loops in this space based at the identity, we get another space $\Omega_{\rm id} \Omega_\star \CQ^d_\CH$. From the globular presentation, this space is precisely $\Omega_\star \CQ^{d-1}_\CH$:
\vspace{3mm}
\begin{mdframed}[backgroundcolor=blue!5]
\begin{theorem}\label{thmomegaspectrum}
    The space $\Omega_\star \mathcal{Q}^d_\CH$ of stable $d$-QCA for a fixed site Hilbert space $\mathcal{H}$ satisfies
    \[\Omega_{\rm id} \Omega_\star \CQ^d_\CH \cong \Omega_\star\CQ^{d-1}_\CH.\]
    In other words, $\Omega_\star\CQ^d_\CH$ defines an $\Omega$-spectrum.
\end{theorem}
\end{mdframed}
\vspace{3mm}
\noindent This means in particular that the homotopy classes of maps from a space $X$ to $\mathcal{Q}^d$ satisfies the axioms of a (reduced) generalized cohomology theory $\mathcal{Q}^d(X)$. In particular, they form an abelian group, and the stable blend anomaly is an element of the group $\mathcal{Q}^d(BG)$. This is closely related to the $\Omega$-spectrum conjecture for invertible gapped phases, a topic we return to in \hyperref[subsecrelationtoinvertiblestates]{Section \ref{subsecrelationtoinvertiblestates}}. 

\begin{figure}
    \centering
    \includegraphics[width=8cm]{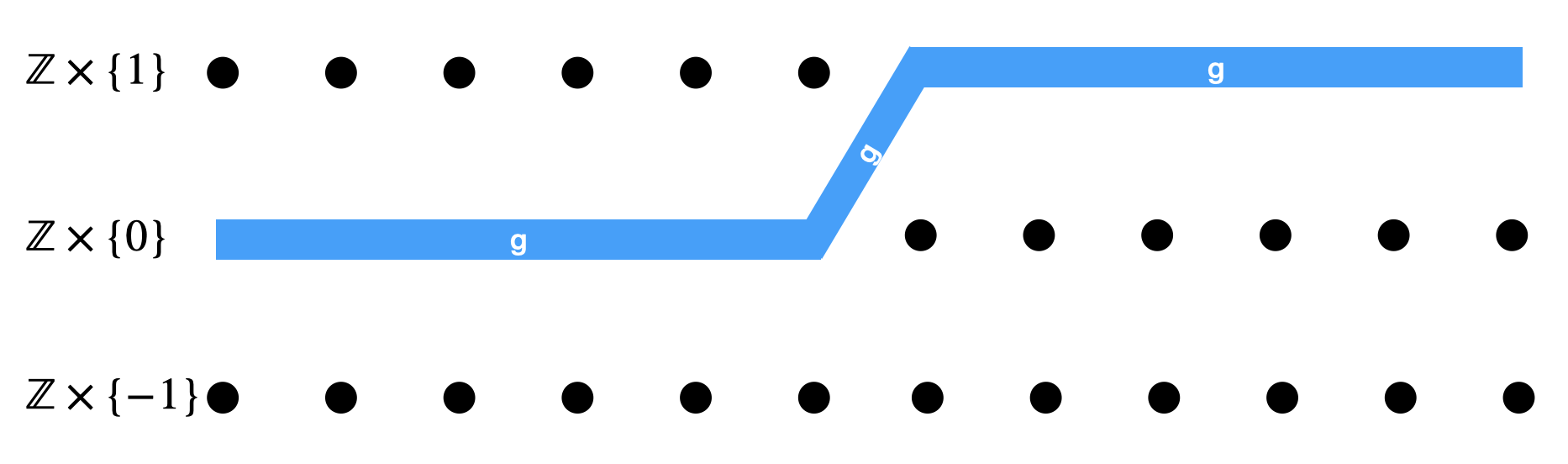}
    \caption{A ``layer shifting'' blend of $G$-representations along the first axis (other axes not shown), may be constructed by conjugating a QCA $G$-representation by layer swaps between layer $0$ and $1$ on a half-space of $\IZ^d$. The result is shown. By construction, these blends satisfy the group law. By \hyperref[lemmaGblend]{Lemma \ref{lemmaGblend}}, this defines a homotopy of the corresponding maps $BG \to \CQ^d_\CH$.}
    \label{figshiftinghomotopy}
\end{figure}

\begin{figure}
    \centering
    \includegraphics[width=8cm]{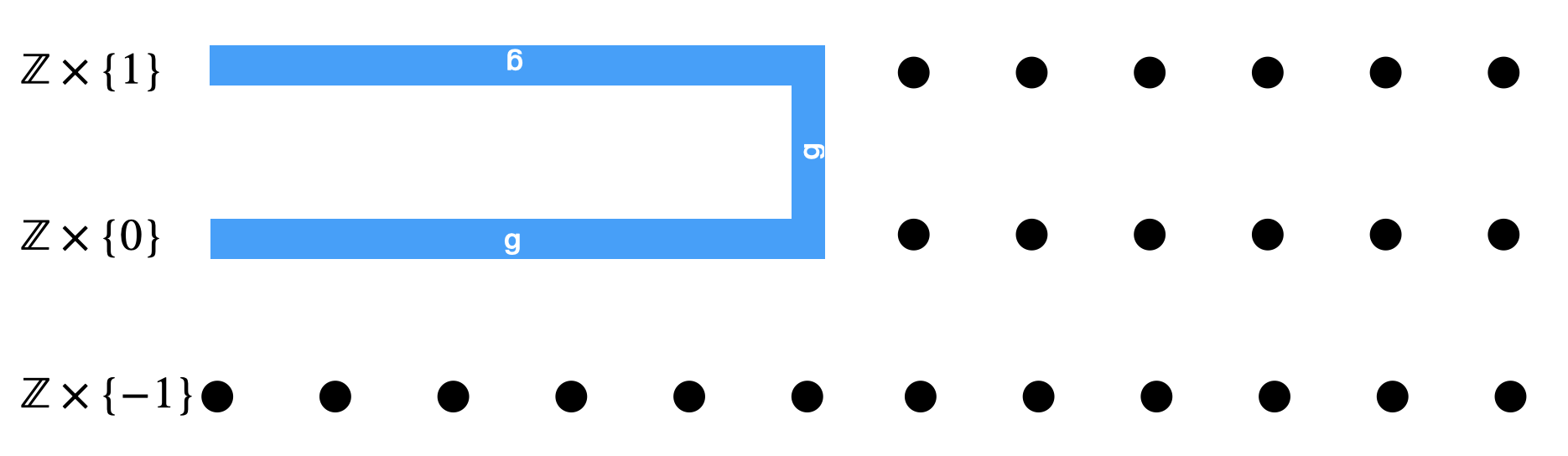}
    \caption{A ``folding'' blend from the QCA $G$-representation $\alpha(g) \otimes \alpha^\text{rev}(g)$ to the identity, which again may be constructed by a suitable permutation of sites. By \hyperref[lemmaGblend]{Lemma \ref{lemmaGblend}}, this defines a homotopy of the corresponding maps $BG \to \CQ^d_\CH$.}
    \label{figsflippinghomotopy}
\end{figure}

We find the following:
\vspace{3mm}
\begin{mdframed}[backgroundcolor=blue!5]
\begin{cor}\label{corollaryabelaingroup}
    The stable blend anomaly for a stable $d$-dimensional QCA $G$-representation $\alpha$ is an element of an abelian group
    \[ [\alpha] \in \CQ^d_\CH(BG).\]
    If we have a second such representation $\beta$, and consider the tensor product representation $[\alpha \otimes \beta]$, then
    \[ [\alpha \otimes \beta] = [\alpha] + [\beta].\]
    Let $\alpha^{\rm rev}$ be $\alpha$ reflected anywhere along the first axis. Then
    \[ [\alpha^{\rm rev}] = - [\alpha].\]
\end{cor}
\end{mdframed}
\vspace{3mm}
\begin{proof}

    The additive structure of the maps is abstractly defined by
    \[(\alpha+\beta): BG \xrightarrow{\Delta} BG \times BG \xrightarrow{(\alpha,\beta)} \CQ^d \times \CQ^d \xrightarrow{*} \CQ^d,\]
    where $\Delta$ is the diagonal map, and $*$ is the product on $\CQ^d$, which comes from $\CQ^d$ being equivalent to the full subcategory of the identity object (constant loop) in $\Omega_\star\CQ^{d+1}$, or equivalently $B \Omega_{id} \Omega_\star \CQ^{d+1}$. Since $\Omega_\star \CQ^{d+1}$ is an infinite loop space (it is $\Omega_{\rm id}^n \Omega_\star \CQ^{d+n}_\CH$ for all $n \ge 0$), this product gives an abelian group structure to the homotopy classes of maps $BG \to \CQ^d$, since loops may be concatenated \cite{may1999concise}.
    
    To be more concrete, on the 1-morphism level the product on $\CQ^d$ is given by the composition of blends $1 \equiv_1 1$, which is stable $d$-QCA composition. So the map $(\alpha+\beta)$ sends the 1-morphism $\star \xrightarrow{g} \star$ of $BG$ to $\alpha(g) \beta(g)$.
    
    In general, $\alpha(g)\beta(g)$ is \emph{not} a QCA $G$-representation unless $\alpha(g)$ and $\beta(h)$ commute for all $g, h \in G$. Indeed, in this case the higher morphisms of $BG$ will activate non-trivial QCA with the combinatorial structure of the diagonal map $\Delta$ encoding the non-trivial commutators. So although $\alpha(g)\beta(g)$ is not a QCA $G$-representation, this higher data makes it a QCA $G$-representation \emph{up to (coherent) homotopy}.

    In the special case that all the commutators are trivial, $\alpha(g)\beta(g)$ is a QCA $G$-representation, and all the higher data collapses to yield identity morphisms. This occurs for example if $\alpha$ and $\beta$ have disjoint domains. In this case, the operation of composition is the same as stacking, and we have $(\alpha+\beta) = \alpha \otimes \beta$.
    
    We can in fact always choose a homotopy representative $[\alpha'] = [\alpha]$ such that $\alpha'$ and $\beta$ have disjoint domains, by a ``layer shifting homotopy'' as shown in Fig. \ref{figshiftinghomotopy}. This produces blends of $G$-representations which themselves are $G$-representations, so by \hyperref[lemmaGblend]{Lemma \ref{lemmaGblend}}, we get homotopies. Since the homotopy class $[\alpha + \beta]$ only depends on the homotopy classes $[\alpha]$, $[\beta]$,
    \[ [\alpha + \beta] = [\alpha'+\beta] = [\alpha'\beta] = [\alpha \otimes \beta].\]

    To show $[\alpha^\text{rev}] = -[\alpha]$, we show $[\alpha] + [\alpha^\text{rev}] = 0$. For this, we observe that we can blend $\alpha \otimes \alpha^\text{rev}$ to the identity representation through blends which are $G$-representations by folding, as in Fig. \ref{figsflippinghomotopy}. The result then follows from \hyperref[lemmaGblend]{Lemma \ref{lemmaGblend}}.

\end{proof}

From now on, we will mostly focus on stable anomalies, and say ``unstable'' when referring to the non-stabilized version built on $\CQ(\CA)$.

\subsection{Anomaly Indices and the Else-Nayak Index}\label{subseccomptheanom}

This abstract structure is very satisfying, but it leaves us with a practical problem. Given a QCA $G$-representation, how do we compute its blend anomaly?

From the homotopy point of view, we want to ask whether the map
\[\alpha:BG \to \CQ^d_\CH\]
(or the unstable version) is null-homotopic. There is a method in algebraic topology known as obstruction theory which allows us to answer questions like this. We will present the version based on a ``homotopy lifting problem''. This version is simpler to reason about abstractly, although for computations in \hyperref[subsecelsenayak]{Section \ref{subsecelsenayak}} we will also discuss the equivalent ``homotopy extension problem'' \cite{whitehead2012elements,hatcher2005algebraic,may1999concise}, which more closely resembles the construction of Else and Nayak \cite{Else_2014}.

We begin by introducing the ``Whitehead tower'' of $\CQ^d_\CH$, which is a sequence of fibrations (in $\infty$-groupoids \cite{goerss1999simplicial})
\[\label{eqnwhiteheadtower}\cdots \to \CQ^{d,2}_\CH \xrightarrow{p_2} \CQ^{d,1}_\CH \xrightarrow{p_1} \CQ^{d,0}_\CH=\CQ^d_\CH.\]
The first fibration is the universal cover, and the next are generalizations of these, which we write as
\begin{equation}\label{eqnwhiteheadfibration}
\begin{tikzcd}
B^{k-1} \pi_k(\CQ^d_\CH) \arrow[r] & \CQ^{d,k}_\CH \arrow[d, "p^k"] & \\
& \CQ^{d,k-1}_\CH \arrow[r,"c^{k-1}"] & B^k \pi_k(\CQ^d_\CH)
\end{tikzcd}    
\end{equation}
where $B^{k-1}$ is the $k-1$-fold delooping of $\pi_k(\CQ^d_\CH)$ and $c^{k-1}$ is the classifying map of this fibration \cite{goerss1999simplicial}. In particular, $\CQ^{d,k}_\CH$ is $k$-connected, ie. $\pi_{\le k}\CQ^{d,k}_\CH = 0$. It is a special property of $\CQ^d_\CH$ that this tower tops out in finitely many steps, with
\[\CQ^{d,d+2}_\CH \cong B^{d+2} U(1)_{disc} \\
\CQ^{d, \ge d+3}_\CH \cong \star,\]
where $U(1)_{disc}$ indicates $U(1)$ with the discrete topology\footnote{It is possible to consider $\infty$-groupoids enriched in topological spaces, by giving a non-trivial topology to the set of globes. There is a natural topology on stable local unitaries which replaces the $U(1)_{disc}$ at the top with a $\IZ$ one step higher. It is likely necessary to consider this topology when studying anomalies of topological groups $G$, such as Lie groups, but we do not pursue it here.}. See \hyperref[prophomotopycalc]{Proposition \ref{prophomotopycalc}}. It is thus a homotopy $d+2$-type.

Since these are fibrations, there is a homotopy-theoretic obstruction to the ``lifting problem'' for maps into these spaces, such as the $G$-representation map $\alpha$. This motivates the following definition (an analogous definition exists for the unstable anomalies using $\CQ(\CA)$).
\vspace{3mm}
\begin{mdframed}[backgroundcolor=blue!5]
\begin{definition}
    Beginning with $\alpha^0 = \alpha$ and $k = 1$, the \textbf{1st anomaly index} is the obstruction to lifting $\alpha^0$ to
    \[\alpha^1:BG \to \CQ^{d,1}_\CH,\]
    meaning that
    \[p_1 \circ \alpha^1 = \alpha^0\]
    up to homotopy. The obstruction to the existence of this lift is a cohomology class
    \[ [c^0 \circ \alpha^0] \in H^1(BG,\pi_1\CQ^d_\CH),\]
    where
    \[c^0:\CQ^d_\CH \to B \pi_1 \CQ^d_\CH\]
    is the classifying map of the universal cover $\CQ^{d,1}_\CH$ (see \eqref{eqnwhiteheadfibration}).
    
    Continuing this way, after having chosen lifts up to $\alpha^{k-1}$, the \textbf{$k$th anomaly index} is defined as the obstruction to lifting
    \[\alpha^{k-1}:BG \to \CQ^{d,k-1}_\CH\]
    to
    \[\alpha^{k}:BG \to \CQ^{d,k}_\CH,\]
    meaning $p^k \circ \alpha^{k}$ is homotopy equivalent to $\alpha^{k-1}$. The $k$th anomaly index is a cohomology class
    \[ [c^{k-1} \circ \alpha^{k-1}] \in H^k(BG,\pi_k\CQ^d_\CH),\]
    where $c^{k-1}: \CQ^{d,k-1}_\CH \to B^k \pi_k \CQ^d_\CH$ is the classifying map of the Whitehead fibration \eqref{eqnwhiteheadfibration}.
\end{definition}
\end{mdframed}
\vspace{3mm}
Note that because of the $\Omega$-spectrum property (\hyperref[thmomegaspectrum]{Theorem \ref{thmomegaspectrum}}),
\[\pi_{k} \CQ^d_\CH = \pi_1 \CQ^{d-k+1}_\CH,\]
which is a group of stable blend equivalence classes of $d-k+1$-dimensional QCA.

The construction of the first lift corresponds to choosing for each $g$, a blend $\beta(g)$ from $\alpha(g)$ to the identity. The 1st anomaly index is simply the blend equivalence class of each $\alpha(g)$ which obstructs this. The next lift corresponds to choosing a blend from $\beta(g) \beta(h)$ to $\beta(gh)$, and so on. At the last stage we need to choose a local operator, and end up getting an obstruction cocycle in $H^{d+2}(BG,U(1))$, which is the index discussed by Else and Nayak \cite{Else_2014}. We give a generalization of their construction in \hyperref[subsecelsenayak]{Section \ref{subsecelsenayak}}, showing how to compute the anomaly indices on the lattice.

The use of $\infty$-groupoids is actually necessary for implementing the Else-Nayak proposal, especially beyond one dimension, since the non-abelian structure of QCAs prevents one from considering, eg. a group 2-cocyle valued in QCAs. One needs a definition of \emph{non-abelian cohomology}, which $\infty$-groupoids naturally provide. To appreciate some of the complexity of the higher dimensional anomaly indices, see the recent works \cite{kapustin2025higher,kawagoe2025anomalydiagnosissymmetryrestriction}, which gave formulas for the Else-Nayak index in $d = 2$ (corresponding to what we call the fourth anomaly index).

A subtle point is that the anomaly indices in general \emph{depend on the choice of lifts}. We explore this in detail in \hyperref[appdependenceonlifts]{Appendix \ref{appdependenceonlifts}}. Thus, they may not be well-defined given just a QCA $G$-representation. The correct statement from obstruction theory \cite{hatcher2005algebraic,may1999concise} is as follows.
\vspace{3mm}
\begin{mdframed}[backgroundcolor=blue!5]
\begin{theorem}\label{homotopyanomalytheorem}
    The (stable) blend anomaly is trivial if and only if there exists a sequence of lifts $\alpha^k$ such that each anomaly index vanishes.
\end{theorem}
\end{mdframed}
\vspace{3mm}

\begin{proof}
If $\alpha:BG \to \CQ^d$ is null-homotopic, then the lifting problem is trivial since we can take all $\alpha^{k}$ for $k > 0$ to be constant maps. Thus, all anomaly indices vanish.

Conversely, if all anomaly indices vanish for a certain sequence of lifts, we have shown that $\alpha$ is homotopy equivalent to a constant map $BG \to \star$ composed with the sequence of $p^k$'s, which altogether is null-homotopic.
\end{proof}

Note that the 1st and 2nd anomaly indices are always well-defined (see \hyperref[propanomalyindexdependence]{Proposition \ref{propanomalyindexdependence}} and Appendix \ref{appdependenceonlifts}).

\section{Spectra of Invertible States and SRE Anomalies}\label{secspectrasreanom}

\subsection{Spectra of Invertible States}\label{subsecrelationtoinvertiblestates}

In this section we make contact with invertible states. We construct a $\Omega$-spectrum for a subclass of invertible states we call FDQC-invertible, which is closely related to our QCA spectrum, addressing the well-known conjecture \cite{kitaev2013sre} (see also \cite{kitaevIPAMtalk,Freed_2021,Beaudry_2024,kubota2025stablehomotopytheoryinvertible}).

States may be defined in terms of expectations of local operators, and so we represent them as maps\footnote{A linear map on the local algebra $\CA$ extends to the quasi-local algebra iff it is bounded. This will automatically be true for all FDQC-invertible states.} $\psi:\CA \to \IC$. This way, we get an action of QCA $\alpha$ on states $\psi$ by composing $\psi \mapsto \psi \circ \alpha$.
\vspace{3mm}
\begin{definition}
    A state $\psi:\CA\to \IC$ is a \textbf{product state} if for all operators $a,b \in \CA$ with disjoint support, $\psi(ab) = \psi(a) \psi(b)$. We say $\psi$ is \textbf{FDQC-invertible} if it has an inverse state $\tilde \psi:\CA \to \IC$ such that
    \[\psi \otimes \tilde \psi = \psi_0 \circ \alpha_C\]
    for a product state $\psi_0:\CA \otimes  \CA \to \IC$ and an FDQC $C$ with associated QCA $\alpha_C$.\footnote{We could also consider the inverse state to live on a different algebra $\tilde \CA$, but then we could enlarge both algebras to $\CA \otimes \tilde \CA$ by extending by product states.}
\end{definition}
\vspace{3mm}

In the literature, when discussing invertible states people typically have in mind states $\psi$ such that $\psi \otimes \tilde\psi$ is connected to a product state by some time 1 evolution by a local Hamiltonian with bounded terms. FDQC-invertible states are also invertible in this sense, since we may obtain FDQC by time-dependent Hamiltonian evolution giving each circuit element. However, the classification of this kind of invertible state may be different from that of FDQC-invertible states. It seems likely that one can generalize our methods to include arbitrary Hamiltonian evolution, beginning with a suitable definition of approximate QCA and a notion of blend. Then we would apply the same construction we outline in this section.

The close relationship between QCA and FDQC-invertible states comes from the following observation.
\vspace{3mm}
\begin{proposition}\label{propinvertibilityqcaentangler}
    If $\alpha$ is a QCA on $\CA$, then $\alpha \otimes \alpha^{-1}$ is a QCA on $\CA \otimes \CA$ and can be represented as an FDQC
    \[\alpha \otimes \alpha^{-1} =(\prod_n S_n) (\alpha^{-1} \otimes 1)(\prod_x S_x)(\alpha \otimes 1) = \prod_x S_x \prod_x (\alpha \otimes 1)(S_x)\]
    where $S_x$ is the local swap gate at site $x$, which swaps the site algebras $\CA_x$ in each of the two tensor factors.
\end{proposition}
\begin{proof}
    This argument is well-known \cite{arrighi2009unitaritypluscausalityimplies}. The $S_x$ are all commuting, so the same is true for $(\alpha \otimes 1)(S_x)$. Since $\alpha \otimes 1$ has bounded spread, we can thus stagger these gates in a finite depth manner to get a circuit.
\end{proof}
\begin{cor}\label{propinvstatessameasqcaentanglablestates}
     If $\psi_0:\CA \to \IC$ is a product state, then $\psi_0 \circ \alpha$ is an FDQC invertible state, with inverse $\psi_0 \circ \alpha^{-1}$.
\end{cor}
\vspace{3mm}
\begin{definition}
    A state $\psi:\CA \to \IC$ is \textbf{QCA-entanglable} if it is of the form $\psi_0 \circ \alpha$ for some product state $\psi_0$.
\end{definition}
\vspace{3mm}
\noindent It is not clear if all FDQC-invertible states are QCA-entanglable. For instance, $\psi_0$ admits a commuting projector Hamiltonian $H_0$ which is made of single-site projectors onto $\psi_0$. This Hamiltonian is gapped and has $\psi_0$ as its unique ground state, and so we call it a parent Hamiltonian. Applying a QCA $\alpha$, we obtain $\alpha^{-1}(H_0)$, which is a commuting projector parent Hamiltonian for $\psi_0 \circ \alpha$. It is known in the broader context of invertible states that some invertible states, namely those with non-zero thermal Hall conductance, such as a Chern insulator, do not admit almost-local commuting projector parent Hamiltonians \cite{Zhang_2022}. They are thus not almost-local-unitary-entanglable. We will see when we include symmetries that there are likely examples of states which are symmetrically FDQC-invertible but not symmetrically QCA-entanglable.\footnote{In particular, QCA-entanglable SPTs come with a bulk boundary correspondence, see \hyperref[secsptandbulkboundary]{Section \ref{secsptandbulkboundary}}, and there is a mismatch between the $\IZ_8$ group of 2+1d $\IZ_2$ SPTs \cite{Tarantino_2016,kapustin2014symmetryprotectedtopologicalphases} and 1d $\IZ_2$ blend anomalies, by our calculations in \hyperref[secanomalousferms1d]{Section \ref{secanomalousferms1d}}. If these 2d SPTs are really FDQC-invertible (which has not been shown, but seems likely), then they are thus not QCA-entanglable.}

\begin{figure}
    \centering
    \includegraphics[width=8cm]{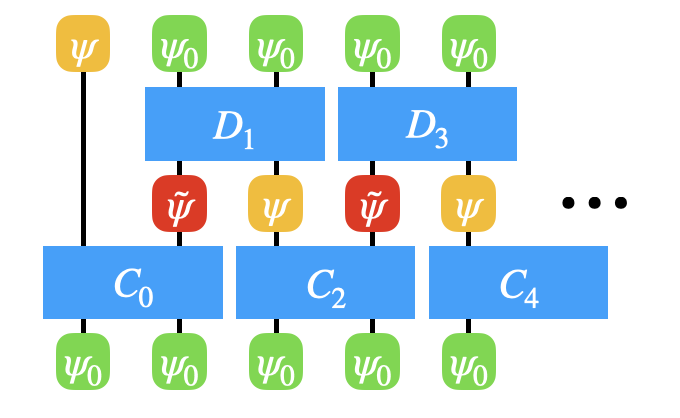}
    \caption{The swindle circuit which acting on a product state $\psi_0 \otimes \psi_0 \otimes \psi_0 \otimes \cdots$ produces $\psi \otimes \psi_0 \otimes \psi_0 \otimes \cdots$ where $\psi$ is an arbitrary FDQC-invertible state. This circuit allows us to study FDQC-invertible states which are not necessarily QCA entanglable. If we extend this circuit infinitely in both directions, we get a circuit which fixes the product state $\psi_0$, but which does not admit a truncation fixing $\psi_0$ (unless $\psi$ is blend equivalent to $\psi_0$, see \hyperref[thminvertiblespectrum]{Theorem \ref{thminvertiblespectrum}}). This is the basis for our cofiber construction of the $\Omega$-spectrum of FDQC-invertible states.}
    \label{figswindlecircuit}
\end{figure}

On the other hand, we can exploit the following well-known trick:
\vspace{3mm}
\begin{proposition}\label{propswindle}
    \textbf{(the swindle)} Let $\psi:\CA \to \IC$ be an FDQC-invertible state and let $\psi_0:\CA \to \IC$ be any product state. There is an FDQC $\alpha$ on
    \[\tilde\CA=\bigotimes_{n \in \IZ_{\ge 0}} \CA\]
    such that
    \[\label{eqnswindle}(\psi_0 \otimes \psi_0 \otimes \cdots) \circ\alpha = (\psi \otimes \psi_0 \otimes \psi_0 \otimes \cdots).\]
\end{proposition}
\begin{proof}
    This mimics another well-known argument called the Eilenberg-Mazur swindle, adapted to the setting of invertible states by Kitaev \cite{kitaev2013sre}. Since $\psi$ is FDQC-invertible, for each $n$, we can find an FDQC $C_n$ acting on the $n$th and $(n+1)$st tensor factors of $\tilde\CA$, and takes $\psi_0 \otimes \psi_0$ to $\psi \otimes \tilde \psi$, where $\tilde \psi$ is some inverse of $\psi$. We can also find an FDQC $D_n$ acting on the $n$th and $(n+1)$st tensor factors taking $\tilde \psi \otimes \psi$ to $\psi_0 \otimes \psi_0$. We let $\alpha$ be defined by the FDQC (see \hyperref[figswindlecircuit]{Figure \ref{figswindlecircuit}})
    \[\label{eqnswindleqca}\prod_{n \ge 0} D_{2n+1} \prod_{n \ge 0} C_{2n}.\]
    By construction, it satisfies \eqref{eqnswindle}.
\end{proof}

This motivates the following definition.
\vspace{3mm}
\begin{mdframed}[backgroundcolor=blue!5]
\begin{definition}\label{defqcamodfixingprodstate}
    Let $\CH$ be a fixed site Hilbert space, and consider stable $d$-QCA acting on the lattice $\IZ^\omega$ built from $\CH$. Let $\psi_0$ be a fixed state in $\CH$, which defines a product state $\bar \psi_0$ on $\IZ^\omega$. We can define the $\Omega$-spectrum
    \[\CQ^d_{\CH,\psi_0}\]
    of stable $d$-QCA, blends, and so on, all of which are required to fix\footnote{Local unitaries are required to fix the state exactly, not just up to phases.} the chosen product state $\bar \psi_0$. This has a natural map of spectra
    \[\Omega_\star \CQ^d_{\CH,\psi_0} \to \Omega_\star \CQ^d_\CH,\]
    and we let
    \[\CQ^d_{\CH,inv} = \lim_{k \to \infty} \Omega^k (\Omega_\star \CQ^{d+k}_\CH/\Omega_\star \CQ^{d+k}_{\CH,\psi_0})\]
    be the cofiber $\Omega$-spectrum. We will show in \hyperref[thminvertiblespectrum]{Theorem \ref{thminvertiblespectrum}} how $\CQ^d_{inv}$ can be regarded an $\Omega$-spectrum of FDQC-invertible states.
\end{definition}
\end{mdframed}
\vspace{3mm}
Under this definition, the swindle FDQC constructed in \hyperref[propswindle]{Proposition \ref{propswindle}} can be regarded as a blend from the identity to something equivalent to the identity, since its ``tail'' in \eqref{eqnswindleqca} fixes $\psi_0$ for $n>1$ (see also \hyperref[figswindlecircuit]{Figure \ref{figswindlecircuit}}). Thus, it corresponds to a loop in $\Omega_\star \CQ^{d+1}$ relative to $\Omega_\star \CQ^{d+1}_{\psi_0}$. By the definition of the cofiber spectrum, this yields an element of $\CQ^{d}_{inv}$.\footnote{Since the swindle constructs an FDQC, one may be worried that the construction can blend to the identity, and so we have made something trivial. However, we are not guaranteed to be able to blend the ``tail'' of \eqref{eqnswindleqca} along any axis in a way that still fixes $\psi_0$. Therefore, these naive blends do not produce lower dimensional blends in the cofiber spectrum, even though we did get a lower dimensional QCA. From a homotopy point of view, this blend may yield a non-trivial element of the relative homotopy group $\pi_1(\Omega_\star \CQ^{d+1}_\CH,\Omega_\star \CQ^{d+1}_{\CH,\psi_0},1)$.}. This construction also does not depend on the product state $\psi_0$, because any two choices are related by an on-site change of basis, which gives an automorphism of $\CQ^d$ respecting blending, and hence the $\infty$-groupoid and spectrum structure.

To recap, we obtain FDQC-invertible states by applying QCA to $\bar \psi_0$. By taking the cofiber spectrum $\CQ^d_{inv}$ we quotient out (in the correct spectrum sense) by precisely those QCA which send $\bar \psi_0$ to itself. Because of the swindle, this simultaneously gives us access to all FDQC-invertible states. This seems to be good evidence that $\CQ^d_{inv}$ can be interpreted as a $\Omega$-spectrum of FDQC-invertible states, which we now pursue.

To prove a correspondence with FDQC-invertible states, we need a definition of state which is suitable for the stable setting where the QCA spectrum exists.
\vspace{3mm}
\begin{definition}
    We consider the algebra $\CA^\omega_\CH$ on $\IZ^\omega$ with site Hilbert space $\CH$. A \textbf{stable $d$-state} is a state $\psi:\CA^\omega_\CH \to \IC$ such that there exists a triple of non-negative integers $a,b,l$ defining the \textbf{domain of $\psi$}
     \[D(\psi) = \IZ^d \times [a,b]^l \times 0 \times \cdots\]
     such that
     \[\psi = \psi_{D} \otimes \bar\psi_0,\]
     where $\psi_{D}$ is a state on $\CA^\omega_\CH(D(\psi))$ and $\bar \psi_0$ is the product state $\psi_0$ on all other sites. $\psi$ is thus determined by the state $\psi_D$ on its domain, which is a finitely-thickened $d$-dimensional lattice. A stable $d$-state is \textbf{FDQC-invertible} if $\psi_D$ is FDQC-invertible.

     A \textbf{blend of stable $d$-states} $\psi \equiv_i \psi'$ is a stable $d$-state $\psi''$ whose domain contains both the domains of $\psi$ and $\psi'$, and which, considered as a map $\psi'':\CA^\omega_\CH(D(\psi'')) \to \IC$, equals $\psi \otimes \bar\psi_0$ to the left of the blend interface, and $\psi' \otimes \bar \psi_0$ to the right of the blend interface. Here $\bar \psi_0$ denotes padding with the product state $\psi_0$ on sites of $D(\psi'') -D(\psi)$ and $D(\psi'')-D(\psi')$ in each respective case.
\end{definition}

\vspace{3mm}
\begin{mdframed}[backgroundcolor=blue!5]
\begin{theorem}\label{thminvertiblespectrum}
    The path components of $\CQ^d_{\CH,inv}$ can be identified as
    \[\label{eqpathcomponentsinvertible}\pi_0 \CQ^d_{\CH,inv} = \left\{\text{FDQC-invertible stable $d$-states}\right\}/\sim,\]
    where $\sim$ is blend equivalence along the $d$th axis. These equivalence classes form an abelian group under stacking. Furthermore, there is a long exact sequence for each $n$:
   \begin{equation} \label{eqinvcofibersequence}
    \begin{tikzcd} [column sep=0.3in, row sep=0.3in] 
	 & 
 &
 \cdots
 &
 \pi_{n+1}\CQ^{d-1}_{\CH,\psi_0}
 \\
 & \pi_{n+1}\CQ^{d}_{\CH,\psi_0} & \pi_{n+1} \CQ^{d}_{\CH} 
 & 
 \pi_n \CQ^{d}_{\CH,inv}
 \\
 & \pi_{n+1}\CQ^{d+1}_{\CH,\psi_0}
 & \cdots
 \arrow["i",from=3-2, to=3-3]
 \arrow["\psi_0",from=1-3, to=1-4]
	\arrow[from=2-3, to=2-4,"\psi_0"]
	\arrow[from=2-2, to=2-3,"i"]
	\arrow[from=1-4, to=2-2, out=-20, in=160,"s"]
    \arrow[from=2-4, to=3-2, out=-20, in=160,"s"]
\end{tikzcd}
\end{equation}
where for $n = 0$, $i$ is the inclusion of QCA fixing $\psi_0$ into all QCA; $\psi_0:\alpha \mapsto \bar\psi_0 \circ \alpha$ is the map from QCA to FDQC-invertible states; and $s$ is the swindle construction of \hyperref[propswindle]{Proposition \ref{propswindle}}, which from an FDQC-invertible state produces an FDQC in one higher dimension which fixes $\psi_0$ and creates $\psi$ when it is appropriately truncated.
\end{theorem}
\end{mdframed}
\vspace{3mm}
\noindent This proof is somewhat involved, and uses some of the details of the construction of the QCA spectrum we develop in \hyperref[secconstruction]{Section \ref{secconstruction}}. The proof can be found in \hyperref[appinvstateproof]{Appendix \ref{appinvstateproof}}.

\subsection{Anomalies as obstructions to trivial symmetric states}\label{subsecsreanom}

Let us return to the question of anomalies. We have so far studied the homotopy class of maps
\[\alpha:BG \to \CQ^d_{\CH}\]
as obstructions to stably disentangling QCA $G$-representations. However, we may also be interested in whether such a symmetry admits a symmetric short-range-entangled (SRE) state, meaning a state which is created by an FDQC from a product state. If it does not, then all symmetric states must be long-range-entangled (LRE) (i.e. not SRE). From the point of view of predicting the ground state of a symmetric Hamiltonian, this is typically the more interesting question. A classic lattice theorem along these lines is the Lieb-Schultz-Mattis theorem \cite{lieb1961two}. This has been interpreted as an anomaly by many authors, see eg. \cite{cho2017anomaly,else2020topological,Cheng_2023}. From the point of view of quantum field theory, we expect that 't Hooft anomalies play a dual role as obstructions to gauging as well as obstructions to trivial symmetric states. As emphasized by \cite{tu2025anomaliesglobalsymmetrieslattice,shirley2025anomalyfreesymmetriesobstructionsgauging}, these concepts are different on the lattice, since there are symmetries with non-trivial blend anomalies which admit symmetric SRE states (see \hyperref[examplechiralfloquetcircuit]{Example \ref{examplechiralfloquetcircuit}} below).

We would thus like to use homotopy theory to formulate another kind of lattice anomaly, which we will call the ``SRE anomaly'', which is an obstruction to having a symmetric SRE state. This motivates the following definition.
\vspace{3mm}
\begin{mdframed}[backgroundcolor=blue!5]
\begin{definition}\label{defdynamicalanomaly}
    Let $\CH$ be a fixed site Hilbert space, and $\alpha:BG \to \CQ^d_\CH$ a stable $d$-QCA $G$-representation. We define its \textbf{SRE anomaly} as the (based) homotopy class of the induced map
    \[BG \xrightarrow{\alpha} \CQ^d_\CH \to B\CQ^d_{\CH,inv},\]
    given by delooping the cofiber map
    \[\Omega_\star \CQ^d_{\CH} \to \CQ^d_{\CH,inv}.\]
\end{definition}
\end{mdframed}
\vspace{3mm}
\noindent This definition satisfies many of the nice properties that the stable blend anomaly satisfied, but now applicable to the problem of finding symmetric SRE states.
\vspace{3mm}
\begin{mdframed}[backgroundcolor=blue!5]
\begin{theorem}\label{thmdynamicalanomaly}
    With the notation as in \hyperref[defdynamicalanomaly]{Definition \ref{defdynamicalanomaly}}, the SRE anomalies of stable $d$-QCA $G$-representations form an abelian group
    \[ \CQ^{d+1}_{\CH,inv}(BG,\star)\]
    (a reduced generalized cohomology theory) such that
    \begin{enumerate}
        \item The group structure corresponds to stacking
        \[ [\alpha \otimes \beta] = [\alpha] + [\beta],\]
        \item Any stably-disentanglable $G$-representation $\alpha$ has $[\alpha] = 0$
        \item The reversed $G$-representation $\alpha^\text{rev}$ obtained by reflection along the $d$th axis (as in \hyperref[corollaryabelaingroup]{Corollary \ref{corollaryabelaingroup}}) has
        \[ [\alpha^\text{rev}] = -[\alpha].\]
        \item Furthermore, if there is a stable $d$-state of the form
    \[\psi=\bar \psi_0 \circ \alpha_C\]
    for some FDQC $C$ (such a state may be called \textbf{FDQC-SRE}) which is moreover fixed by the action of $G$, so that for each $g$,
    \[\psi \circ \alpha(g) = \psi,\]
    then the SRE anomaly vanishes, i.e.
    \[ [\alpha] = 0 \in \CQ^{d+1}_{\CH,inv}(BG,\star).\]
    In other words, if the SRE anomaly is non-vanishing, $\alpha$ does not admit symmetric FDQC-SRE states.
        \item
        Finally, we have a long exact sequence
    \begin{equation} \label{eqnlongexactsequenceSRE}
    \hspace*{-0.3in}
    \begin{tikzcd} [column sep=0.3in, row sep=0.3in] 
	 & 
 &
 \cdots
 &
 \CQ^{d}_{\CH,inv}(BG,\star)
 \\
 & \CQ^{d}_{\CH,\psi_0}(BG) & \CQ^{d}_{\CH}(BG)
 & 
  \CQ^{d+1}_{\CH,inv}(BG,\star)
 \\
 & \CQ^{d+1}_{\CH,\psi_0}(BG)
 & \cdots
 \arrow["i",from=3-2, to=3-3]
 \arrow["SRE",from=1-3, to=1-4]
	\arrow[from=2-3, to=2-4,"SRE"]
	\arrow[from=2-2, to=2-3,"i"]
	\arrow[from=1-4, to=2-2, out=-20, in=160,"s"]
    \arrow[from=2-4, to=3-2, out=-20, in=160,"s"]
\end{tikzcd}
\end{equation}
    where $SRE(\alpha)$ gives the SRE anomaly of a QCA $G$-representation. Its kernel is the image under $i$ of those QCA $G$-representations fixing a product state $\bar\psi_0$ and its cokernel are SRE anomalies not realizable by QCA $G$-representations, see \hyperref[secanomalousferms1d]{Section \ref{secanomalousferms1d}}. The connecting homomorphism $s$ is given by the swindle circuit of \hyperref[propswindle]{Proposition \ref{propswindle}}, which fixes the product state and has $G$-symmetric gates so it is in the kernel of $i$.
    \end{enumerate}    
\end{theorem}
\end{mdframed}
\begin{proof}
    The abelian group structure comes because $\CQ^d_{\CH,inv}$ is an $\Omega$-spectrum, so $B\CQ^d_{\CH,inv}$ is an infinite loop space. We have used this structure to identify the SRE anomalies, which are the homotopy classes of maps $BG \to B\CQ^d_{\CH,inv}$, with the reduced cohomology in one higher dimension (written as cohomology relative to the point $\star$):
    \[\CQ^{d+1}_{\CH,inv}(BG,\star).\]
    This is an indication of a bulk-boundary correspondence we return to in \hyperref[secsptandbulkboundary]{Section \ref{secsptandbulkboundary}}. The three properties of the group structure follow because $\alpha:BG \to B\CQ^d_{\CH,inv}$ factors through $\alpha:BG \to \CQ^d_\CH$. See the analogous properties for the non-symmetric case discussed in \hyperref[subsecstabilization]{Section \ref{subsecstabilization}}.

    Now suppose there is a symmetric FDQC-SRE state $\psi = \bar \psi_0 \circ \alpha_C$. The $G$-representation $\alpha_C \alpha(g) \alpha_C^{-1}$ thus fixes the product state $\bar \psi_0$. Since $C$ is a FDQC, $\alpha_C \alpha(g) \alpha_C^{-1}$ as a blend of $G$-representations to $\alpha(g)$, so they are homotopic by the argument in \hyperref[lemmaGblend]{Lemma 1}. This means we can homotope $\alpha:BG \to \CQ^d_{\CH}$ into $\CQ^d_{\CH,\psi_0}$, so $\alpha$ is null-homotopic in the cofiber.

    Finally, the long exact sequence is the cofiber long exact sequence for
    \[B \Omega_\star \CQ^{d+1}_{\CH,\psi_0} \to B \Omega_\star \CQ^{d+1}_{\CH} \to B\CQ^d_{\CH,inv}\]
    combined with $B \Omega_\star \CQ^{d+1}_\CH = \CQ^d_\CH$, and likewise for $\CQ^d_{\CH,\psi_0}$. Recall this is equivalent to the spectrum property $\Omega_{id} \Omega_\star \CQ^{d+1}_\CH = \Omega_\star \CQ^d_\CH$.
\end{proof}

\begin{example}\label{examplechiralfloquetcircuit}
    Consistent with this theory, it has recently been observed that there are non-stably-disentanglable symmetries which nonetheless admit symmetric product states \cite{shirley2025anomalyfreesymmetriesobstructionsgauging,tu2025anomaliesglobalsymmetrieslattice}. These should be regarded as being in the kernel of the map $SRE$. One example, constructed in \cite{shirley2025anomalyfreesymmetriesobstructionsgauging}, is an FDQC $\IZ_2$-representation $\alpha(0)=1$, $\alpha(1)=\alpha_C$, where $C$ is a particular FDQC satisfying $\alpha_C^2 = 1$ but having a truncation $C'$ to a half-space that has $\alpha_{C'}^2$ equal to a unit translation along the boundary of the half-space composed with an FDQC.
    
    We can regard the truncated circuit as a blend $\beta(1)=\alpha_{C'}:\alpha(1) \equiv_2 1$. The 1st anomaly index of \hyperref[subseccomptheanom]{Section \ref{subseccomptheanom}} vanishes because we are able to choose this blend ($\alpha(0)$ is the identity so we can use itself as a blend, $\beta(0) = \alpha(0)$). The 2nd anomaly index measures whether these blends satisfy the group law up to 2-blends (see \hyperref[subsecelsenayak]{Section \ref{subsecelsenayak}}). The two-cocycle of interest is
    \[\beta_2(1,1) = \beta(1) \beta(1) \beta(0)^{-1} = \beta(1)^2,\]
    which by its construction involves a translation along the first axis. Thus, it does not admit a blend to the identity along this axis.
    
    We can of course choose different truncations, so the obstruction is the cohomology class of
    \[ [\beta_2] \in H^2(B\IZ_2,\IZ) = \IZ_2.\]
    In particular, we can choose different $\beta(1)$ and see if one of them does result in a blendable $\beta(1)^2$. We should not have to change $\beta(0)$ because we can work with normalized group 2-cocycles \cite{brown2012cohomology}. We can for instance add a translation in the blend region to $\beta(1)$, but this changes $\beta(1)^2$ by a double translation, which is why we have a $\IZ_2$ invariant above. In this case, since $\beta(1)^2$ is a unit translation, $[\beta_2] \neq 0$. This anomaly-index does not depend on choices (see \hyperref[propanomalyindexdependence]{Proposition \ref{propanomalyindexdependence}}), so it follows that
    \[ [\alpha] \neq 0 \in \CQ^2_\CH(BG),\]
    so $\alpha$ is not stably-disentanglable.
    
    On the other hand, $\alpha$ does admit symmetric product states \cite{shirley2025anomalyfreesymmetriesobstructionsgauging,tu2025anomaliesglobalsymmetrieslattice} so it should have vanishing SRE anomaly. We can at least show that the 2nd anomaly index computed above does not yield an SRE anomaly. We could in fact define a whole sequence of \textbf{SRE anomaly indices} by studying the Whitehead tower of $B\CQ^d_{\CH,inv}$, exactly as in \hyperref[subseccomptheanom]{Section \ref{subseccomptheanom}} and \hyperref[subsecelsenayak]{Section \ref{subsecelsenayak}}. We would find $[\beta_2]$ as above, but we are only interested in its image under the map induced by the cofiber sequence \eqref{eqinvcofibersequence}:
    \[H^2(B\IZ_2,\pi_2 \CQ^2_\CH) \to H^2(B\IZ_2,\pi_1 \CQ^2_{\CH,inv}).\]
    $\pi_2 \CQ^2_\CH=\pi_1\CQ^1_\CH$ is generated by a translation along the first axis, which fixes $\bar \psi_0$, so it is in the image of
    \[\pi_2 \CQ^2_{\CH,\psi_0} \to \pi_2 \CQ^2_\CH\]
    and therefore the map
    \[\pi_2 \CQ^2_\CH \to \pi_1 \CQ^2_{\CH,inv}\]
    is zero. In fact, with a little more work using the Atiyah-Hirzebruch spectral sequence, one can show $\CQ^2_{\CH,inv}(B\IZ_2) = 0$ so the SRE anomaly is trivial.
\end{example}

\subsection{Classifying SPTs and Bulk-Boundary Correspondence}\label{secsptandbulkboundary}

In this section, we add global on-site symmetries to the discussion of \hyperref[subsecrelationtoinvertiblestates]{Section \ref{subsecrelationtoinvertiblestates}}, giving a spectrum of invertible $G$-symmetric states, and discussing bulk-boundary correspondences connecting SPTs with blend and SRE anomalies. However, we will see the bulk-boundary correspondences on the lattice are likely not one-to-one.

Let $\CH$ be a site Hilbert space carrying a representation of a group $G$. We consider QCAs on the local algebra built from $\CH$, which inherits an on-site $G$ action. We would like to classify QCAs $\alpha$ commuting with $G$, up to blends commuting with $G$, as well as $G$-symmetric FDQC-invertible states, up to blends of such states. These will have associated $\Omega$-spectra that are related in much the same way as we saw in \hyperref[subsecrelationtoinvertiblestates]{Section \ref{subsecrelationtoinvertiblestates}}. In fact these spectra will be embedded inside the previous ones as the fixed points of a $G$ action.

Indeed, when we construct the space $\CQ^d_\CH$, it inherits an action of $G$, since given a stable $d$-QCA $\alpha$, we obtain another stable $d$-QCA $g \circ \alpha \circ g^{-1}$ by taking $g \in G$ to act diagonally on all the copies $\CH$ in the domain of $\alpha$. More precisely, suppose we have a stable $d$-QCA $\alpha$ with a domain $D(\alpha)$ which describes an algebra with site Hilbert space $\CH$ on a thickened $d$-dimensional lattice. We obtain a $G$ action on $D(\alpha)$ by tensor product
\[g_{D(\alpha)} = \bigotimes_{x \in D(\alpha)} g_x,\]
where $g_x$ is the action on the copy of $\CH$ at site $x$. Then
\[g_{D(\alpha)} \circ \alpha \circ g_{D(\alpha)}^{-1}\]
is another stable $d$-QCA with the same domain.

This action is compatible with blends and their composition, including those implemented by local unitaries. Thus, we get a cellular action of $G$ on $\CQ^d_\CH$ which has the property that if $g$ fixes a cell, it fixes it pointwise. This makes it a $G$-CW complex \cite{greenlees1995equivariant}. We denote the fixed points of this $G$-action as
\[(\CQ^d_\CH)^G = \{\sigma \in \CQ^d_\CH\ |\ \forall g,\ g(\sigma) = g\},\]
where $\sigma$ denotes any cell of $\CQ^d_\CH$. Because of the nice $G$ action, this is a sub-simplicial complex of $\CQ^d_\CH$. It is exactly what we would build from our construction of $\CQ^d_\CH$ if we instead only include QCAs and blends which commute with the on-site $G$ action. We encode this in the following:

\vspace{3mm}
\begin{mdframed}[backgroundcolor=blue!5]
\begin{definition}
    Let $\CH$ be a site Hilbert space carrying a representation of a group $G$.
    \begin{enumerate}
        \item A \textbf{stable $G$-$d$-QCA} is a stable $d$-QCA $\alpha$ on a domain $D(\alpha) \subset \IZ^\omega$, commuting with the induced tensor product $G$ action on $\CA^\omega_\CH(D(\alpha))$. 
        \item A \textbf{$G$-blend of stable $G$-$d$-QCA} is a blend which is itself a stable $G$-$d$-QCA.
        \item We obtain the \textbf{space of stable $G$-$d$-QCA} $(\CQ^d_\CH)^G$ as the fixed points of the induced $G$ action on $\CQ^d_\CH$.
    \end{enumerate}
\end{definition}
\end{mdframed}
\vspace{3mm}
\vspace{3mm}
\begin{mdframed}[backgroundcolor=blue!5]
\begin{proposition}
    Let $\CH$ be a site Hilbert space carrying a representation of a group $G$.
    \begin{enumerate}
        \item The $G$-action on $\CQ^d_\CH$ commutes with the $\Omega$-spectrum maps (which are inclusions) and so $\Omega_\star\CQ^d_\CH$ defines a (naive) $G$-$\Omega$-spectrum in the sense of \cite{greenlees1995equivariant}.
        \item The $G$-fixed points $\Omega_\star(\CQ^d_\CH)^G$ define a sub-$\Omega$-spectrum of $\CQ^d_\CH$.
        \item $\pi_1((\CQ^d_\CH)^G)$ is the group of stable $G$-$d$-QCAs up to $G$-blend equivalence along the $d$th axis, with the group operation given by stacking.
        \item The inclusion of the fixed points induces a ``forgetful map''
        \[\label{eqnforgetfulmap}F_G:\pi_1((\CQ^d_\CH)^G) \to \pi_1(\CQ^d_\CH)\]
        which takes a $G$-blend equivalence class of $G$-QCA to the blend equivalence class of the QCA, forgetting the $G$ structure.
    \end{enumerate}
\end{proposition}
\end{mdframed}
\vspace{3mm}

As mentioned above, we can consider stable $G$-$d$-QCA as entanglers for certain $G$-symmetric stable $d$-states when applied to product states, analogous to how we considered QCA-entanglable states above. These states will be invertible in a $G$-symmetric sense as follows:
\vspace{3mm}
\begin{mdframed}[backgroundcolor=blue!5]
    \begin{definition}
    Let $\CH$ be a site Hilbert space carrying a representation of a group $G$ with a vector $\psi_0$ with $g \cdot \psi_0 = \psi_0$. We consider stable $d$-states on the algebra $\CA^\omega_\CH$, which inherits an action of $G$. A $G$-symmetric stable $d$-state $\psi$ is \textbf{$G$-FDQC invertible} if there is another $G$-symmetric stable $d$-state $\tilde \psi$ such that
    \[\psi \otimes \tilde \psi = \bar \psi_0 \circ \alpha_C,\]
    where $\alpha_C$ is a QCA associated to a FDQC $C$ with $G$-symmetric gates, which we call a \textbf{$G$-FDQC} and $\bar \psi_0$ is a product state made from $\psi_0$ on a suitable domain.
\end{definition}
\end{mdframed}
\vspace{3mm}
It is important to require the individual gates of the FDQC being each $G$-symmetric, rather then just the FDQC commuting with $G$ as a whole. One reason is that FDQC with $G$-symmetric gates admit $G$-blends by truncating the circuit. It may be that an FDQC which commutes with $G$ as a QCA defines a non-trivial $G$-QCA. These will represent non-trivial elements of the kernel of the forgetful map $F_G$ in \eqref{eqnforgetfulmap}. These are the entanglers of ``true SPTs'' which are trivial invertible states only once the symmetry is forgotten, and we want to consider true SPTs as non-trivial.

With this definition we have the analog of \hyperref[thminvertiblespectrum]{Theorem \ref{thminvertiblespectrum}}, giving us a spectrum of $G$-FDQC invertible stable $d$-states:
\vspace{3mm}
\begin{mdframed}[backgroundcolor=blue!5]
\begin{theorem}\label{thmsptclassification}
    Let $\CH$ be a site Hilbert space carrying a representation of a group $G$ with a vector $\psi_0$ with $g \cdot \psi_0 = \psi_0$.
    \begin{enumerate}
        \item If $\alpha$ is a $G$-QCA, then $\alpha \otimes \alpha^{-1}$ can be written as a $G$-FDQC.
        \item If $\psi$ is a $G$-FDQC invertible stable $d$-state, then there is a $G$-FDQC on a $d+1$-dimensional half-space which produces $\psi$ at its boundary when applied to $\bar \psi_0$.
        \item We obtain a $G$-action on the space $\CQ^d_{\CH,\psi_0}$ of stable $d$-QCAs $\alpha$ fixing the product state $\bar\psi_0 = \bigotimes_{x \in D(\alpha)} \psi_0$ on their domain, and this defines a naive $G$-$\Omega$-spectrum, with the inclusion $\Omega_\star\CQ^d_{\CH,\psi_0} \to \Omega_\star\CQ^d_{\CH}$ an equivariant map of naive $G$-$\Omega$-spectra.
        \item Let $\CQ^{d,G}_{\CH,inv}$ be the cofiber spectrum of the inclusion
        \[\Omega_\star(\CQ^{d}_{\CH,\psi_0})^G \to \Omega_\star (\CQ^{d}_{\CH})^G \to \CQ^{d,G}_{\CH,inv}.\]
        This is equivalent to the $G$-fixed points of the induced $G$-action on $\CQ^d_{\CH,inv}$:
        \[\CQ^{d,G}_{\CH,inv} = (\CQ^d_{\CH,inv})^G.\]
        We henceforth use the notation on the LHS for fixed points of these spectra since it is unambiguous.
        \item $\pi_0(\CQ^{d,G}_{\CH,G})$ is the abelian group of $G$-FDQC invertible stable $d$-states up to blend equivalence along the first axis, with the group structure given by stacking.
        \item There is a long exact sequence
        \begin{equation} \label{eqequivinvcofibersequence}
        \hspace{-0.3in}
    \begin{tikzcd} [column sep=0.3in, row sep=0.3in] 
	 & 
 &
 \cdots
 &
 \pi_{n+1}\CQ^{d-1,G}_{\CH,\psi_0}
 \\
 & \pi_{n+1}\CQ^{d,G}_{\CH,\psi_0} & \pi_{n+1} \CQ^{d,G}_{\CH} 
 & 
 \pi_n \CQ^{d,G}_{\CH,inv}
 \\
 & \pi_{n+1}\CQ^{d+1,G}_{\CH,\psi_0}
 & \cdots
 \arrow["i",from=3-2, to=3-3]
 \arrow["\psi_0",from=1-3, to=1-4]
	\arrow[from=2-3, to=2-4,"\psi_0"]
	\arrow[from=2-2, to=2-3,"i"]
	\arrow[from=1-4, to=2-2, out=-20, in=160,"s"]
    \arrow[from=2-4, to=3-2, out=-20, in=160,"s"]
\end{tikzcd}
\end{equation}
            as in \eqref{eqinvcofibersequence}.
        \item The inclusion of fixed points induces a forgetful map
        \[F_G:\pi_0(\CQ^{d,G}_{\CH,inv}) \to \pi_0(\CQ^{d}_{\CH,inv}).\]
        Elements of the kernel of this map are equivalence classes of true SPTs up to blend equivalence along the $d$th axis (compare \eqref{eqnforgetfulmap}).
        
    \end{enumerate}
\end{theorem}
\end{mdframed}
\begin{proof}
The argument for the first statement is the same as in \hyperref[propinvertibilityqcaentangler]{Proposition \ref{propinvertibilityqcaentangler}}, once noting that the local swap gates are $G$-symmetric, and $\alpha \otimes 1$ sends $G$-symmetric gates to $G$-symmetric gates since it commutes with the $G$ action. The second statement follows from the argument of \hyperref[propswindle]{Proposition \ref{propswindle}} exactly, noting now that the $C$ and $D$ circuits are $G$-FDQC. The third and fourth points follow from the nice cellular action of $G$, which makes the inclusion a $G$-cofibration \cite{greenlees1995equivariant}. The proof of the fifth point is exactly as in the proof of \hyperref[thminvertiblespectrum]{Theorem \ref{thminvertiblespectrum}}, described in \hyperref[appinvstateproof]{Appendix \ref{appinvstateproof}}, except now we use the swindle $G$-circuit. For the sixth point, this is the cofiber long exact sequence. For the last point, the map is induced by the commutative square of inclusions:
\begin{equation}
    \begin{tikzcd}
        \CQ_{\CH,\psi_0}^{d,G} \arrow[r] \arrow[d] & \CQ_{\CH}^{d,G} \arrow[d] \\
        \CQ_{\CH,\psi_0} \arrow[r] & \CQ_{\CH}^d
    \end{tikzcd}
\end{equation}
\end{proof}

It is generally expected that true SPTs (meaning those which realize trivial invertible phases when $G$ is forgotten) in $d$ dimensions should correspond to anomalies in $d-1$ dimensions. However, it is not so obvious how to define this bulk-boundary correspondence on the lattice. Furthermore, our calculation of $\CQ^1_{f,inv}(B\IZ_2) = \IZ_2 \times \IZ_4$ in \hyperref[secanomalousferms1d]{Section \ref{secanomalousferms1d}} does not match the expected $\IZ_8$ classification of 2+1d fermionic $\IZ_2$ true SPTs (see \cite{Tarantino_2016}). It thus seems likely that there is no general bulk-boundary correspondence relating true SPTs to SRE anomalies.

However, there is a way to relate SPT entanglers to blend anomalies. Suppose our SPT $\psi$ admits a QCA entangler, meaning there is a $G$-QCA $\alpha$ such that $\psi = \psi_0 \circ \alpha$. Suppose also that $\alpha$ admits a blend to the identity, so that $\psi$ is blend-trivial as an invertible state. It is thus a pure SPT. For these states we can express a bulk-boundary correspondence, along the lines of \cite{Bultinck_2019,zhang2022topologicalinvariantssptentanglers,shirley2025anomalyfreesymmetriesobstructionsgauging}. We phrase it in terms of an anomaly-free QCA $G$-representation $\sigma(g)$ and a blendable QCA $\alpha$ which commutes with it, yielding a map
\[(\sigma,\alpha):BG \times B\IZ \to \CQ^d_\CH.\]
We can study the homotopy class $[\sigma,\alpha]$ of this map. We find there is a ``bulk-boundary correspondence'' putting these homotopy classes in one to one correspondence with stable blend anomalies of QCA $G$-representations in $d-1$-dimensions:
\vspace{3mm}
\begin{mdframed}[backgroundcolor=blue!5]
\begin{theorem}\label{thmbulkboundarycorresp}
    The homotopy classes of maps
    \[ [\sigma,\alpha]:BG \times B\IZ \to \CQ^d_\CH\]
    such that $[\sigma]$ and $[\alpha]$ separately are homotopically trivial, are in one-to-one correspondence with homotopy classes of maps
    \[\beta:BG \to \CQ^{d-1}_\CH.\]
\end{theorem}
\end{mdframed}
\begin{proof}
    Maps $BG \times B\IZ \to \CQ^d_\CH$ are equivalent to maps
    \[BG \to \text{Maps}(B\IZ,\CQ^d_\CH).\]
    Since $B\IZ = S^1$, the latter is the free loop space $L \CQ^d_\CH$. It sits in a fibration
    \begin{equation}
        \begin{tikzcd}
            \Omega_\star \CQ^d_\CH \arrow[r] & L\CQ^d_\CH \arrow[d] \\ & \CQ^d_\CH
        \end{tikzcd}
    \end{equation}
    Since $\Omega_\star \CQ^d_\CH$ is an infinite loop space, this fibration splits (see \cite{malkiewich2023spectra} exercise 2.24), giving us
    \[L\CQ^d_\CH = \Omega_\star \CQ^d_\CH \times \CQ^d_\CH.\]
    Thus, homotopy classes of maps $BG \times B\IZ \to \CQ^d_\CH$ are given by a pair of homotopy classes of a map
    \[ \sigma:BG \to \CQ^d_\CH,\]
    representing the stable blend anomaly of $\sigma$, which is assumed to be trivial, and a homotopy class
    \[ \phi:BG \to \Omega_\star \CQ^d_\CH.\]
    By construction, $\phi$ sends the basepoint of $BG$ to
    \[\alpha:B\IZ = S^1 \to \CQ^d_\CH.\]
    As this is assumed to be null-homotopic as well, $\phi$ lands in the component of the identity $d$-QCA, and that component is homotopy equivalent to $\CQ^{d-1}_\CH$. The result follows.
\end{proof}

To get some intuition for this theorem, we can represent the free loop space $L\CQ^d$ as an $\infty$-groupoid whose objects are diagrams in $\CQ^d_\CH$
\[\star \xrightarrow{\alpha} \star\]
and whose 1-morphisms are diagrams
\begin{equation}
    \text{Hom}(\star \xrightarrow{\alpha} \star,\star \xrightarrow{\beta} \star) = \begin{tikzcd}
\star \arrow[r, "\alpha"] \arrow[d, "\gamma"'] & \star \arrow[d, "\delta"] \\
\star \arrow[r, "\beta"'] \arrow[ru, Leftarrow, shorten <=8pt, shorten >=8pt] & \star
\end{tikzcd}
\end{equation}
with composition by vertical pasting. Note this differs from the based loop space $\Omega_\star \CQ^d_\CH$ for which the 1-morphisms must have $\gamma = \delta = 1$ and the picture reduces to a 2-globe, as in \eqref{eqn2globe} (so a morphism only exists between $\alpha$ and $\beta$ if they admit a blend along the first axis). 2-morphisms are cubes with suitable decorations and so on.

The map $BG \to L\CQ^d_\CH$ sends its base point $\star$ to $\star \xrightarrow{\alpha} \star$ and its 1-morphisms $\star \xrightarrow{g} \star$ to the 1-morphisms
\begin{equation}
    \begin{tikzcd}
\star \arrow[r, "\sigma(g)"] \arrow[d, "\alpha"'] & \star \arrow[d, "\alpha"] \\
\star \arrow[r, "\sigma(g)"'] \arrow[ru, Leftarrow, shorten <=8pt, shorten >=8pt] & \star
\end{tikzcd} \qquad \in \text{Hom}(\star \xrightarrow{\alpha} \star,\star \xrightarrow{\alpha} \star)
\end{equation}
where the blend is the identity, encoding the commutation $\sigma(g)\alpha = \alpha\sigma(g)$. Higher morphisms may be filled in with identities as well since $\sigma(g)$ satisfy the group laws.

The issue is that when we try to homotope this map into $\Omega_\star \CQ^d_\CH$ by blending $\gamma:\alpha \equiv_1 1$, we may not be able to choose $\gamma$ to commute with $\sigma(g)$ (i.e. to be a blend of $G$-QCA) so that these higher morphisms will need to be filled in with non-trivial data, representing the lower dimensional anomaly.

\begin{figure}
    \centering
    \includegraphics[width=8cm]{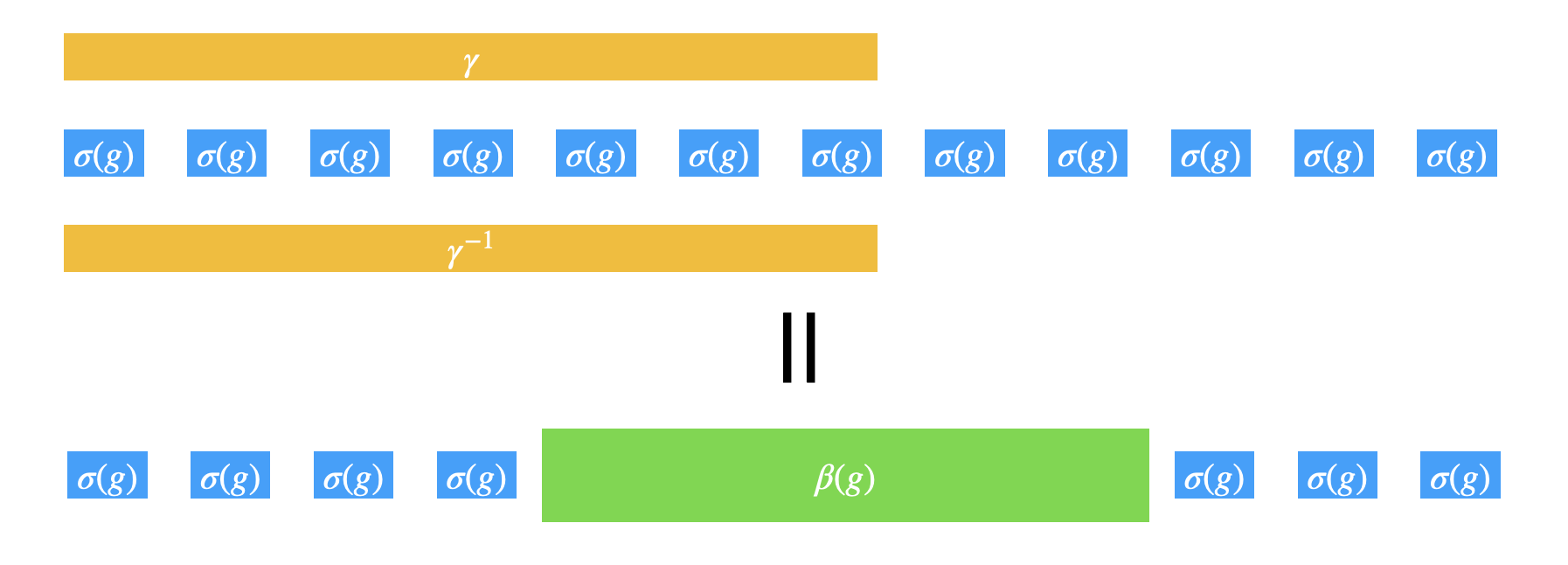}
    \caption{Given an on-site symmetry $\sigma(g)$ and a blend $\gamma:\alpha \equiv_1 1$, where $\alpha$ commutes with $\sigma(g)$, we can construct a QCA $G$-representation $\beta(g)$ in one lower dimension, which serves as an obstruction to finding a blend $\gamma':\alpha \equiv_1 1$ which commutes with $\sigma(g)$. Considering $\alpha$ as an SPT entangler, this construction yields the ``anomalous boundary symmetry'' of the SPT.}
    \label{figbulkboundaryanom}
\end{figure}

One way to construct the obstruction is as follows. Let $\gamma:\alpha \equiv_1 1$. Consider
\[\sigma'(g) = \gamma \sigma(g) \gamma^{-1}: \sigma(g) \equiv_1 \sigma(g).\]
Since $\alpha$ commutes with $\sigma(g)$, these are blends $\sigma(g) \equiv_1 \sigma(g)$, and they satisfy the group law since $\sigma(g)$ does. Let us now suppose that $\sigma(g)$ is on-site (or admits blends of $G$-representations), so that we can truncate the action of $\sigma' (g)$ to a finite region to form
\[\beta(g):1\equiv_1 \sigma(g) \equiv_1 \sigma(g) \equiv_1 1,\]
which also satisfy the group law, by construction. See \hyperref[figbulkboundaryanom]{Fig. \ref{figbulkboundaryanom}}. The homotopy class of $\beta:BG \to \CQ^{d-1}_\CH$ is equivalent to the homotopy class of $[\sigma,\alpha]$, and represents an obstruction to finding a blend of $G$-QCA from $\alpha$ to the identity. We note that this is precisely the usual construction of the boundary anomalous symmetry from a bulk SPT entangler \cite{Bultinck_2019,zhang2022topologicalinvariantssptentanglers,shirley2025anomalyfreesymmetriesobstructionsgauging}.

\section{Constructing a Space of QCA}\label{secconstruction}

\subsection{Globular Picture and Homotopy Groups}\label{subsecglobular}

The first step to constructing the $\infty$-groupoid $\CQ(\CA)$ will be to construct it as a ``globular set'', formalizing the diagrams which were outlined in \hyperref[subsecspacefirstpass]{Section \ref{subsecspacefirstpass}}.

\vspace{3mm}
\begin{mdframed}[backgroundcolor=blue!5]
\begin{definition}[Globular Set of QCA]\label{defglobularset}
    Let $\CA$ be a local algebra on $\IZ^d$. Let $1 \le k \le d+1$. We define a globular set $\CQ(\CA) =\{\CQ(\CA)_n\}_{0 \le n \le d+2}$ for which the set of $n$-globes $\CQ(\CA)_n$ is defined as follows
    \begin{enumerate}
        \item There is a unique 0-globe called $\star$.
        \item A 1-globe is a QCA.
        \item A 2-globe is a pair of QCAs $\varphi_1,\varphi_1'$ and a blend
        \[\varphi_2:1\equiv_1 \varphi_1' \varphi_1^{-1}.\]
        We write this as
        \[(\varphi_2:\varphi_1 \to_1 \varphi_1'),\]
        where we introduce the notation $\alpha:\beta \to_1 \gamma$, meaning $\alpha:1 \equiv_1 \gamma \beta^{-1}$.
        \item For $3 \le k \le d+1$, a $k$-globe consists of an array
        \[
        \begin{pmatrix}
            \varphi_{k}&:& \varphi_{k-1}&\rightarrow_{k-1}&\varphi_{k-1}^\prime \\
            && &\vdots& \\
            &:& \varphi_2&\rightarrow_2&\varphi_2^\prime \\
            &:& \varphi_1&\rightarrow_1&\varphi_1^\prime
        \end{pmatrix},\]
        where we introduce the notation $\alpha:\beta \to_n \gamma$ to mean
        \[\alpha:1\equiv_m 1 \quad \forall m<n \\
        \alpha: 1 \equiv_n \gamma \beta^{-1}.\]
        The array is shorthand to mean for each $1 \le n \le k$, our $k$-globe has
        \[\varphi_{n+1}:\varphi_n \to_n \varphi_n,\] and likewise for each $1 \le n < k$,
        \[\varphi_{n+1}':\varphi_n \to_n \varphi_{n}'.\]
        Note that $\varphi_{n+1}$ and $\varphi_{n+1}'$ are supported in $[a,b]^{n-1} \times \IZ_{\ge c} \times \IZ^{d-n}$, where $[a,b]$ is some interval and $c \in \IZ$. In particular, for $k = d+1$, $\varphi_{d+1}$ is supported in a finite region $[a,b]^d$.

    \item For $k = d+2$ a $d+2$-globe is an array
    \[
    \begin{pmatrix}
        U&:& \varphi_{d+1}&\rightarrow_{d+1}&\varphi_{d+1}^\prime \\
    && &\vdots& \\
    &:& \varphi_2&\rightarrow_2&\varphi_2^\prime \\
    &:& \varphi_1&\rightarrow_1&\varphi_1^\prime
    \end{pmatrix},\]
    where for a local unitary $U$,
    \[\label{eqntopmorphism}U:\varphi_{d+1} \to_{d+1} \varphi_{d+1}'\]
    has the special meaning
    \[\text{Ad}\ U = \varphi_{d+1}' \varphi_{d+1}^{-1}\]
    (note the RHS is supported in a finite region). To simplify the notation we also let
    \[U:\alpha \equiv_{d+1} \beta\]
    mean ${\rm Ad}\ U= \beta \alpha^{-1}$.
    \end{enumerate}
    To be a globular set we must define maps (which can be thought of as ``source'' and ``target'', respectively)
    \[\sigma_k,\tau_k:\CQ(\CA)_{k+1} \to \CQ(\CA)_{k}\]
    satisfying certain axioms. We let $\sigma_0,\tau_0$ be the constant maps. For $1 \le k \le d$, we let
    \[\sigma_k
        \begin{pmatrix}
            \varphi_{k+1}&:& \varphi_{k}&\rightarrow_{k}&\varphi_{k}^\prime \\
            && &\vdots& \\
            &:& \varphi_2&\rightarrow_2&\varphi_2^\prime \\
            &:& \varphi_1&\rightarrow_1&\varphi_1^\prime
        \end{pmatrix} = 
        \begin{pmatrix}
            \varphi_{k}&:& \varphi_{k-1}&\rightarrow_{k-1}&\varphi_{k-1}^\prime \\
            && &\vdots& \\
            &:& \varphi_2&\rightarrow_2&\varphi_2^\prime \\
            &:& \varphi_1&\rightarrow_1&\varphi_1^\prime
        \end{pmatrix} \\
        \tau_k
        \begin{pmatrix}
            \varphi_{k+1}&:& \varphi_{k}&\rightarrow_{k}&\varphi_{k}^\prime \\
            && &\vdots& \\
            &:& \varphi_2&\rightarrow_2&\varphi_2^\prime \\
            &:& \varphi_1&\rightarrow_1&\varphi_1^\prime
        \end{pmatrix} = 
        \begin{pmatrix}
            \varphi_{k}'&:& \varphi_{k-1}&\rightarrow_{k-1}&\varphi_{k-1}^\prime \\
            && &\vdots& \\
            &:& \varphi_2&\rightarrow_2&\varphi_2^\prime \\
            &:& \varphi_1&\rightarrow_1&\varphi_1^\prime
        \end{pmatrix}\]
    The case $k = d+1$ is defined analogously by taking the appropriate sub-array. It is easy to check these definitions satisfy the ``globular identities''
    \[\sigma_k \circ \sigma_{k+1} = \sigma_k \circ \tau_{k+1} \\
    \tau_k \circ \sigma_{k+1} = \tau_k \circ \tau_{k+1}\]
    so that this is indeed a globular set.

    We extend $\sigma_k$, $\tau_k$ to maps $\forall l \ge 0$
    \[\sigma_k,\tau_k:\CQ(\CA)_{k+l+1} \to \CQ(\CA)_k\]
    by taking the appropriate subarrays as above.

    It is useful to also define ``component functions''. If
    \[ (X) = 
        \begin{pmatrix}
            \varphi_{k+1}&:& \varphi_{k}&\rightarrow_{k}&\varphi_{k}^\prime \\
            && &\vdots& \\
            &:& \varphi_2&\rightarrow_2&\varphi_2^\prime \\
            &:& \varphi_1&\rightarrow_1&\varphi_1^\prime
        \end{pmatrix}\]
    is a $k+1$-globe, then
    \[s_n(X) = \varphi_n \\
    t_n(X) = \varphi_n'.\]
    In this case we write, with a slight abuse of notation
    \[ (X) = 
        \begin{pmatrix}
            X&:& s_k(X)&\rightarrow_{k}&t_k(X) \\
            && &\vdots& \\
            &:& s_1(X)&\rightarrow_1&t_1(X)
        \end{pmatrix}\]
    where $X$ represents either the QCA $\varphi_{k+1}$ when $k+1 \le d+1$ or the unitary $U$ when $(X)$ is a $d+2$-globe.
\end{definition}
\end{mdframed}
\vspace{3mm}

Next we will need to define compositions of $n$-globes. In particular, two $n$-globes will have $n$ different composition actions depending along which direction we glue them. Compare \hyperref[subsecspacefirstpass]{Section \ref{subsecspacefirstpass}}, especially Eqs. \eqref{eqnhorizontal2globecomposition} and \eqref{eqnvertical2globecomposition} which depict the horizontal and vertical compositions of 2-globes.

\vspace{3mm}
\begin{mdframed}[backgroundcolor=blue!5]
\begin{definition}[Globular Composition of QCA]\label{defcomposition}

    Let $k,l \ge 0$, $(X),(Y) \in \CQ(\CA)_{k+l+1}$. We say $(X)$ and $(Y)$ are $k$-composable as $(X) \circ_k (Y)$ if $\sigma_k(X) = \tau_k(Y)$. Equivalently, in terms of their components, $X$ and $Y$ are $k$-composable iff for all $j \le k$,
    \[s_j(X) = s_j(Y) \\
    t_j(X) = t_j(Y).\]

    Composition is defined as follows.
    \begin{enumerate}
        \item Two 1-globes $(X)$ and $(Y)$ are always 0-composable since there is a unique 0-globe $\star$, and composition corresponds to QCA composition
        \[(X) \circ_0 (Y) = (XY).\]
        \item Two 2-globes
        \[(X) = (X:s_1(X) \to_1 t_1(X)) \\
        (Y) = (Y:s_1(Y) \to_1 t_1(Y))\]
        are 1-composable if $s_1(X) = t_1(Y)$. In this case, their 1-composition is (compare \eqref{eqnvertical2globecomposition})
        \[(X) \circ_1 (Y) = (XY: s_1(Y) \to_1 t_1(X)).\]
        Note this works because
        \[X Y: 1 \equiv_1 t_1(X) s_1(X)^{-1} t_1(Y) s_1(Y)^{-1}\]
        and the 1-composability condition means $s_1(X)^{-1} t_1(Y) = 1$, so $(X) \circ_1 (Y)$ is a 2-globe.
        \item In general, two $k+1$-globes $(X)$, $(Y)$ are $k$-composable as $(X) \circ_k (Y)$ if
        \[s_n(X) = t_n(Y) \quad 1 \le n \le {k-1} \\
        s_k(X) = t_k(Y).\]
        Their composition is defined to be
        \[(X) \circ_k (Y) = \begin{pmatrix}
            XY&:& s_k(Y)&\rightarrow_{k}&t_k(X) \\
            && &\vdots& \\
            &:& s_1(Y)&\rightarrow_1&t_1(X)
        \end{pmatrix}\]
        This is a $k+1$-globe, since for $2 \le n \le k$
        \[s_{n}(Y): 1 \equiv_{n-1} t_{n-1}(Y) s_{n-1}(Y)^{-1},\]
        but $t_{n-1}(Y) = t_{n-1}(X)$, so
        \[s_{n}(Y): 1 \equiv_{n-1} t_{n-1}(X) s_{n-1}(Y)^{-1},\]
        and likewise for $t_n(X)$. At the top,
        \[XY: 1 \equiv_k t_k(X) s_k(X)^{-1} t_k(Y) s_k(Y)^{-1},\]
        but $s_k(X)^{-1}t_k(Y) = 1$ by composability. Thus, $(X) \circ_k (Y)$ defined above is a $k+1$-globe.

        \item We need to also define lower compositions between globes. Two 2-globes
        \[(X) = (X:s_1(X) \to_1 t_1(X)) \\
        (Y) = (Y:s_1(Y) \to_1 t_1(Y))\]
        are \emph{always} 0-composable, since there is a unique 0-globe $\star$, and their 0-composition is defined to be (compare \eqref{eqnhorizontal2globecomposition})
        \[(X) \circ_0 (Y) = (X\  {}^{s_1(X)}(Y):s_1(X) s_1(Y) \to_1 t_1(X) t_1(Y)),\]
        where
        \[{}^{\alpha}(\beta) = \alpha\beta \alpha^{-1}\]
        is the conjugation action. $(X) \circ_0 (Y)$ is a 2-globe because
        \[X\ {}^{s_1(X)}(Y): 1 \equiv_1 t_1(X) s_1(X)^{-1} s_1(X) t_1(Y) s_1(Y)^{-1} s_1(X)^{-1} \\
        t_1(X) s_1(X)^{-1} s_1(X) t_1(Y) s_1(Y)^{-1} s_1(X)^{-1} = t_1(X) t_1(Y) (s_1(X) s_1(Y))^{-1}.\]
    
    \item Now we express the general case. Let $(X), (Y) \in \CQ(\CA)_{k+l+1}$ be $k$-composable $k+l+1$-globes, $l,k \ge 0$. We define $(X) \circ_k (Y)$ to be
    \[\label{eqncomposition1}X\ {}^{s_{k+l}(X) \cdots s_{k+1}(X)}(Y)\]
    at the top level, and
     \begin{align}\label{eqncomposition2}
        s_j(X\circ_k Y) &= \begin{cases}
            s_j(Y) & j \leqslant k \\
            s_j(X)\ {}^{s_{j-1}(X) \cdots s_{k+1}(X)}(s_j(Y)) & j > k
        \end{cases} \\
        t_j(X\circ_k Y) &= \begin{cases}
            t_j(X) & j\leqslant k \\
            t_j(X)\ {}^{s_{j-1}(X) \cdots s_{k+1}(X)}(t_j(Y)) & j > k
        \end{cases}
    \end{align}
    We show this defines a $k+l+1$-globe in \hyperref[propwelldefined]{Proposition \ref{propwelldefined}}.
    
    \end{enumerate}

\end{definition}
\end{mdframed}
\vspace{3mm}

Let us turn to the proof that \eqref{eqncomposition1} and \eqref{eqncomposition2} define a $k+l+1$-globe, so the compositions above are well-defined. \\

\begin{mdframed}[backgroundcolor=blue!5]
    \begin{proposition}[Composition Operations Are Well-Defined]\label{propwelldefined}
    Given two $k+l+1$-globes $(X), (Y) \in \CQ(\CA)_{k+l+1}$ which are $k$-composable. The data in \eqref{eqncomposition1} and \eqref{eqncomposition2} define a $k+l+1$-globe $(X) \circ_k (Y)$.
\end{proposition}
\end{mdframed}

\begin{proof}

    We have already shown that $(X) \circ_k (Y)$ is a $k+l+1$-globe when $l = 0$. Therefore, we proceed by induction.

    Suppose that the operation $\circ_k$ is well defined on all $k$-composable pairs of $k+l$-globes. We will show that it is also well defined for composable pairs $(X), (Y)$ of $k+l+1$-globes. If $(X)$ and $(Y)$ are $k$-composable, then $\sigma_{k+1}(X)$ and $\sigma_{k+l}(Y)$ are a composable pair of $k+l$-globes, as are $\tau_{k+1}(X)$ and $\tau_{k+1}(Y)$. We observe that under the definition,
    \[\sigma_{k+l}(X \circ_k Y) = \sigma_{k+1}(X) \circ_k \sigma_{k+l}(Y)\]
    and likewise for $\tau_{k+1}$. Therefore, by the inductive hypothesis $(X) \circ_k (Y)$ satisfies
    \[s_n(X \circ_k Y), t_n(X \circ_k Y): s_{n-1}(X \circ_k Y) \to_{n-1} t_{n-1}(X \circ_k Y)\]
    for all $n \le k+l+1$. The only remaining thing to show is that
    \[X\ {}^{s_{k+l}(X) \cdots s_{k+1}(X)}(Y): s_{k+l}(X \circ_k Y) \to_{k+l} t_{k+1}(X \circ_k Y) \quad (?).\]
    Recall we are interested in $l > 0$, so we can write (using \eqref{eqncomposition2})
    \[s_{k+l}(X \circ_k Y) = s_j(X)\ {}^{s_j(X) \cdots s_{k+1}(X)}(s_j(Y)) \\
    t_{k+l}(X \circ_k Y) = t_j(X)\ {}^{s_j(X) \cdots s_{k+1}(X)}(t_j(Y)).\]
    Now we compute
    \[X\ {}^{s_{k+l}(X) \cdots s_{k+1}(X)}(Y): 1 \equiv_{k+l} t_{k+l}(X) s_{k+l}(X)^{-1}\ {}^{s_{k+l}(X) \cdots s_{k+1}(X)}( t_{k+l}(Y) s_{k+1}(Y)^{-1}).\]
    We need to check the RHS is correct:
    \[
    t_{k+l}(X) s_{k+l}(X)^{-1}\ {}^{s_{k+l}(X) \cdots s_{k+1}(X)}( t_{k+l}(Y) s_{k+l}(Y)^{-1}) \\
    = t_{k+l}(X)\ {}^{s_{k+l-1}(X) \cdots s_{k+1}(X)}( t_{k+l}(Y) s_{k+l}(Y)^{-1}) s_{k+l}(X)^{-1} \\
    = t_{k+l}(X)\ {}^{s_{k+l-1}(X) \cdots s_{k+1}(X)}( t_{k+l}(Y)) \left(s_{k+l}(X)\ {}^{s_{k+l-1}(X) \cdots s_{k+1}(X)}( s_{k+l}(Y)) \right)^{-1},\]
    which agrees with \eqref{eqncomposition2}.

\end{proof}

For computations, it is sometimes useful to express the arrays of blends in the more familiar balanced form. That is, given a $k+1$-globe $(X)$, we can define an array
\[\label{eqnbalancedpresentation}\begin{Bmatrix}
            X s_k(X) \cdots s_1(X) &:& s_k(X) s_{k-1}(X) \cdots s_1(X) &\equiv_{k}&t_k(X) s_{k-1}(X) \cdots s_1(X) \\
            && &\vdots& \\
            &:& s_2(X) s_1(X)&\equiv_2&t_2(X) s_1(X) \\
            &:& s_1(X)&\equiv_1&t_1(X)
        \end{Bmatrix}\]
This is an equivalent presentation of the $k+1$-globe $(X)$, which matches the pictures in \hyperref[subsecspacefirstpass]{Section \ref{subsecspacefirstpass}}. As noted above, any two $k+1$-globes $(X)$ and $(Y)$ are 0-composable. Their zero composition has a simple form in this balanced presentation. Let $A_j(X) = s_j(X) \cdots s_1(X)$, $A_{k+1}(X) = X s_k(X) \cdots s_1(X)$, $B_j(X) = t_j(X) s_{j-1}(X) \cdots s_1(X)$. The 0-composition of $(X)$ and $(Y)$ may be written
\[\label{eqnzerocomposition}(X) \circ_0(Y) = \begin{Bmatrix}
            A_{k+1}(X) A_{k+1}(Y) &:& A_k(X) A_k(Y) &\equiv_{k}& B_k(X) B_k(Y) \\
            && &\vdots& \\
            &:& A_2(X) A_2(Y)&\equiv_2& B_2(X) B_2(Y) \\
            &:& A_1(X) A_1(Y) &\equiv_1& B_1(X) B_1(Y)
        \end{Bmatrix}\]

So far we have constructed a globular set $\{\CQ(\CA)_n\}_n$ with compositions defined for globes. It is easy to show each of these compositions is in fact associative and invertible, so each $\CQ(\CA)_n$ becomes a group in $n$ ways. However, one finds that these different group structures on $\CQ(\CA)_n$ do not satisfy the ``interchange law'', and so this structure \emph{does not} define a strict $d+2$-group\footnote{It is easy to show, however, that the interchange law is always satisfied \textit{up to a higher morphism}. This is because the two orders of evaluation are related by commutators of QCA, which are expressible as circuits by an argument similar to \hyperref[propinvertibilityqcaentangler]{Propositon \ref{propinvertibilityqcaentangler}}, and so both orders of evaluation are blend equivalent. All of the interchange laws involving the vertical composition of unitary cells are satisfied on-the-nose.}. However, it does define a ``weak'' $d+2$-group (and hence a homotopy $d+2$-type). Checking this is extremely tedious, and so we postpone it to a follow-up work \cite{followup}. Once this is checked, we can apply all of the usual methods for studying topological spaces up to homotopy to the study of $\CQ(\CA)$ and lattice anomalies.

The above definitions are already sufficient for many calculations. Here is one important one:
\vspace{3mm}
\begin{mdframed}[backgroundcolor=blue!5]
\begin{proposition}\label{prophomotopycalc}
    Let $\CA$ be a local algebra on $\IZ^d$. $\CQ(\CA)$ has the following homotopy groups
    \[\pi_0 \CQ(\CA) = \star \\
    \pi_n\CQ(\CA) = \frac{\{\text{QCAs supported on }[a,b]^{n-1} \times \IZ^{d-n+1}\text{ for some $a$, $b$}\}}{\text{blends along the $n$th axis}} \quad 1 \le n \le d \\
    \pi_{d+1}\CQ(\CA) = \frac{\text{QCAs supported on }[a,b]^d\text{ for some $a,b$}}{\text{local unitaries}} \\
    \pi_{d+2}\CQ(\CA) = U(1) \quad \text{(with discrete topology)}\\
    \pi_{>d+2} \CQ(\CA) = 0.\]
\end{proposition}
\end{mdframed}
\vspace{3mm}
\begin{proof}
    By construction $\CQ(\CA)$ is connected since it has a single 0-globe. Its homotopy groups $\pi_n \CQ(\CA)$ may be computed\footnote{We are claiming here with proof to be postponed to \cite{followup} that the ``obvious'' definition of homotopy groups in this globular set will be the right one, applying a theorem of Ara \cite{Ara_2013}.} by considering the set of $n$-globes $(X) \in \CQ(\CA)_n$ satisfying
    \[s_m(X) = t_m(X) = 1 \quad \forall m \le n,\]
    i.e.
    \[(X) = \begin{pmatrix}
            X&:& 1&\rightarrow_{n-1}&1 \\
            && &\vdots& \\
            &:& 1&\rightarrow_1&1
        \end{pmatrix}\]
    which form a group $C_n$ under any $k$-composition for $k < n$, which are all isomorphic. Let $1_n$ be the identity in this group, which can be written
    \[1_n = \begin{pmatrix}
            1&:& 1&\rightarrow_{n-1}&1 \\
            && &\vdots& \\
            &:& 1&\rightarrow_1&1
        \end{pmatrix}\]
    We consider the quotient of $C_n$ by the (normal) subgroup $C_{n,0}$ of those $n$-globes $(X)$ as above for which there further exists an $n+1$-globe $(Y)$ such that $\sigma_n(Y) = 1_n$ and $\tau_n(Y) = (X)$. We have
    \[\pi_n\CQ(\CA) = C_n/C_{n,0}.\]
    Unpacking the definition, there are four cases:
    
    1. Let $n \le d$. An $n$-globe $(X) \in C_n$ is determined by a QCA $X$ which is supported in $[a,b]^{n-1} \times \IZ^{d-n+1}$, and it is in $C_{n,0}$ iff it admits a blend to the identity along the $n$th axis.

    2. If $n = d+1$, an $d+1$-globe $(X) \in C_{d+1}$ is determined by a QCA $X$ supported in $[a,b]^d$ for some $a,b$. It is in $C_{d+1,0}$ iff it is $\text{Ad }U$ for some local unitary $U$. In a tensor product bosonic Hilbert space, $\pi_{d+1} = 0$. However, in a fermionic superalgebra we are interested in taking QCAs which commute with fermion parity as QCAs, meaning up to a phase, but taking only local unitaries which commute with fermion parity on the nose. Thus, $\pi_{d+1} = \IZ_2$. See also \hyperref[secanomalousferms1d]{Section \ref{secanomalousferms1d}}.

    3. If $n = d+2$, $C_{d+2}$ is determined by a local unitary operator $X$, but that local unitary has to satisfy $\text{Ad }X = 1$ as a QCA, so $X$ is a phase. Therefore $\pi_{d+2} = U(1)$.

    4. If $n > d+2$, $C_n = 0$ and $\pi_n = 0$.
\end{proof}

To extend the above definitions to the case of stable $d$-QCA as in \hyperref[subsecstabilization]{Section \ref{subsecstabilization}}, we simply redefine
\[\alpha: \beta \to_n \gamma\]
to mean that $\alpha, \beta,\gamma$ are stable $d$-QCAs, and $\alpha$ defines a blend of stable $d$-QCAs
\[\alpha:1 \equiv_{d-m} 1 \quad m < n, \\
\alpha: 1 \equiv_{d-n} \beta \gamma.\]
The construction also clearly extends to QCAs satisfying various conditions. For example, we can restrict all QCA to preserve a chosen product state $\psi_0$, as in \hyperref[subsecrelationtoinvertiblestates]{Section \ref{subsecrelationtoinvertiblestates}}. Or we can restrict to those commuting with an on-site symmetry as in \hyperref[secsptandbulkboundary]{Section \ref{secsptandbulkboundary}}.

\subsection{Else-Nayak Revisited and Computing the Anomaly Indices}\label{subsecelsenayak}

Now we want to express the obstruction-theoretic anomaly indices introduced in \hyperref[subseccomptheanom]{Section \ref{subseccomptheanom}} in terms of the $\infty$-groupoid data expressed above and describe how to obtain explicit cocycle formulas for them. Our construction gives an implementation of the proposal of Else and Nayak \cite{Else_2014}. We will outline a construction which in principle works in every dimension, which needs to be computed once per dimension. We give calculations up to the third anomaly index, leaving an algorithmic determination of higher indices to future work.

We begin with a QCA $G$-representation $\alpha$, which defines a map
\[\alpha:BG \to \CQ(\CA).\]
We want to know whether this map is null-homotopic. In other words, we want to know if there exists a map
\[\beta:BG \times[0,1] \to \CQ(\CA)\]
such that $\beta|_{BG \times 0}$ is a constant map and $\beta|_{BG\times1} = \alpha$.

To phrase this in a cellular way, we can think of the simplicial structure of $BG$ giving us a cell decomposition of $BG\times[0,1]$ in terms of prisms $\Delta^n \times [0,1]$. The map $\beta$ defines a labeling of these prisms such that
\begin{enumerate}
    \item Every vertex is labeled by the unique object $\star$ of $\CQ(\CA)$.
    \item Every $k$-simplex, $k>0$ of $BG \times 0$ is labeled with an identity $k$-morphism.
    \item The 1-simplices of $BG \times 1$, which are determined by elements $g \in G$, are labeled with $\alpha_g$.
    \item The $k$-simplices, $k > 1$ of $BG \times 1$ are labeled with identity $k$-morphisms.
    \item The 1-simplex $\star \times [0,1]$ is labeled with the identity 1-morphism.
    \item All other ``interior'' cells will be determined by $\beta$.
\end{enumerate}
This suggest an iterative procedure to construct the null-homotopy $\beta$ where we first build the skeleton $X_0$ where only the first five types of cells are filled in, and then we add the remaining ``interior'' $k$-cells one level at a time, defining a sequence of spaces
\[X_0 \subset X_1 \subset \cdots  \\
\bigcup_{n=0}^\infty X_n = BG \times [0,1].\]
This gives a dual presentation (homotopy extension problem rather than homotopy lifting problem \cite{hatcher2005algebraic}) of the same obstruction sequence outlined in \hyperref[subseccomptheanom]{Section \ref{subseccomptheanom}}.

Suppose we have constructed the map $\beta_k:X_k \to \CQ(\CA)$ and we now want to extend it to $X_{k+1}$. The $k+1$-cells of $X_{k+1}$ are prisms $\Delta^k \times [0,1]$ where $\Delta^k$ is a $k$-simplex of $BG$, which is labeled by a sequence of $k$-group elements $g_1,\ldots,g_k$. The boundary of the prism is topologically a $k$-sphere with a based map to $\CQ(\CA)$ defined by $\beta_k$, and we can measure its class in $\pi_k\CQ(\CA)$. These combine into a group cohomology class \cite{hatcher2005algebraic}
\[ [\beta_k] \in H^k(BG,\pi_k \CQ(\CA)),\]
which is equal to the $k$th anomaly index $[c^{k-1} \circ \alpha^{k-1}]$ defined in \hyperref[subseccomptheanom]{Section \ref{subseccomptheanom}}. We will see that this presentation of the obstruction theory closely matches the proposal of Else and Nayak \cite{Else_2014}.

To have an explicit cocycle for the $k$th anomaly index, all we need is a way of computing the homotopy class of the boundary of a prism $\Delta^k \times[0,1]$ with labels given by $\beta_k$. Homotopy groups in the globular picture appear naturally as the obstruction to filling in a hollow globe (see \hyperref[prophomotopycalc]{Proposition \ref{prophomotopycalc}}). Thus, we will need to evaluate hollow prisms into hollow globes. There is no canonical way of doing this, although quite general algorithms exist (see eg. ``excision of extremals'' algorithm in \cite{street1991parity}). Coherence conditions imply that any two evaluations are homotopy equivalent, so any method of evaluating the diagram will give equivalent anomaly indices. Furthermore, every prism has the same shape, and is labeled by $k$ abstract group elements, so we need only do one computation for each $k$ to have a formula for $[\beta_k]$ for every group $G$, every dimension $d$, and every local algebra $\CA$ at once.

For example, an interior $2$-cell is determined by a single element $g \in G$ and must be filled in with a diagram of $\CQ(\CA)$ as follows:
\begin{center}
        \begin{tikzpicture}[scale=0.5]
        \node (A) at (-2, 0) {$*$};
        \node (B) at (2, 0) {$*$};
        \node (A1) at (-2, -4) {$*$};
        \node (B1) at (2, -4) {$*$};

        \draw [double distance=2pt, arrows={-Classical TikZ Rightarrow} ] 
        (0, -3.5) -- node[left] {$\beta_{g}$} (0, -0.2);

        \path[commutative diagrams/.cd, every arrow, every label]
        (A) edge node[above] {$\alpha_{g}$} (B)

        (A1) edge node[above] {$1$} (B1)

        (A1) edge node[left] {$1$} (A)
        (B1) edge node[right] {$1$} (B);

    \end{tikzpicture}

    \end{center}
Here we draw $BG \times 1$ on the top and $BG \times 0$ on the bottom, so the top 1-simplex is labeled by our QCA representation $\alpha_g$ and other 1-simplices are labeled by the identity. To be a diagram in $\CQ(\CA)$, $\beta_g$ must be an appropriate 2-morphism. In this diagram, it goes from $1 \circ 1$ on the right and bottom sides to $\alpha_g \circ 1$ on the top and left sides. It is thus equivalent to the 2-globe
\begin{center}
\label{eqn2globe}
\begin{tikzcd}[column sep=huge]
\star \arrow[r, bend left=40, "\alpha_g"{name=f}] 
  \arrow[r, bend right=40, "1"'{name=g}] 
  & \star
  \arrow[Rightarrow, from=g, to=f, "\beta_g"]
\end{tikzcd}
\end{center}
We recognize $\beta_g$ as defining a blend $\beta_g:1 \equiv_1 \alpha_g$. The obstruction for the existence of this blend is precisely $[\alpha_g] \in \pi_1 \CQ(\CA)$. Therefore we find
\[ [\beta_1] = [\alpha_g] \in H^1(BG,\pi_1 \CQ(\CA))\]
is the obstruction to choosing blends along the first axis for each of the QCAs $\alpha_g$, as argued in \hyperref[subseccomptheanom]{Section \ref{subseccomptheanom}} from the homotopy lifting problem.

Interior 3-cells take the following form
\begin{center}
    \includegraphics[width=5cm]{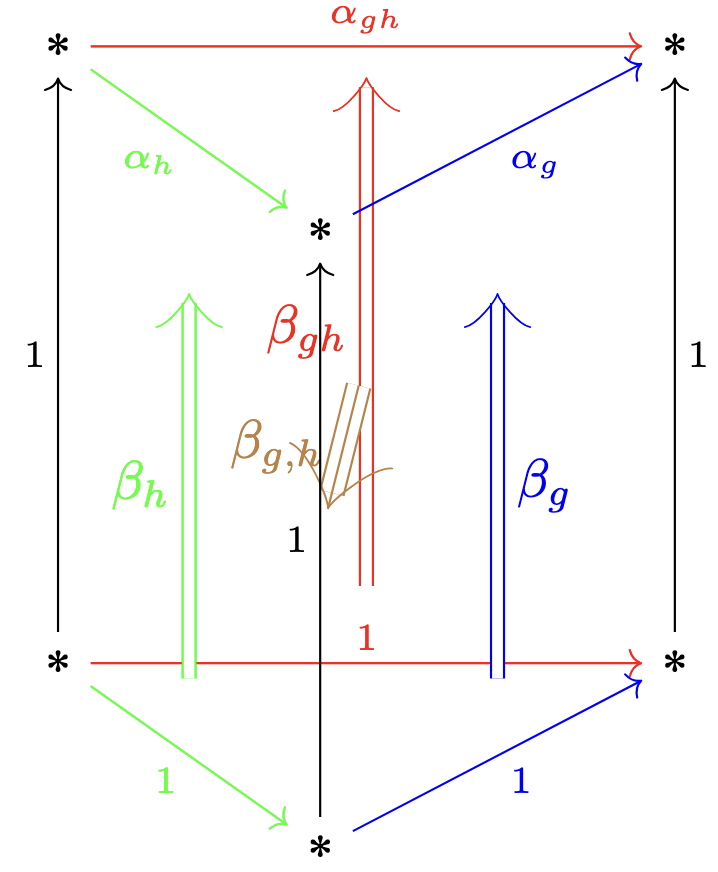}
\end{center}

These need to be filled in with a 3-morphism $\beta_{g,h}$ as shown in orange. One strategy to evaluate the diagram is to evaluate the ``front'' and ``back'' as 2-globes. Then the hollow prism is the hollow 3-globe with these 2-globes as source and target. The front of the diagram consists of
\begin{center}
    \includegraphics[width=5cm]{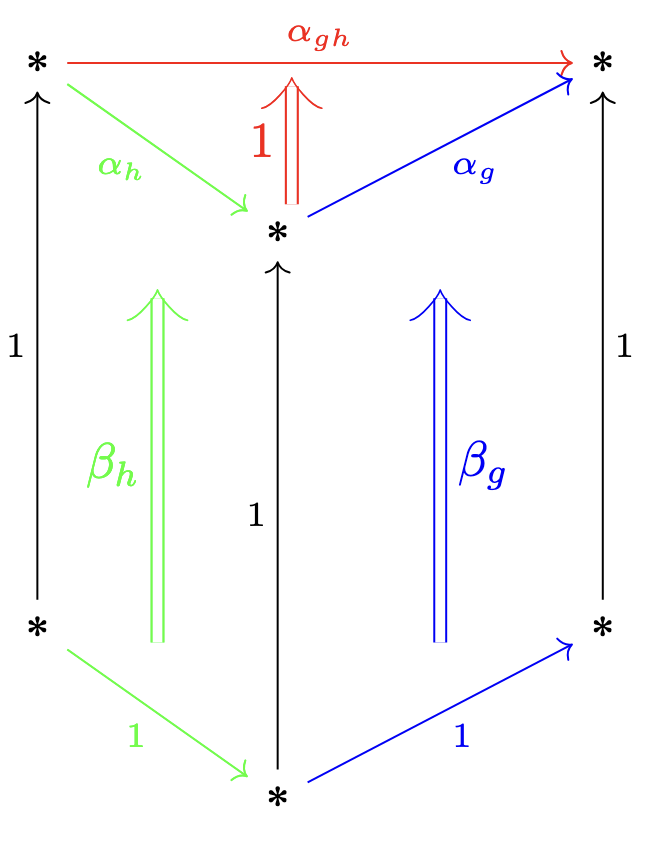}
\end{center}
(the top triangle has the identity 2-morphism which was not shown above for simplicity, since it lies along $BG \times 1$). Since they are joined along an identity morphism, we can collapse the quadrilateral to 2-globes and evaluate them:
\begin{equation}\label{eqnhorizontal2globecomposition}
\begin{tikzcd}[column sep=huge]
\star \arrow[r, bend left=40, "\alpha_h"{name=f}] 
  \arrow[r, bend right=40, "1"'{name=g}] 
  & \star
  \arrow[Rightarrow, from=g, to=f, "\beta_h"] \arrow[r, bend left=40, "\alpha_g"{name=f2}] 
  \arrow[r, bend right=40, "1"'{name=g2}] &  \arrow[Rightarrow, from=g2, to=f2, "\beta_g"]\star
\end{tikzcd} = \begin{tikzcd}[column sep=huge]
\star \arrow[r, bend left=40, "\alpha_g\alpha_h"{name=f}] 
  \arrow[r, bend right=40, "1"'{name=g}] 
  & \star
  \arrow[Rightarrow, from=g, to=f, "\beta_g \beta_h"]
\end{tikzcd}
\end{equation}
where we used the rules for 0-composition of 2-globes. When we paste the identity morphism on top, this becomes simply
\begin{equation}\label{eqntarget2globe}\begin{tikzcd}[column sep=huge]
\star \arrow[r, bend left=40, "\alpha_{gh}"{name=f}] 
  \arrow[r, bend right=40, "1"'{name=g}] 
  & \star
  \arrow[Rightarrow, from=g, to=f, "\beta_g \beta_h"]
\end{tikzcd}\end{equation}
Meanwhile, the back of the diagram evaluates to
\begin{equation}\label{eqnsource2globe}\begin{tikzcd}[column sep=huge]
\star \arrow[r, bend left=40, "\alpha_{gh}"{name=f}] 
  \arrow[r, bend right=40, "1"'{name=g}] 
  & \star
  \arrow[Rightarrow, from=g, to=f, "\beta_{gh}"]
\end{tikzcd}\end{equation}
Thus, $\beta_{g,h}$ is a 3-globe whose source is \eqref{eqntarget2globe} and whose target is \eqref{eqnsource2globe}. We can express it as an array
\[(\beta_{g,h}) = \begin{pmatrix}
            \beta_{g,h}&:& \beta_{gh}&\rightarrow_{2}&\beta_{g} \beta_h \\
            &:& 1&\rightarrow_1& \alpha_{gh}
        \end{pmatrix}\]
In particular, it defines a blend
\[\beta_{g,h}:1 \equiv_2 \beta_g \beta_h\beta_{gh}^{-1}.\]
This exists iff $\beta_g \beta_h \beta_{gh}^{-1}$ is trivial in $\pi_2 \CQ(\CA)$. Thus,
\[ [\beta_2] = [\beta_g \beta_h \beta_{gh}^{-1}] \in H^2(BG,\pi_2\CQ(\CA)).\]
This invariant was recently discussed in \cite{shirley2025anomalyfreesymmetriesobstructionsgauging,tu2025anomaliesglobalsymmetrieslattice} as an obstruction to on-siteability in two dimensions. Here we see how it appears from homotopy theory.

Note that this calculation does not explicitly refer to the dimension $d$ of the lattice. It looks the same in all dimensions $d \ge 0$. In $d = 0$, it has a special interpretation. $\beta_g:1 \to_1 \alpha_g$ in $d = 0$ correspond to unitary operators $U_g$ implementing $\alpha_g = \text{Ad}\ U_g$ (see \eqref{eqntopmorphism}). In this dimension $\pi_2 \CQ(\CA) = U(1)$ and the class $[\beta_2] = U_g U_h U_{gh}^{-1} \in H^2(BG,U(1))$ represents the class of the projective representation $\alpha_g$.

Interior 4-cells are four dimensional prisms whose ``front'' and ``back'' faces look like the following

\begin{center}
    \includegraphics[width=12cm]{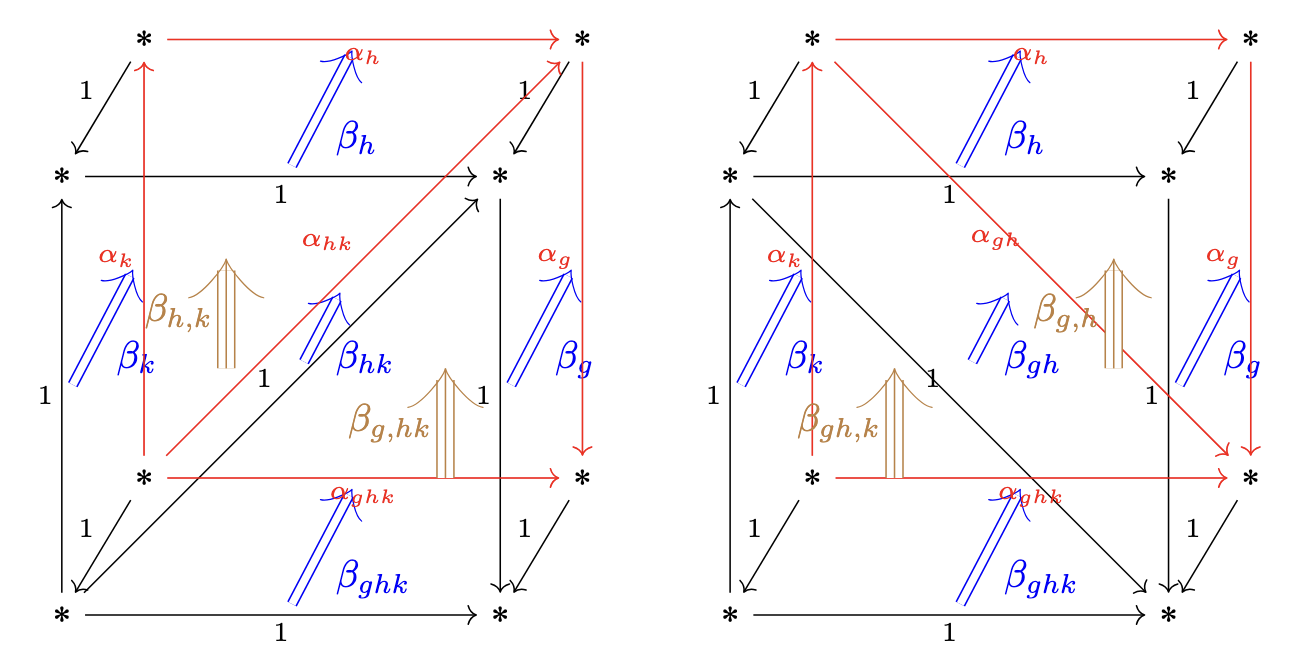}
\end{center}
This prism is to be filled with the 4-cell $\beta_{g,h,k}$. As in $k=2$, we just need to evaluate the front and back of the diagram as 3-globes to get the 3rd anomaly index.

At this point, it becomes convenient to draw our prisms, which always look like $\Delta^n \times [0,1]$ for some $n$, as simplices $\Delta^n$. 1-simplices are labeled by $\beta_g$, 2-simplices by $\beta_{g,h}$, and so on. The front and back of the 4-prism above become
\begin{equation}
\begin{tikzpicture}[>=latex, scale=1.2]
  \node (A) at (0,2) {$\star$};
  \node (B) at (2,2) {$\star$};
  \node (C) at (0,0)   {$\star$};
  \node (D) at (2,0)   {$\star$};

  \draw[->] (A) -- node[above] {$\beta_h$} (B);
  \draw[->] (C) -- node[left]  {$\beta_k$} (A);
  \draw[->] (B) -- node[right] {$\beta_g$} (D);
  \draw[->] (C) -- node[below] {$\beta_{ghk}$} (D);

  \draw[->] (C) -- (B);
  
  \node at (1.6,1.2) {$\beta_{hk}$};

  \node at (0.7,1.5) {$\Uparrow\beta_{h,k}$};

  \node at (1.3,0.5) {$\Uparrow\beta_{g,hk}$};
\end{tikzpicture}
\end{equation}

\begin{equation}
    \begin{tikzpicture}[>=latex, scale=1.2]
  
  \node (A) at (0,2) {$\star$};
  \node (B) at (2,2) {$\star$};
  \node (C) at (0,0)   {$\star$};
  \node (D) at (2,0)   {$\star$};

  \draw[->] (A) -- node[above] {$\beta_h$} (B);
  \draw[->] (C) -- node[left]  {$\beta_k$} (A);
  \draw[->] (B) -- node[right] {$\beta_g$} (D);
  \draw[->] (C) -- node[below] {$\beta_{ghk}$} (D);

  \draw[->] (A) -- (D);
  
  \node at (0.5,1.2) {$\beta_{gh}$};

  \node at (1.5,1.5) {$\Uparrow\beta_{g,h}$};

  \node at (0.7,0.5) {$\Uparrow\beta_{gh,k}$};
\end{tikzpicture}
\end{equation}
We have labeled the cells with the corresponding cells from $\beta_3$. However, we have to express this composition in terms of globes. For example, we have to write
\begin{equation}
\begin{tikzpicture}[>=latex, scale=1,  baseline=(current bounding box.center)]
  
  \node (B) at (2,0) {$\star$};
  \node (C) at (0,-2)   {$\star$};
  \node (D) at (2,-2)   {$\star$};

  \draw[->] (B) -- node[right] {$\beta_g$} (D);
  \draw[->] (C) -- node[below] {$\beta_{ghk}$} (D);

  \draw[->] (C) -- (B);
  
  \node at (0.8,-0.8) {$\beta_{hk}$};

  \node at (1.3,-1.5) {$\Uparrow\beta_{g,hk}$};
\end{tikzpicture} = \begin{tikzcd}[column sep=huge]
\star \arrow[r, bend left=40, "\beta_g\beta_{hk}"{name=f}] 
  \arrow[r, bend right=40, "\beta_{ghk}"'{name=g}] 
  & \star
  \arrow[Rightarrow, from=g, to=f, "\beta_{g,hk}"]
\end{tikzcd}
\end{equation}
Here the RHS is drawn as a 2-globe, but it is shorthand for the 3-globe
\[\label{eqnfirst3globe}(\beta_{g,hk}) = \begin{pmatrix}
            \beta_{g,hk}&:& \beta_{ghk}&\rightarrow_{2}&\beta_{g} \beta_{hk} \\
            &:& 1&\rightarrow_1& \alpha_{ghk}
        \end{pmatrix}\]
To compose with the triangle above, they must have the same source and target, so we have to take the ``whiskered'' globe
\begin{equation}
\begin{tikzpicture}[>=latex, scale=1,  baseline=(current bounding box.center)]
  
  \node (A) at (0,0) {$\star$};
  \node (B) at (2,0) {$\star$};
  \node (C) at (0,-2)   {$\star$};
  \node (D) at (2,-2)   {$\star$};

  \draw[->] (A) -- node[above] {$\beta_h$} (B);
  \draw[->] (C) -- node[left]  {$\beta_k$} (A);
  \draw[->] (B) -- node[right] {$\beta_g$} (D);

  \draw[->] (C) -- (B);
  
  \node at (1.3,-1.2) {$\beta_{hk}$};

  \node at (0.7,-0.5) {$\Uparrow\beta_{h,k}$};

\end{tikzpicture}
=
\begin{tikzcd}[column sep=huge]
\star \arrow[r, bend left=40, "\beta_{h}\beta_{k}"{name=f}] 
  \arrow[r, bend right=40, "\beta_{hk}"'{name=g}] 
  & \star
  \arrow[Rightarrow, from=g, to=f, "\beta_{h,k}"] \arrow[r,"\beta_g"] & \star
\end{tikzcd}
=
\begin{tikzcd}[column sep=huge]
\star \arrow[r, bend left=40, "\beta_g\beta_{h}\beta_{k}"{name=f}] 
  \arrow[r, bend right=40, "\beta_g\beta_{hk}"'{name=g}] 
  & \star
  \arrow[Rightarrow, from=g, to=f, "{}^{\beta_g}\beta_{h,k}"]
\end{tikzcd}
\end{equation}
where the RHS is computing using the 0-composition of 3-globes, giving (compare \eqref{eqncomposition1})
\[\label{eqnsecond3globe}\begin{pmatrix}
            1&:& \beta_{g}&\rightarrow_{2}&\beta_{g}  \\
            &:& 1&\rightarrow_1& \alpha_{g}
        \end{pmatrix} \circ_0 \begin{pmatrix}
            \beta_{h,k}&:& \beta_{hk}&\rightarrow_{2}& \beta_{hk} \\
            &:& 1&\rightarrow_1& \alpha_{hk}
        \end{pmatrix} =\begin{pmatrix}
            {}^{\beta_g}\beta_{h,k}&:& \beta_{g}\beta_{hk}&\rightarrow_{2}&\beta_{g} \beta_{hk} \\
            &:& 1&\rightarrow_1& \alpha_{ghk}
        \end{pmatrix} = ({}^{\beta_g}\beta_{h,k})\]
where ${}^{\beta_g}\beta_{h,k} = \beta_g\beta_{h,k}\beta_g^{-1}$. Then the 2-composition of the two 3-globes \eqref{eqnfirst3globe} and \eqref{eqnsecond3globe} gives
\[\label{eqntarget}(\beta_{g,hk}) \circ_2 ({}^{\beta_g}\beta_{h,k}) = \begin{pmatrix}
            {}^{\beta_g}\beta_{h,k} \beta_{g,hk}&:& \beta_{g}\beta_{h} \beta_k&\rightarrow_{2}&\beta_{ghk} \\
            &:& 1&\rightarrow_1& \alpha_{ghk}
        \end{pmatrix}\]
The evaluation of the other square is similar. The interesting piece to obtain is
\begin{equation}
    \begin{tikzpicture}[>=latex, scale=1,baseline=(current bounding box.center)]
  \node (A) at (0,2) {$\star$};
  \node (B) at (2,2) {$\star$};
  \node (C) at (0,0)   {$\star$};
  \node (D) at (2,0)   {$\star$};

  \draw[->] (A) -- node[above] {$\beta_h$} (B);
  \draw[->] (C) -- node[left]  {$\beta_k$} (A);
  \draw[->] (B) -- node[right] {$\beta_g$} (D);

  \draw[->] (A) -- (D);
  
  \node at (0.5,1.2) {$\beta_{gh}$};

  \node at (1.5,1.5) {$\Uparrow\beta_{g,h}$};

\end{tikzpicture}
=
\begin{tikzcd}[column sep=huge]
\star \arrow[r,"\beta_k"] & \star \arrow[r, bend left=40, "\beta_{g}\beta_{h}"{name=f}] 
  \arrow[r, bend right=40, "\beta_{gh}"'{name=g}] 
  & \star
  \arrow[Rightarrow, from=g, to=f, "\beta_{g,h}"]
\end{tikzcd}
=
\begin{tikzcd}[column sep=huge]
\star \arrow[r, bend left=40, "\beta_g\beta_{h}\beta_{k}"{name=f}] 
  \arrow[r, bend right=40, "\beta_g\beta_{hk}"'{name=g}] 
  & \star
  \arrow[Rightarrow, from=g, to=f, "\beta_{g,h}"]
\end{tikzcd}
\end{equation}
On this side the whiskering does not modify $\beta_{g,h}$ because of the asymmetry of \eqref{eqncomposition1}. Combining with the other triangle, we obtain
\[\label{eqnsource}(\beta_{g,h}) \circ_2 (\beta_{gh,k}) = \begin{pmatrix}
            \beta_{g,h} \beta_{gh,k}&:& \beta_{g}\beta_{h} \beta_k&\rightarrow_{2}&\beta_{ghk} \\
            &:& 1&\rightarrow_1& \alpha_{ghk}
        \end{pmatrix}\]
The 3rd anomaly index is therefore
\[\label{eqnelsenayakh3index} [\beta_3] = [{}^{\beta_g}(\beta_{h,k}) \beta_{g,hk} (\beta_{g,h} \beta_{gh,k})^{-1}] \in H^3(BG,\pi_3\CQ(\CA)).\]
This gives rise to the familiar $H^3(BG,U(1))$ Else-Nayak index when $d = 1$. In higher dimensions, we see it is valued in blend equivalence classes of QCA, and defines a new obstruction to on-siteability.

\section{Anomalous 1d $\IZ_2$ Symmetries in Fermionic Systems and the Homotopy Type of $\CQ^1$, $\CQ^1_{inv}$}\label{secanomalousferms1d}

In this section, we will compute the homotopy type of $\CQ^1$ and $\CQ^1_{inv}$ for a 1d lattice of bosons and fermions. This will demonstrate the computational techniques homotopy theory makes available to us for analyzing lattice anomalies. We find some curious differences with the lore from the continuum and the Anderson dual spin cobordism. However, we show that the results of our calculations are physically sensible and give correct predictions.

It is expected based on the spin cobordism group $\Omega^3_{\rm spin}(B\IZ_2) = \IZ_8$ that there is a $\IZ_8$ classification of $\IZ_2 \times \IZ_2^F$ SPTs in 2d as well as anomalies of $\IZ_2 \times \IZ_2^F$ symmetries in 1d fermionic systems \cite{kapustin2014symmetryprotectedtopologicalphases}. We will show in fact that the stable lattice anomalies of QCA $\IZ_2$-representations is
\[\CQ^1_f(B\IZ_2) = \IZ_4,\]
where the subscript refers to the local Hilbert space being a two-state Fermionic Fock space. This one calculation actually allows us to completely characterize the homotopy type of $\CQ^1_f$.

We will also show that
\[\CQ^1_{f,inv}(B\IZ_2) = \IZ_4 \times \IZ_2.\]
We find in particular that $\CQ^1_f$ is \emph{not} equal to the expected cobordism spectrum from \cite{kapustin2014symmetryprotectedtopologicalphases,Freed_2021}. However, we will show that the group above correctly classifies the SRE anomalies of symmetries $\alpha$ which are $\IZ_2$ in the sense that their square $\alpha^2$ admits symmetric SRE states. Therefore, this group just has a different interpretation.

The absence of the generators of $\IZ_8$ has been anticipated in several places, see eg. Appendix G of \cite{Ellison_2019} and \cite{jones2021one}. The basic reason is that such anomalous $\IZ_2$ symmetries should map a product state to the Kitaev chain state, a particular invertible 1d fermionic state. However, any such entangler must be a nontrivial 1d QCA since this fermionic state does not blend to a product state (it has a Majorana edge mode). The group of 1d QCA is known to be non-torsion \cite{Fidkowski_2019}, and therefore no $\IZ_2$ symmetry can by a non-trivial QCA. Considering the bulk-boundary correspondence in \hyperref[secsptandbulkboundary]{Section \ref{secsptandbulkboundary}}, this also means that the corresponding 2+1d SPTs do not admit symmetric circuit disentanglers \cite{Ellison_2019}.
=

\begin{figure}
    \centering
    \includegraphics[width=0.7\linewidth]{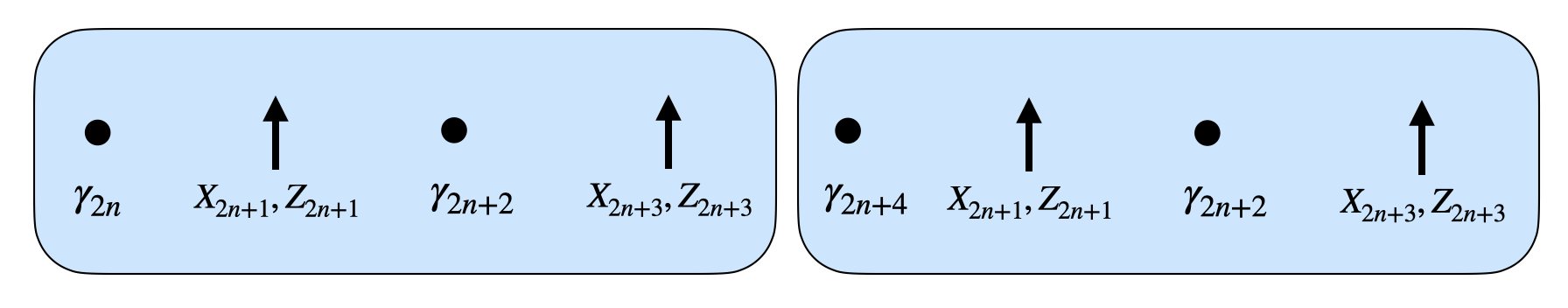}
    \caption{This figure shows the generators of a 1d local fermionic superalgebra with Majorana operators interspersed between qubit operators. The blue boxes show one possible grouping into unit cells each defining a matrix superalgebra. This local algebra is used in Section \ref{secanomalousferms1d} to construct anomalous $\IZ_2$ symmetries.}
    \label{fig1dfermionicchain}
\end{figure}

\vspace{3mm}
  \begin{mdframed}[backgroundcolor=blue!5]
\begin{proposition}\label{propnonzeropostnikovinvariant}
    Let $\CQ^1_f$ be the space of stable 1-QCA on the local superalgebra built from two-state fermion Fock spaces. We have
    \[\pi_1 \CQ^f_1 = \IZ \\
    \pi_2 \CQ^f_1 = \IZ_2 \\
    \pi_3 \CQ^f_1 = U(1) \\
    \pi_{>3} \CQ^f_1 = 0.\]
    We also compute
    \[\CQ^f_1(B\IZ_2) = \IZ_4.\]
    It follows that in general dimensions there is a non-trivial Postnikov invariant between $\pi_{d+2} \CQ_f^d = U(1)$ and $\pi_{d+1} \CQ_f^d = \IZ_2$ given by
    \[i \circ Sq^2: B^{d+1}\IZ_2 \to B^{d+3}U(1)_{disc}\]
    where $disc$ denotes the discrete topology on $U(1)$ and $i$ is the inclusion map $\IZ_2 \to U(1)$. For $d = 1$, we thus completely characterize the homotopy type $\CQ^1_f$.
\end{proposition}
\end{mdframed}
\vspace{3mm}
\begin{proof}
    The computation of the homotopy groups of $\CQ^1_f$ are a special case of \hyperref[prophomotopycalc]{Proposition \ref{prophomotopycalc}}. An interesting one in arbitrary dimensions is
    \[\pi_{d+1} \CQ^d_f = \pi_2 \CQ^1_f = \pi_1 Q^0_f = \IZ_2.\]
    Indeed, we require fermionic stable 0-QCA to commute with fermion parity \emph{as QCA}. Thus, when we write them as local unitary operators they are either parity even or parity odd. If they are parity odd, then they cannot be expressed as local bosonic unitaries, and represent a non-trivial element of $\pi_1 Q_0^f$. On the other hand, if we square such an element, it becomes parity even, and can be thus represented. Therefore, $\pi_1 Q^0_f = \IZ_2$.

    Furthermore, $\pi_1 \CQ_1^f = \IZ$,  given by a fermionic version of the GNVW invariant \cite{Gross_2012,Fidkowski_2019}. The generator of this group is given by the ``Majorana translation'', defined as follows. We write the Majorana generators of the Fermion algebra at site $i$ as $\gamma_i$, $\tilde \gamma_i$. The Majorana translation acts by $\gamma_i \mapsto \tilde \gamma_i$, $\tilde \gamma_i \mapsto \gamma_{i+1}$ (see \hyperref[figmajtrans]{Fig. \ref{figmajtrans}}). This has a GNVW invariant half that of a qubit translation, since the single Majorana algebra $Cl(1)$ is dimension 2 and the qubit algebra $M(2,\IC)$ is dimension 4.

    The Postnikov tower of $\CQ^1_f$ thus takes the form
    \begin{equation}
        \begin{tikzcd}
            B^3 U(1)_{disc} \arrow[r] & \CQ^1_f \arrow[d] \\
            B^2 \IZ_2  \arrow[r] & \CQ^1_{f,2} = B^2 \IZ_2 \times B\IZ \arrow[d]\\
            & \CQ^1_{f,1} = B\IZ
        \end{tikzcd}
    \end{equation}
    where $\CQ^1_{f,2} = B^2 \IZ_2 \times B\IZ$ since $B\IZ = S^1$ is one-dimensional, and doesn't admit non-trivial $B^2 \IZ_2$ fibrations.

    The homotopy type of $\CQ^1_f$ is thus specified by a classifying map
    \[B^2 \IZ_2 \times B\IZ \to B^5 \IZ\]
    giving the structure of the top fibration. Equivalently, the class of this fibration is given by an element of
    \[H^4(B^2\IZ_2 \times B\IZ,U(1)_{disc}) = \IZ_2.\]
    This element corresponds to
    \[i \circ Sq^2:B^2 \IZ_2 \to B^4 U(1)_{disc},\]
    where $i$ is the inclusion $\IZ_2 \to U(1)$ and $Sq^2$ is the Steenrod square $B^2 \IZ_2 \to B^4 \IZ_2$. Note that $B\IZ$ does not enter into this fibration, so
    \[\CQ^1_f = E \times B\IZ,\]
    where $E$ sits in the fibration
    \begin{equation}
        \begin{tikzcd}
            B^3 U(1)_{disc} \arrow[r] & E \arrow[d] \\
             & B^2\IZ_2
        \end{tikzcd}
    \end{equation}
    which is yet to be determined of the two possible choices.

    A short calculation shows that we can distinguish the two possibilities by studying
    \[\CQ^1_f(B\IZ_2) = \begin{cases}
        \IZ_2 \times \IZ_2 & \text{trivial fibration} \\
        \IZ_4 & \text{non-trivial fibration (correct)}
    \end{cases}\]
    We will show that the second holds, and therefore that the fibration is non-trivial.

    The result for general dimensions will then follow. Indeed, since $\CQ_f^d$ is an infinite loop space, its Postnikov invariants must be stable cohomology operations. $i \circ Sq^2$ is the only stable cohomology operation that has the right source and target, and if this Postnikov invariant is present for $d = 1$ it must be present for all $d \ge 1$ as well.

    To show $\CQ^1_f(B\IZ_2) = \IZ_4$, we construct an FDQC $\IZ_2$-representation, in other words an FDQC $U$ (with bosonic unitary gates) satisfying $U^2 = 1$. This defines a map
    \[U:B\IZ_2 \to \CQ^1_f.\]
    We will show that $U \otimes U$ has a non-trivial Else-Nayak index, so
    \[U\otimes U:B\IZ_2 \to \CQ^1_f\]
    factors through $B^4 \IZ \to \CQ^1$, which is non-contractible in either case above. Thus, $U$ represents a $\IZ_4$ element in the group of maps $B\IZ_2 \to \CQ^1_f$, showing out of the two possibilities we must have $\CQ^1_f(B\IZ_2) = \IZ_4$, and the non-trivial Postnikov invariant above.

    We will observe also that $U$ has a truncation $\tilde U$ such that $\tilde U^2$ acts like a local fermionic operator, and cannot be represented as a local bosonic unitary. Thus, $U$ already gives a nontrivial map
    \[B\IZ_2 \to B^2 \pi_2\CQ^1_f= B^2 \IZ_2\]
    representing a nontrivial 2nd anomaly index (the 1st automatically vanishes because $U$ is blendable, being a FDQC). Note that although in general the anomaly indices depend on choices of lifts (see Appendix \ref{appdependenceonlifts} and \hyperref[propanomalyindexdependence]{Proposition \ref{propanomalyindexdependence}}), the 1st and 2nd anomaly indices are always well-defined.
    
    We collect the construction of $U$ in Example \ref{Z4anomalyforfermionic1dcircuit}. We will actually use a local algebra built from fermions and qubits (see Fig. \ref{fig1dfermionicchain}), but the qubit algebra can be regarded as a subalgebra of two fermions and the circuit $U$ trivially extends with the same properties.

\end{proof}

\begin{example}\label{Z4anomalyforfermionic1dcircuit}
We will construct a fermionic FDQC $\IZ_2$-representation with a $\IZ_4$ stable blend anomaly, demonstrating $\CQ^1_f(B\IZ_2) = \IZ_4$. There are many related constructions in the literature, especially the works of \cite{Ellison_2019,tantivasadakarn2018full}, which constructed circuit disentanglers for the corresponding 2+1d $\IZ_2$ SPT. Using the bulk-boundary correspondence (see \hyperref[secsptandbulkboundary]{Section \ref{secsptandbulkboundary}}), one can produce corresponding QCA representations in 1d. Here we give a simple and direct construction in terms of a finite depth circuit.

Let us consider a 1d lattice $\IZ$ with bosonic qubits at each odd site, with algebra generators $X_{2n+1}$, $Z_{2n+1}$, and a single Majorana operator $\gamma_{2n}$ at each even site (see Fig. \ref{fig1dfermionicchain}).
Let us define the ``controlled parity gate''
\[CP_{2n+1} = \exp\left(i\pi \left(\frac{1-Z_{2n+1}}{2} \right) \left(\frac{1-i\gamma_{2n} \gamma_{2n+2}}{2} \right)\right).\]
In the $Z_{2n+1}$ basis, this operator acts as the identity if $Z_{2n+1} = 1$, and acts as $i\gamma_{2n} \gamma_{2n+2}$ if $Z_{2n+1} = -1$. It satisfies
\[CP_{2n+1}^2 = 1 \\
CP_{2n+1} X_{2n+1} CP_{2n+1} = i\gamma_{2n} X_{2n+1} \gamma_{2n+2} \\
CP_{2n+1} \gamma_{2n} CP_{2n+1} = \gamma_{2n} Z_{2n+1} \\
CP_{2n+1} \gamma_{2n+2} CP_{2n+1} = Z_{2n+1}\gamma_{2n+2} \\
CP_{2n+1} CP_{2n+3} CP_{2n+1} CP_{2n+3} = \exp\left(i\pi \left(\frac{1-Z_{2n+1}}{2} \right) \left(\frac{1-Z_{2n+3}}{2} \right)\right) = CZ_{2n+1}.\]
Other commutation relations with generators are trivial. The last relation can be checked in the $Z_{2n+1}$, $Z_{2n+3}$ basis. If $Z_{2n+1} = 1$ or $Z_{2n+3} = 1$, then the operators commute, since one acts as the identity. However, if $Z_{2n+1} = Z_{2n+3} = -1$, they act as $i\gamma_{2n} \gamma_{2n+2}$ and $i \gamma_{2n+2} \gamma_{2n+4}$, which anti-commute.

Now let us consider the circuit
\[C =\prod_n CP_{4n+3} \prod_n CP_{4n+1}.\]
We have
\[C^2 = \prod_n CZ_{2n+1} = \prod_n e^{(-1)^n\frac{i\pi}{4}  Z_{2n+1} Z_{2n+3}} \qquad \text{(up to phases)} \\
C^\dagger X_{4n+1} C = i \gamma_{4n} X_{4n+1} \gamma_{4n+2} \\
C^\dagger X_{4n+3} C = i Z_{4n+1} \gamma_{4n+2}  X_{4n+3}  \gamma_{4n+4} Z_{4n+5} \\
C^\dagger \prod_n X_{2n} C = \prod_n X_{2n} \qquad \text{(up to phases)}.\]
Finally we construct the depth four circuit
\[U =  \left(\prod_n e^{(-1)^ni \frac{\pi}{8} Z_{2n+1} Z_{2n+3}}\right) C \left(\prod_n X_{2n+1} \right) \\
= \prod_n e^{(-1)^n i \frac{\pi}{8} Z_{2n+1} Z_{2n+3}} \prod_n CP_{4n+3} \prod_n CP_{4n+1} \prod_n X_{2n+1}.\]
A small computation shows $U^2 = 1$ (up to phases), so $U$ defines an FDQC $\IZ_2$-representation on this fermionic algebra.

We wish to compute the anomaly indices of $U$. First, we define a truncation of the circuit
\[\tilde U = \prod_{n \ge 0} e^{(-1)^n i \frac{\pi}{8} Z_{2n+1} Z_{2n+3}} \prod_{n \ge 0} CP_{4n+3} \prod_{n \ge 0} CP_{4n+1} \prod_{n \ge 0} X_{2n+1}.\]
A short calculation shows
\[\tilde U^2 = \gamma_0 e^{\frac{i \pi}{4} Z_1} \qquad \text{(up to phases)}.\]
Since this is a fermionic operator, $\tilde U^2$ is not realizable as a local unitary. Therefore, we have encountered a non-trivial anomaly index.

Suppose now we take two tensor copies of our 1d lattice, call the layers $A$ and $B$, and consider the symmetry $U \otimes U = U_A U_B$. We will have
\[(\tilde U_A \tilde U_B)^2 = \gamma_{A,0} e^{\frac{i\pi}{4} Z_{A,1}} \gamma_{B,0} e^{\frac{i\pi}{4} Z_{B,1}} \qquad \text{(up to phases)}.\]
Unlike the case with a single layer, now the RHS is representable by the local unitary operator
\[N = \gamma_{A,0} e^{\frac{i\pi}{4} Z_{A,1}} \gamma_{B,0} e^{\frac{i\pi}{4} Z_{B,1}}.\]
We can proceed to compute the Else-Nayak index $\omega = \beta_3$:
\[e^{i \omega(1,1,1)} =N \left((\tilde U_A \tilde U_B) N (\tilde U_A \tilde U_B)^\dagger\right)^\dagger\]
(see \eqref{eqnelsenayakh3index}). We find
\[\tilde U \gamma_0 \tilde U^\dagger = \gamma_0 Z,\]
so
\[(\tilde U_A \tilde U_B) N (\tilde U_A \tilde U_B)^\dagger = \gamma_{A,0} e^{-\frac{i\pi}{4}Z_{A,1}} Z_{A,1} \gamma_{B,0} e^{-\frac{i\pi}{4}Z_{B,1}} Z_{B,1},\]
so
\[e^{i\omega(1,1,1)} = \gamma_{A,0}^2 e^{\frac{i\pi}{2}Z_{A,1}} Z_{A,1} \gamma_{B,0}^2 e^{\frac{i\pi}{2}Z_{B,1}} Z_{B,1} = i Z_{A,1}^2 i Z_{B,1}^2 = -1.\]
Thus, we find the non-trivial Else-Nayak index $[\omega] \in H^3(B\IZ_2,U(1))$ for two tensor copies of the fermionic FDQC $\IZ_2$-representation $U$.
\end{example}

\vspace{3mm}
\begin{mdframed}[backgroundcolor=blue!5]
\begin{proposition}
Let $\CQ^1_{f,inv}$ be the space of FDQC-invertible stable $1$-states in the same fermionic algebra. We have
    \[\pi_0 \CQ^1_{f,inv} = \IZ_2 \\
    \pi_1 \CQ^1_{f,inv} = \IZ_2 \\
    \pi_2 \CQ^1_{f,inv} = U(1) \\
    \pi_{>2} \CQ^1_{f,inv} = 0.\]
    We also compute
    \[\CQ^1_{f,inv}(B\IZ_2) = \IZ_4 \times \IZ_2.\]
\end{proposition}
\end{mdframed}
\begin{proof}
    The homotopy groups follow from the long exact sequence \eqref{eqinvcofibersequence}. This long exact sequence is convenient to represent in an array with connecting maps going between rows from $\pi_n \CQ^1_{f,inv} \to \pi_n \CQ^1_{f,\psi_0}$. In this case, we have
    \begin{equation}
        \begin{tikzcd}
            n & \pi_{n+1} \CQ^1_{f,\psi_0} & \pi_{n+1} \CQ^1_f & \pi_n \CQ^1_{f,inv} \\
            3 & 0 & 0 & 0 \\
            2 & 0 & U(1) \arrow[r] & U(1) \\
            1 & 0 & \IZ_2 \arrow[r] & \IZ_2 \\
            0 & \IZ \arrow[r] & \IZ \arrow[r] & \IZ_2
        \end{tikzcd}
    \end{equation}
    The $n=3$ row is zero because $\CQ^1_{f,\psi_0}$and $\CQ^1_f$ are homotopy 2-types by construction.
    
    For $n=2$, consider first $\pi_3 \CQ^1_{f,\psi_0}$. These correspond to local unitary operators $U$ with $\text{Ad}\ U = 1$ satisfying $U \psi_0 = \psi_0$. Such $U$ must be a phase operator since $\text{Ad}\ U = 1$ and therefore must in fact have $U = 1$ to fix the product state.
    
    For $n = 1$, consider $\pi_{2} \CQ^1_{f,\psi_0} = \pi_1 \CQ^0_{f,\psi_0}$. These are automorphisms $\alpha$ of a superalgebra commuting with the fermion parity and fixing $\psi_0$. Since $\psi_0$ has definite fermion parity, such an $\alpha$ must be representable by a bosonic unitary operator, and thus $\pi_1 \CQ^0_{f,\psi_0} = 0$. (Compare \hyperref[prophomotopycalc]{Proposition \ref{prophomotopycalc}}). The previous row now completely follows by exactness.

    Finally, for the short exact sequence of $n = 0$, the Majorana translation $\alpha$ (see \hyperref[figmajtrans]{Fig. \ref{figmajtrans}}) the generator of the group $\CQ^1_f$ \cite{Gross_2012,Fidkowski_2019}. It takes the product state $\psi_0$ to the Kitaev chain $\psi_{Kitaev} = \psi_0 \circ \alpha$, which does not admit a blend to a product state, so $\alpha$ mapsto a non-trivial element of $\pi_0 \CQ^1_{f,inv}$.
    
    It follows from exactness that $\alpha$ cannot fix an SRE state. There is also a nice argument for this directly from the non-triviality of $\psi_{Kitaev}$. Suppose towards a contradiction that $\alpha$ fixes an SRE state, meaning we have $\psi_0 \circ C \circ \alpha = \psi_0 \circ C$ for some FDQC $C$. Then we have $\psi_{Kitaev} = \psi_0 \circ \alpha = \psi_0 \circ (C \alpha^{-1} C^{-1} \alpha)$, but $(C \alpha^{-1} C^{-1} \alpha)$ is a circuit, contradicting the LRE of $\psi_{Kitaev}$.

    On the other hand, it is easy to see $\alpha^2$ fixes the product state (see \hyperref[figmajtrans]{Fig. \ref{figmajtrans}}). Therefore $\alpha^2$ generates the kernel of the map $\psi_0:\pi_1 \CQ^1_f \to \pi_0 \CQ^1_{f,inv}$. The bottom row now follows.

    \begin{figure}
    \centering
    \includegraphics[width=8cm]{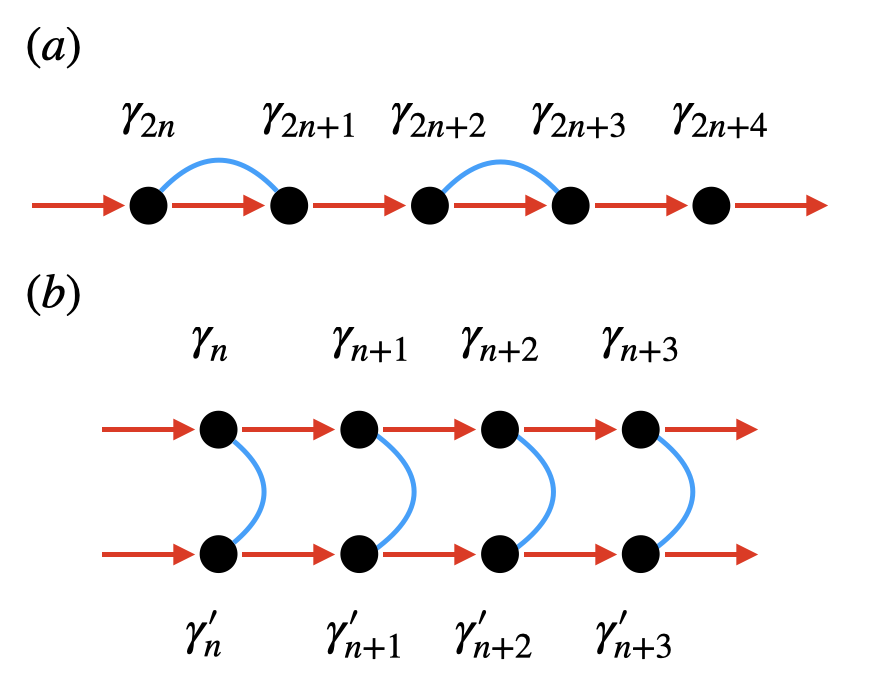}
    \caption{$(a)$ The ``Majorana translation'' QCA $\alpha$, acting on a 1d lattice with a Majorana operator $\gamma_n$ at each site $n$. The QCA acts by $\gamma_n \mapsto \gamma_{n+1}$. To have a (super)tensor product Hilbert space, we can group the Majoranas in pairs, say $\gamma_{2n},\gamma_{2n+1}$. After doing so, one can define a product state $\psi_0$ as the simultaneous eigenvector of all $i\gamma_{2n} \gamma_{2n+1}=1$ (blue bonds). Applying the Majorana translation $\alpha$ to this state yields the famous Kitaev chain state \cite{Kitaev_2001}, a long range entangled 1d state of fermions. However, applying $\alpha^2$ fixes the state $\psi_0$. $(b)$ A product state fixed by $\alpha \otimes \alpha$, which shows that the anomaly of $\alpha$ is order 2.}
    \label{figmajtrans}
\end{figure}

    From the long exact sequence \eqref{eqnlongexactsequenceSRE}, we find that the map $SRE:\CQ^1_{f}(B\IZ_2) = \IZ_4 \to \Omega^1_{f,inv}(B\IZ_2)$ is injective since $\CQ^1_{f,\psi_0} = B\IZ$ so $\CQ^1_{f,\psi_0}(B\IZ_2) = 0$. From the Atiyah-Hirzebruch spectral sequence using the homotopy groups computed above, we further learn that $\CQ^1_{f,inv}(B\IZ_2)$ is one of
    \[\IZ_4, \qquad \IZ_4 \times \IZ_2, \quad \text{or} \quad \IZ_8,\]
    where the $\IZ_4$ subgroup in each case is the image of $\CQ^1_f(B\IZ_2)$.
    
    The Majorana translation $\alpha$ defines a QCA such that $\alpha^2$ fixes the product state $\psi_0$. It thus defines a map
    \[\alpha:B\IZ_2 \to B\CQ^1_{f,inv}\]
    which is not null-homotopic by the first anomaly index, since $[\alpha] \neq 0 \in \pi_0 \CQ^1_{f,inv}$, which we argued above. However, $\alpha \otimes \alpha$ can be easily shown to be null-homotopic (see \hyperref[figmajtrans]{Fig. \ref{figmajtrans}}). Therefore, the homotopy class $[\alpha] \in \CQ^1_f(B\IZ_2)$ is a non-zero, 2-torsion element. It is also not in the image of the map $SRE$, since the generator of that $\IZ_4$ has trivial first anomaly index, being given by an FDQC. Therefore, $\CQ^1_{f,inv}(B\IZ_2) = \IZ_4 \times \IZ_2$.
\end{proof}

We see from this proof that although we did not obtain the $\IZ_8$ anomaly group expected from the continuum, the $\IZ_2$ factor of $\CQ^1_{f,inv}(B\IZ_2)=\IZ_4 \times \IZ_2$ has a natural interpretation on the lattice: it is the anomaly of a symmetry which defines a $\IZ_2$ group \emph{up to a QCA fixing a product state}.

It is interesting to consider how this is consistent with the continuum reasoning. For instance, it is known that if we take the Hamiltonian
\[H = \sum_n i \gamma_n \gamma_{n+1},\]
this is symmetric under the Majorana translation $\alpha$ and has a single gapless Majorana fermion in the IR. From the point of view of the continuum QFT, $\alpha$ becomes the chiral parity of this Majorana fermion (for some recent perspectives on this, see \cite{seiberg2024majorana}). This anomaly represents the generator of the cobordism group $\tilde\Omega^3_{\rm Spin}(B\IZ_2) = \IZ_8$ (using reduced cohomology here because there is no gravitational anomaly).

If we take two copies of this system, the emanant symmetry is still a $\IZ_2$ symmetry with a non-trivial anomaly $\nu = 2$ mod 8 in $\IZ_8$. However, now we can add to this system the trivial symmetric Hamiltonian in \hyperref[figmajtrans]{Figure \ref{figmajtrans}}. This will eventually drive the system into a trivial phase, which seems paradoxical. However, there is no paradox, because as this perturbation is turned on, the emanant $\IZ_2$ symmetry of the two Majorana fermions, which may be considered as a single Dirac, actually becomes a $\IZ$ subgroup of a chiral $U(1)$ symmetry.

We can see what happens to the anomaly when the $\IZ_2$ symmetry is lifted to $\IZ$ by pulling back along the quotient map $B\IZ \to B\IZ_2$:
\[\tilde\Omega^3_{\rm Spin}(B\IZ_2) \to \tilde\Omega^3_{\rm Spin}(B\IZ) = \IZ_2.\]
Thus, the anomalies where $\nu$ is even all become trivial, and a trivial symmetric phase is now permitted.

Indeed, as the strength of the perturbation is increased, the symmetry emanating from $\alpha$ deforms inside the chiral $U(1)$ symmetry of the Dirac fermion until it reaches a point where a symmetric mass term appears, gapping the system into a trivial symmetric phase. The same mechanism was discussed for translations acting as an emanant anomalous $\IZ_2$ symmetry of bosons in \cite{Metlitski_2018}.

\section{Outlook}

In this work, we have made some first steps towards a topological theory of anomalies on the lattice. We have defined blend anomalies which are obstructions to on-siteability/gauging, and SRE anomalies which are obstructions to having a symmetric SRE state. Both types of anomalies are phrased in terms of a homotopy class of a map to either a classifying space of QCAs or a classifying space of FDQC-invertible states. We have shown in the stable setting these classifying spaces form $\Omega$-spectra, and are in fact closely related by a cofiber sequence. We have shown how to use Else-Nayak-style methods and obstruction theory to calculate the anomalies on the lattice, and have used homotopy theory to compute some of the relevant classifications, uncovering some intriguing (although resolvable) tension with the theory of 't Hooft anomalies of QFT.

There are several important problems which remain open. One is how to include invertible states with a non-zero correlation length, such as Chern insulators, which cannot be FDQC-invertible. One approach would be to study approximate QCA, or QCA with tails, and proceed as we have in \hyperref[subsecstabilization]{Section \ref{subsecstabilization}}. One needs only a suitable notion of blend of such approximate QCA. A difficulty we foresee, which is one reason we have not pursued it here, is that one cannot restrict to finite blend intervals once there are tails. A blend of approximate QCA should more smoothly interpolate from one to another. In particular, a blend from the identity to itself is not a strictly lower dimensional QCA, as we needed to have the $\Omega$-spectrum property, but instead will spread out completely in the extra dimension, but become more and more like the identity from far away. Should we accept this as natural? When can we get something strictly lower-dimensional?

A second important problem concerns a pathology of our blend equivalence, which is defined with respect to a particular axis. If we consider 2d QCA up to blend equivalence along the 1st axis, one can show there are continuum many distinct equivalence classes, since we may produce any rational or irrational density of translations along our axis, spaced in a particular way along the other axis. If there are non-torsion invertible states like Chern insulators, we can do the same thing. We can also form concentric spheres of these states or QCA, which do not admit a blend along any axis, and again seem to have continuum many density parameters. A related problem is to define an equivalence relation such that two invertible states which differ by a rotation are equivalent, which is not obviously true for blend equivalence. For recent progress in this last direction, see \cite{sopenko2025reflectionpositivityrefinedindex}.

A third class of problems concerns whether our anomalies are complete. They are defined in terms of homotopies, so when the anomaly vanishes, we may obtain a null-homotopy of the corresponding map. We would like to know whether this null-homotopy can do something useful for us. For example, if the blend anomaly vanishes, does that mean we can gauge (this is probably too naive)? Or if the SRE anomaly vanishes, does that mean a symmetric SRE state really exists? These are crucial problems for example in constructing chiral gauge theories and realizing symmetric mass generation on the lattice.

Finally, there are questions relating these lattice anomalies to the continuum. We have said some words about this at the end of \hyperref[secanomalousferms1d]{Section \ref{secanomalousferms1d}} in regards to a particular interesting example. But there are many interesting general questions. For instance, is there a map $\CQ^\star_{\CH,inv}$ to the cobordism spectrum? Are $\CQ^\star_{\CH,inv}$ or $\CQ^\star_{\CH}$ spectra of invertible TQFTs of some kind?

\textit{Acknowledgements:} We are very grateful to Lei Gioia, Corey Jones, Anton Kapustin, Alexei Kitaev, Max Metlitski, Daniel Ranard, Nikita Sopenko, Ruben Verresen, and Dominic Williamson for enlightening discussions and collaborations on related works. RT is supported by the Bhaumik Presidential Term Chair. RG acknowledges support by the Mani L. Bhaumik Institute for Theoretical Physics and support from the Simons Center for Geometry and Physics, Stony Brook University at which some of the research for this paper was performed.

\begin{oldappendices}

\section{Proof of \hyperref[thminvertiblespectrum]{Theorem \ref{thminvertiblespectrum}} on the spectrum of FDQC-invertible states}\label{appinvstateproof}

\begin{proof}
    For the abelian group structure on blend equivalence classes of FDQC-invertible states is given by tensor product
    \[ [\psi] + [\psi'] = [\psi \otimes \psi'],\]
    which makes sense for $\psi$ and $\psi'$ which have disjoint domains. We can always use a layer-shifting blend as in \hyperref[figshiftinghomotopy]{Figure \ref{figshiftinghomotopy}} to move the domain of $\psi$ in its blend equivalence class to ensure this is the case. This operation does not depend on the arrangement of the domains of $\psi$ and $\psi'$ up to blending of $\psi \otimes \psi'$. Furthermore, if we have a blend of $\psi$, we can make the domain of this blend disjoint from $D(\psi')$ and thus obtain a blend of $\psi \otimes \psi'$. Therefore, the operation above is well-defined.
    
    The operation is also clearly commutative, associative, and has $[\bar \psi_0]$ as the identity, where $\bar \psi_0$ is the product state obtained by taking tensor products of $\psi_0$ over any valid domain. Inverses are given by the state $\psi^\text{rev}$ reflected along the $d$th axis, with blend $\psi \otimes \psi^\text{rev}$ to $\bar \psi_0$ given by the ``folding blend'', as in \hyperref[figbendingblendofstates]{Figure \ref{figbendingblendofstates}}.
    
    Let $\text{Inv}^d$ denote this group. We want to show it is isomorphic to a certain homotopy group of the cofiber spectrum.

    \begin{figure}
        \centering
        \includegraphics[width=8cm]{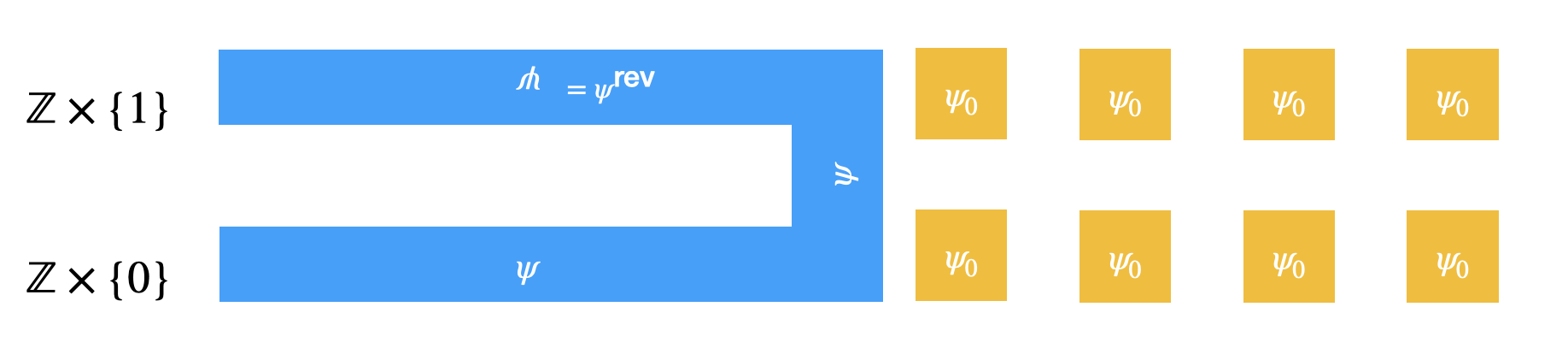}
        \caption{A folding blend of FDQC-invertible stable states gives a blend from $\psi \otimes \psi^\text{rev}$ to $\bar \psi_0$, (compare \hyperref[figsflippinghomotopy]{Figure \ref{figsflippinghomotopy}}). This shows that $\psi^\text{rev}$ is an inverse to $\psi$ in the group of blend-equivalence classes of FDQC-invertible states.}
        \label{figbendingblendofstates}
    \end{figure}

    The homotopy group of interest can be computed as a stable relative homotopy group
    \[\pi_0(\CQ^d_{\CH,inv}) = \lim_{r \to \infty} \pi_{r+1}(\Omega_\star \CQ^{d+r+1}_\CH,\Omega_\star \CQ^{d+r+1}_{\CH,\psi_0},1).\]
    Unpacking this, $\pi_{r+1}(\Omega_\star \CQ^{d+r+1}_\CH,\Omega_\star \CQ^{d+r+1}_{\CH,\psi_0},1)$ is the group of equivalence classes of $r+1$-globes of stable $d+r+1$-QCA (see \hyperref[prophomotopycalc]{Proposition \ref{prophomotopycalc}}, the same caveats apply here), which are expressed as an array of blends (using the balanced presentation \eqref{eqnzerocomposition})
    \[
    \left\{
    \begin{array}{r  c  l  c  l}
    \varphi_{r+1} & \mathrel{:} & \varphi_r & \equiv_{d+1}   & \varphi_r' \\
                  &  & \vdots    &                &            \\
                  & \mathrel{:} & \varphi_2 & \equiv_{d+r}   & \varphi_2' \\
                  & \mathrel{:} & \varphi_1 & \equiv_{d+r+1} & 1 
    \end{array}
    \right\}
    \]
    where $\varphi_1$ is a stable $d+r+1$-QCA which fixes $\bar \psi_0$, and for $2 \le n \le r$, $\varphi_n, \varphi_n'$ are blends of stable $d+r+1$-QCA from $\varphi_{n-1}$ to $\varphi_{n-1}'$ along the $d+r+2-n$th axis, all fixing $\bar \psi_0$, and finally $\varphi_{r+1}:\varphi_r \equiv_{d+1} \varphi_r'$ is just required to be a blend along the $d+1$st axis. Note that $\varphi_{r+1}$ fixes the product state outside of
    \[\IZ^d \times [-w,w]^l \times 0 \times 0 \times \cdots\]
    for some $w,l$. Furthermore, $\varphi_{r+1}$ acts only in a neighborhood of $\IZ^d \times \IZ_{\le 0}^{r+1}$ because it starts with a blend to the identity.
    
    The $r+1$-globes of this form can be checked to be closed under 0-composition of $r+1$-blends, defined by element-wise composition in the array (see \eqref{eqnzerocomposition}, the homotopy group can be constructed using composition along any direction, by the usual Eckman-Hilton argument, the 0-composition is just the easiest to work with in the balanced presentation). In the relative homotopy group, these $r+1$-globes are considered up to blend equivalence along the $d$th axis and up to composition with an $r+1$-globe of the same type where moreover $\varphi_{r+1}$ fixes the product state $\bar \psi_0$ everywhere.

    For the rest of the proof we refer to these $r+1$-globes as ``$r$-blends''.
    
    To compute the stable relative homotopy group, we can consider any such $r$-blend as an $r+1$-blend of stable $d+r+2$ QCA
    \[\left\{
    \begin{array}{r  c  l  c  l}
    \varphi_{r+1} & \mathrel{:} & \varphi_r & \equiv_{d+1}   & \varphi_r' \\
                  &  & \vdots    &                &            \\
                  & \mathrel{:} & \varphi_2 & \equiv_{d+r}   & \varphi_2' \\
                  & \mathrel{:} & \varphi_1 & \equiv_{d+r+1} & 1 
    \end{array}
    \right\} \sim
    \left\{
    \begin{array}{r  c  l  c  l}
    \varphi_{r+1} & \mathrel{:} & \varphi_r & \equiv_{d+1}   & \varphi_r' \\
                  &  & \vdots    &                &            \\
                  & \mathrel{:} & \varphi_2 & \equiv_{d+r}   & \varphi_2' \\
                  & \mathrel{:} & \varphi_1 & \equiv_{d+r+1} & 1          \\
                  & \mathrel{:} & 1         & \equiv_{d+r+2} & 1
    \end{array}
    \right\}
    \]
    and this way take the limit $r \to \infty$.
    
    Given such an $r$-blend, we can apply $\varphi_{r+1}$ to the state $\bar \psi_0$ and produce a stable $d$-state
    \[\psi = \bar \psi_0 \circ \varphi_{r+1}\]
    whose domain is
    \[\IZ^d \times [-w,w]^l \times 0 \times \cdots.\]
    Let us show this state is FDQC-invertible. If we follow the argument of \hyperref[propinvertibilityqcaentangler]{Proposition \ref{propinvertibilityqcaentangler}}, we can construct an FDQC $\varphi_{r+1} \otimes \varphi_{r+1}^{-1}$ which creates
    \[\psi \otimes \psi',\]
    where $\psi' = \bar\psi_0 \circ \varphi_{r+1}^{-1}$. This is not quite satisfactory, because $\varphi_{r+1} \otimes \varphi_{r+1}^{-1}$ acts in a $d+r+1$-dimensional space. However, since it is a circuit, we can truncate it on a large finite $d$-dimensional neighborhood of $\IZ^d$. This creates
    \[\psi \otimes \psi' \otimes \psi_\partial,\] where $\psi_\partial$ is supported near the boundary of this neighborhood, which is a $d$-dimensional tube. This is shown in \hyperref[figtubeinverse]{Figure \ref{figtubeinverse}}. Thus, $\psi$ is FDQC-invertible as a stable $d$-state.

    \begin{figure}
        \centering
        \includegraphics[width=5cm]{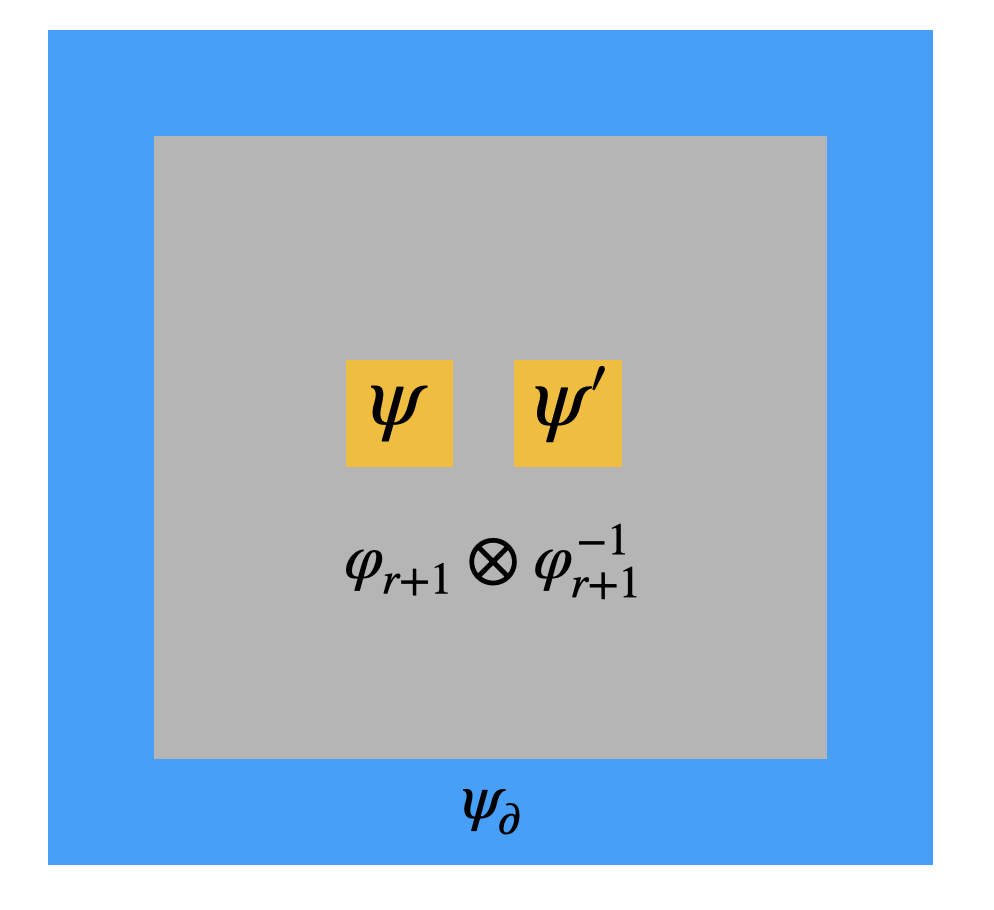}
        \caption{To adapt the argument of \hyperref[propinvertibilityqcaentangler]{Proposition \ref{propinvertibilityqcaentangler}} and \hyperref[propinvstatessameasqcaentanglablestates]{Corollary \ref{propinvstatessameasqcaentanglablestates}} to the stable case, we note that $\varphi_{r+1} \otimes \varphi_{r+1}$ is an FDQC in $d+r+1$ dimensions, and can be arbitrarily truncated. Truncating it in a tubular neighborhood, large enough so that its boundary is causally disconnected from the region where $\psi \otimes \psi'$ is created, we obtain $\psi \otimes \psi' \otimes \psi_\partial$, where $\psi_\partial$ is a state supported near the boundary of the tube. This boundary is $d$-dimensional and so $\psi_\partial$ can be regarded as a stable $d$-state. The FDQC-inverse of $\psi$ is therefore $\psi' \otimes \psi_\partial$.}
        \label{figtubeinverse}
    \end{figure}

    If we have a blend between two $r$-blends, we obtain an FDQC-invertible blend between the two FDQC-invertible states. Furthermore, if we compose two blends, which can always first be blended so to have disjoint domains, the states we obtain by their composition is the tensor product. Thus, we obtain a group homomorphism
    \[f_r:\pi_{r+1}(\Omega_\star \CQ^{d+r+1}_\CH,\Omega_\star \CQ^{d+r+1}_{\CH,\psi_0},1) \to \text{Inv}^d.\]
    This map is clearly compatible with the stabilization from $r$-blends to $r+1$-blends so we also get a group homomorphism
    \[f: \lim_{r \to \infty} \pi_{r+1}(\Omega_\star \CQ^{d+r+1}_\CH,\Omega_\star \CQ^{d+r+1}_{\CH,\psi_0},1) \to \text{Inv}^d.\]
    We want to show $f$ is an isomorphism.

    We first show $f$ is surjective. Given an FDQC-invertible state $\psi$, we can construct the swindle circuit as in \hyperref[propswindle]{Proposition \ref{propswindle}}. Actually we can make a two-sided swindle circuit
    \[S = \prod_{n \in \IZ} D_{2n+1}\prod_{n \in \IZ} C_{2n}\]
    as in the notation of the proposition. This circuit $S$ fixes $\bar \psi_0$ everywhere. We can choose a truncation $S_{\le 0}$ which creates $\psi$ at its boundary. We can regard this as a 1-blend
    \[
    \left\{
    \begin{array}{r  c  l  c  l}
               S_{\le 0}   & \mathrel{:} & S & \equiv_{d+1} & 1 
    \end{array}
    \right\}
    \]
    which is of the type we have been considering. If we apply $f$ to this 1-blend, we get $\psi$ be construction, so $f$ is surjective.

    \begin{figure}
        \centering
        \includegraphics[width=4cm]{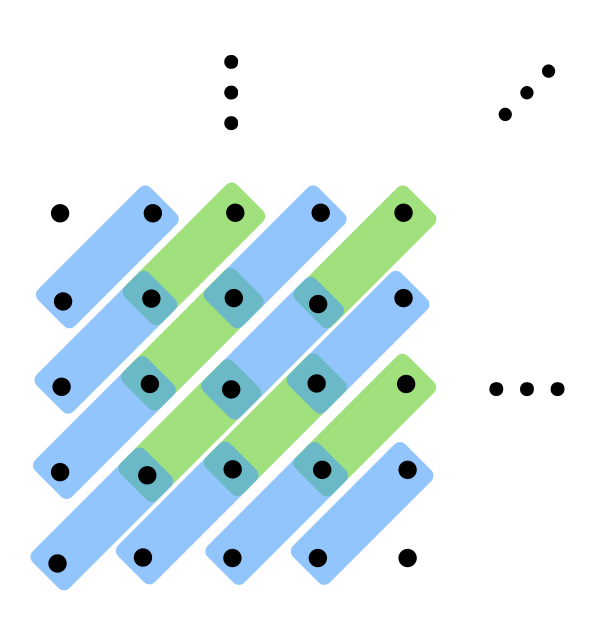} \qquad \includegraphics[width=5cm]{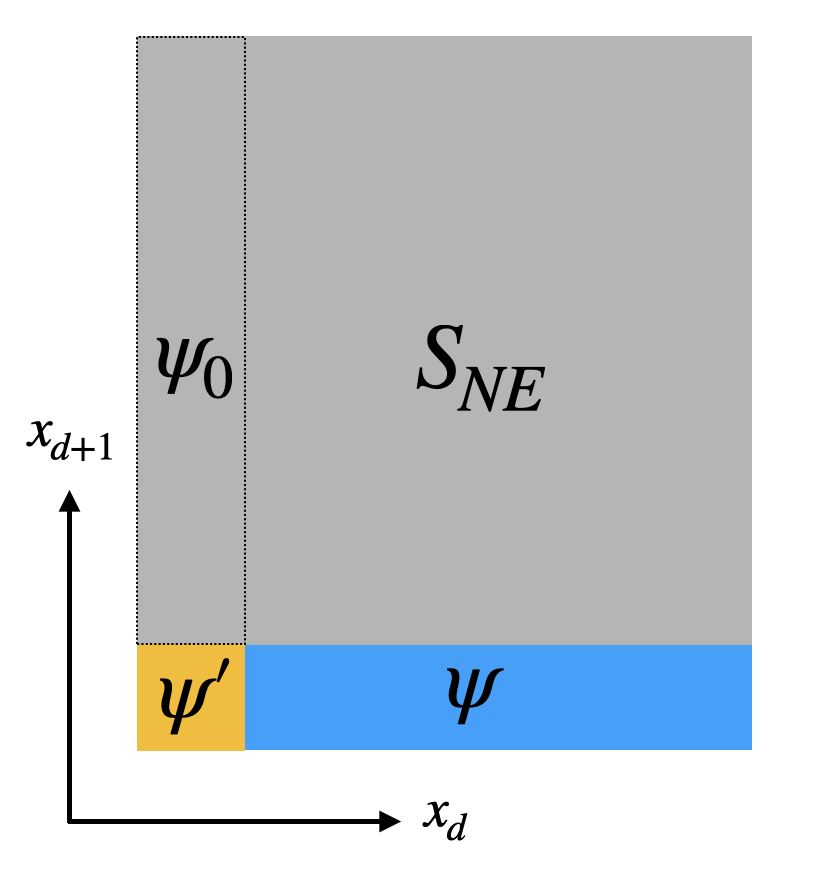}
        \caption{Using a corner variant of the swindle circuit of \hyperref[propswindle]{Proposition \ref{propswindle}}, we can produce FDQC-invertible states along right angled regions using circuits acting on corner regions as shown, where the first layer of the swindle circuit is shown in green and the second layer is shown in blue. (For simplicity we have just depicted the green and blue circuits as two-body gates to show the connectivity.) All sites in the interior of the corner return to the product state $\bar \psi_0$, while sites on the boundary produce the desired FDQC-invertible state. For the proof of the theorem, we want to apply this construction to our blend $\psi'$ of FDQC-invertible states to produce a circuit $S_{NE}$ as shown. This circuit acts on a finite-thickening of $\IZ^{d-1}$ times the NE corner region in the $x_d,x_{d+1}$ plane, such that along the thickened $d$-dimensional half space $\IZ^{d-1} \times \IZ_{\ge 0}$ it produces the state $\psi$ (blue), except for a finite neighborhood of $\IZ^{d-1}$ where the blend region of $\psi'$ is (orange), and elsewhere it produces $\bar \psi_0$ (gray).}
        \label{figcornerswindle}
    \end{figure}
    
    To prove injectivity, suppose we have an $r$-blend as above such that the resulting state $\psi = \bar \psi_0 \circ \varphi_{r+1}$ blends to $\bar \psi_0$ along the $d$th axis by an another FDQC-invertible stable $d$-state state $\psi'$. We want to show that $\varphi_{r+1}$ is equivalent to the identity.

    Since $\psi'$ is FDQC-invertible, it has a swindle circuit $S$ as in \hyperref[propswindle]{Proposition \ref{propswindle}} which acts on a neighborhood of the half-space $\IZ^d \times \IZ_{\ge 0}$. Consider the $x_d,x_{d+1}$-plane. The half space $\IZ \times \IZ_{\ge 0}$ in this plane can be mapped bijectively to the NE corner $\IZ_{\ge 0} \times \IZ_{\ge 0}$ by mapping
    \[f(x_d,x_{d+1}) = \begin{cases}
         (x_d+x_{d+1}, x_{d+1}) & x_d \ge 0 \\
         (x_{d+1}, -x_d+x_{d+1}) & x_d \le 0
    \end{cases}\]
    This mapping can be visualized geometrically as two shear transformations joined along the line $x_{d+1} = x_d$. Therefore, $d(f(\vec x),f(\vec y)) \le C d(\vec x,\vec y)$, where $C$ is a constant independent of $\vec x$ and $\vec y$ and $d$ is the Euclidean distance. Therefore, we obtain another FDQC $f(S)$ (by permuting sites according to $f$) which produces the state $\psi'$ along the $x_{d+1} = 0$ space, which is a right angle
    \[\IZ_{\ge 0} \times 0 \cup 0 \times \IZ_{\ge 0}.\]
    We arrange this so as in \hyperref[figcornerswindle]{Figure \ref{figcornerswindle}} we obtain the ``corner swindle'' circuit $S_{NE}$ that produces a state $\psi'$ along $\IZ_{\ge 0} \times 0$ and elsewhere yields the product state.

    We now consider truncating $S_{NE}$ at a finite but large enough $x_{d+1}$. This produces the blended state $\psi'$ on the lower edge and another state $\psi''$ on its upper edge. These states can be regarded as inverses. Working as in \hyperref[propswindle]{Proposition \ref{propswindle}}, we now build a swindle out of the truncated $S_{NE}$ to obtain a circuit $S_{NE}'$ acting on the corner space which produces $\psi'$ on its boundary.

    The advantage of $S_{NE}'$ now is that we can regard it as a truncation of a circuit $S_E$ acting in the whole right half space $x_d \ge 0$ which preserves $\bar \psi_0$ everywhere by construction. Thus, we may regard $S_{NE}'$ as an $r$-blend. Furthermore, $S_{NE}'$ is equivalent to a trivial $r$-blend, since it admits a blend to the identity along the $d$th axis (the blend is $S_{NE}'$ itself). Thus,
    \[\varphi_{r+1}'=(S_{NE}')^{-1} \circ \varphi_{r+1} \sim \varphi_{r+1}\]
    (equivalence as $r$-blends).
    
    The upshot of this construction is that $\varphi_{r+1}'$ is an equivalent $r$-blend which creates a state $\psi''$ along the thickened half-space
    \[\IZ^{d-1} \times \IZ_{\le 0} \times [-w',w']^{l'}\]
    and elsewhere fixes $\bar \psi_0$.

    \begin{figure}
        \centering
        \includegraphics[width=10cm]{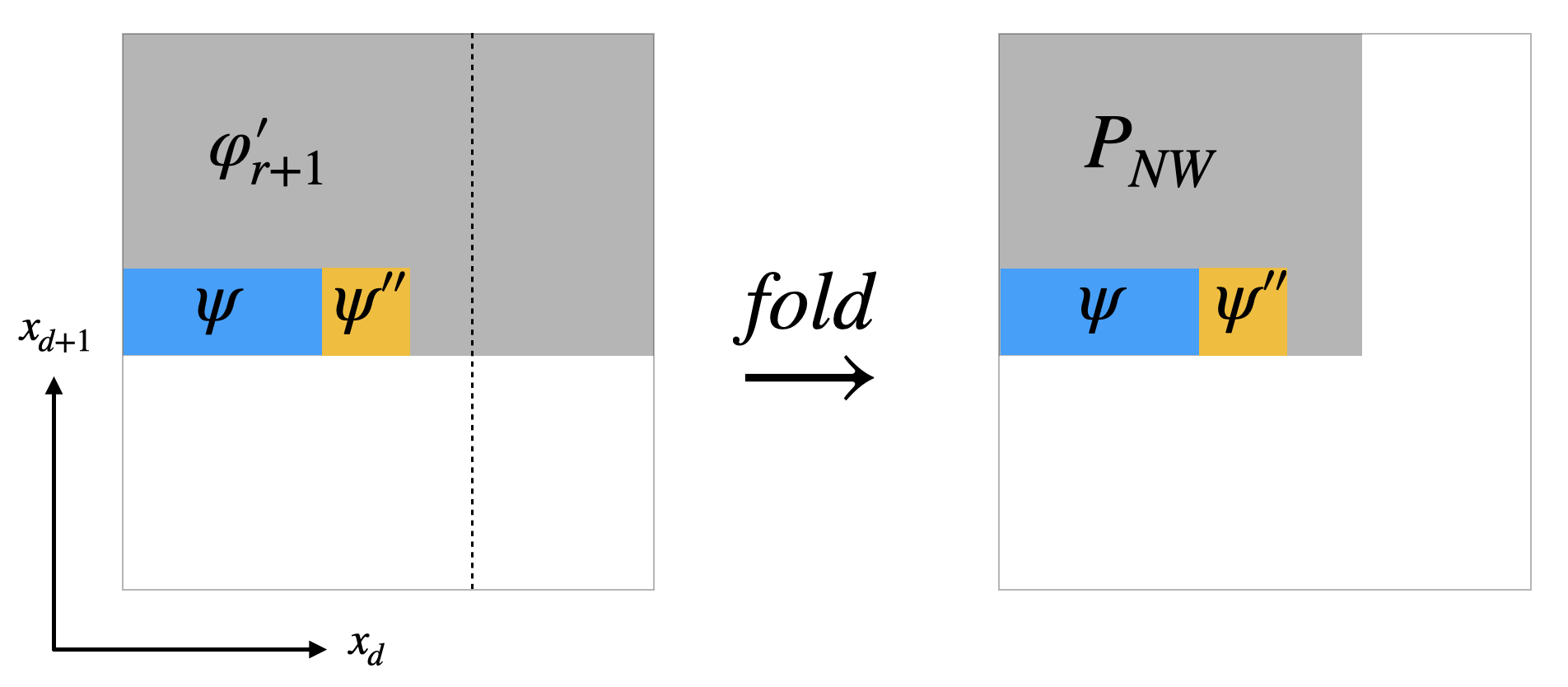}
        \caption{The composition $(S_{NE}')^{-1} \circ \varphi_{r+1} = \varphi_{r+1}'$ produces an $r$-blend which along a thickened half space $H = \IZ^{d-1} \times \IZ_{\le 0}$ we obtain $\psi$ (blue), except for a bounded neighborhood of $\IZ^{d-1}$ (orange), and elsewhere we obtain $\psi_0$. Utilizing a free direction in the infinite dimensional space $\IZ^\omega$, we fold this QCA like a pastry (similar to \hyperref[figsflippinghomotopy]{Figure \ref{figsflippinghomotopy}}), to obtain a new QCA $P_{NW}$ which produces the same state as $\varphi_{r+1}'$ but acts only in a neighborhood of  $\IZ^{d-1} \times \IZ_{\le 0}^{r+2}$. (Note that $\varphi_{r+1}'$ acts only in a neighborhood of $\IZ^d \times \IZ_{\le 0}^{r+1}$ because it starts with a 1-blend to the identity.)}
        \label{figpastry}
    \end{figure}

    For the next step we consider folding $\varphi_{r+1}'$ over a codimension 1 hyperplane at fixed $x_d$, using an unused direction, to produce an $r$-blend $P_{NW}$ as in \hyperref[figpastry]{Figure \ref{figpastry}}. All the maps in the array specifying the $r$-blend are likewise folded, and so we obtain an $r$-blend among blends all of which fix $\bar \psi_0$. Furthermore, $P_{NW}$ is equivalent to a trivial $r$-blend, since it blends along the $d$th axis to the identity, using itself as the blend. Thus, we have
    \[\varphi_{r+1} \sim \varphi_{r+1}' \sim P_{NW}^{-1} \circ \varphi_{r+1}'.\]
    By construction, this last $r$-blend fixes $\bar\psi_0$ everywhere, and is this equivalent to the identity. Finally, we have
    \[\varphi_{r+1} \sim 1,\]
    which is what we wanted to prove for injectivity.

    Finally, the long exact sequence is the usual cofiber long exact sequence, and its interpretation comes from the construction of the map from $r$-blends to FDQC-invertible states.

\end{proof}

\section{FDQC Representations from Group Cohomology}\label{appgroupcohfdqcreps}

Given a class $[\omega] \in H^{d+2}(BG,U(1))$, for finite $G$, we will construct a QCA (actually FDQC) $G$-representation on a lattice composed of on-site Hilbert spaces $\IC[G]$. This follows a well-known construction in \cite{Chen_2013}, reinterpreted in \cite{kapustin2015highersymmetrygappedphases}. We will show it has a series of blends allowing a computation of the Else-Nayak index yielding precisely $[\omega]$.

This calculation works because beyond the first blend $\beta_g$ (see \hyperref[subsecelsenayak]{Section \ref{subsecelsenayak}}), $\beta_{g,h}$ and higher are phase operators and all commute. This makes it a calculation in ordinary group cohomology, not involving the higher non-abelian structure of $\CQ(\CA)$. (See the discussion in \hyperref[subseccomptheanom]{Section \ref{subseccomptheanom}}.)

To proceed with the construction, we must first recall some details of nonabelian cohomology/finite gauge theory \cite{thorngren2018combinatorial}. Let $X$ be a simplicial complex with ordered vertices. We define $G$-valued 1-cocycle $A \in Z^1(X,G)$ to be an assignment of a group element
\[A(xy) \in G\]
for each directed edge $x < y$, such that for each directed triangle $x < y < z$,
\[\label{eqcocyclecondgaugefield}A(xy) A(yz) = A(xz).\]
Given an ``inhomogeneous'' cocycle
\[\omega:G^{d+2} \to \IR/2\pi \IZ\]
for group cohomology \cite{brown2012cohomology,hatcher2005algebraic} and a $G$-valued 1-cocycle $A$, we obtain a $d+2$-cocycle
\[\omega(A) \in C^{d+2}(X,\IR/2\pi \IZ) \\
\int_\Delta \omega(A) := \omega(A(x_0 x_1),A(x_1 x_2),\ldots, A(x_d x_{d+2})\]
where $\Delta$ is any $d+2$-simplex in $X$, with ordered vertices $x_0 < x_1 < \ldots < x_{d+2}$.

We also define a $G$-valued 0-cochain $f \in C^0(X,G)$ to be an assignment $f_x \in G$ for each vertex $x$. $C^0(X,G)$ forms a group and acts on $Z^1(X,G)$ by
\[A \mapsto A^f \\
A^f(xy) = f(x) A(xy) f(y)^{-1}.\]
We think of this as a gauge transformation of $A$.

Consider two simplices $\Delta^n$ and $\Delta^m$ of dimension $n$ and $m$ respectively, with ordered vertices. We define first a triangulation of the product $\Delta^n \times \Delta^m$. Vertices of $\Delta^n \times \Delta^m$ are given by pairs of vertices $(x,y)$, $x$ a vertex of $\Delta^n$ and $y$ a vertex of $\Delta^m$. We give these the lexicographic ordering, meaning $(x_1,y_1) < (x_2,y_2)$ if either $x_1 < x_2$ or $x_1 = x_2$ and $y_1 < y_2$. The $k$-simplices of $\Delta^n \times \Delta^m$ are given by ordered length $k+1$ sequences of these vertices.

\begin{figure}
    \centering
    \includegraphics[width=0.4\linewidth]{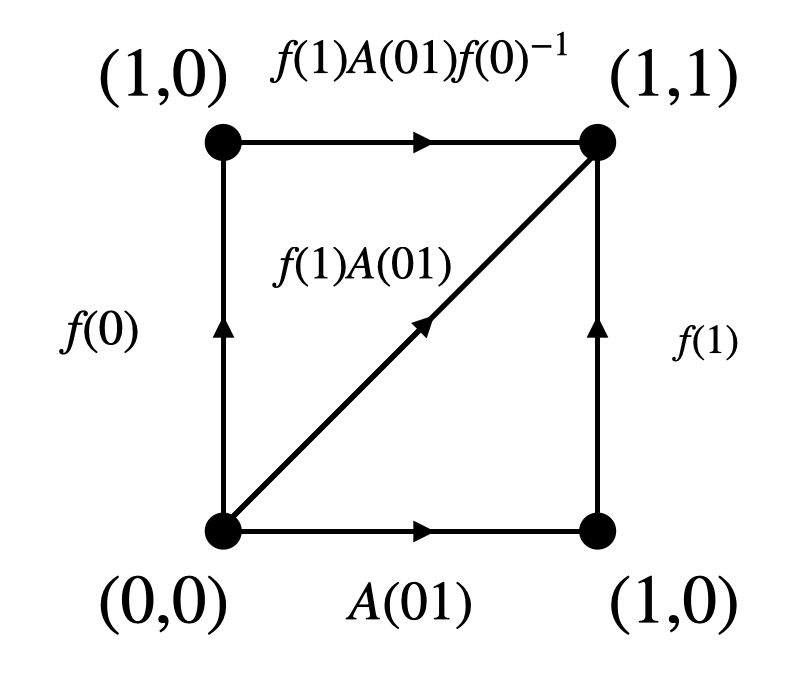}
    \caption{The triangulation of $\Delta^1 \times \Delta^1$ and corresponding $A^{(f)}$.}
    \label{figprism}
\end{figure}

Now suppose we are given $A \in Z^1(\Delta^n,G)$ and $\vec f=(f_1,\ldots,f_m) \in C^0(\Delta^n,G)$. We will define a particular element
\[A^{\vec f} \in Z^1(\Delta^n \times \Delta^m,G)\]
such that $A^{\vec f}$ restricts to $A^{f_j}$ on $\Delta^n \times j$. In particular, for $m = 1$, we get a cocycle $A^{(f)}$ on $\Delta^n \times[0,1]$ which on $\Delta^n \times 0$ restricts to $A$ and which on $\Delta^n \times 1$ restricts to $A^f$. The construction for $n=m=1$ is shown in Fig. \ref{figprism}.

$A^{\vec f}$ is easiest to define in terms of paths. Suppose we are given two vertices $(x_1,y_1) < (x_2,y_2)$ which is associated to an edge $E$. Let $y_0$ be the initial vertex of $\Delta^m$. There is a unique longest ordered path $p_1$ from $(x_1,y_0)$ to $(x_1,y_1)$, increasing only the first coordinate one step at a time. Likewise there is a shortest path $p_2$ from $(x_1,y_1)$ to $(x_2,y_1)$ and also $p_3$ from $(x_2,y_1)$ to $(x_2,y_2)$. Let $p_1$ consist of the edges
\[(x_1,y_0) =(x_1,z_1) < (x_1,z_2) < \cdots < (x_1,z_p) = (x_1,y_1).\]
We define
\[\vec f(p_1) = f_{z_{p-1}}(x_1)\cdots f_{z_2}(x_1) f_{z_1}(x_1),\]
where we have identified the vertices $z_i$ of $\Delta^m$ with $1 < 2 < \cdots < m+1$. Likewise let $p_3$ consist of the edges
\[(x_2,y_1) =(x_2,w_1) < (x_2,w_2) < \cdots < (x_2,w_q) = (x_2,y_2)\]
and define
\[\vec f(p_3) = f_{w_{q-1}}(x_1)\cdots f_{w_2}(x_1) f_{w_1}(x_1).\]
Let $p_2$ consist of edges $e_1,\ldots, e_l$. We also define
\[A(p_1) = A(e_l) A(e_{l-1}) \cdots A(e_1) \in G.\]
Finally let
\[A^{\vec f}(E) = \vec f(p_3) A(p_2) \vec f(p_1)^{-1}.\]
It is not too hard to check that $A^{\vec f}$ satisfies the cocycle equation \eqref{eqcocyclecondgaugefield} as well as the restriction conditions we wanted.

Now let $A \in Z^1(X,G)$. We define, for each $m$, the \textbf{$m$th descendant}
\[\omega_m(A,f_1,\ldots,f_m) \in C^{d+2-m}(X,\IR/2\pi\IZ)\]
where $f_1,\ldots,f_m \in C^0(X,G)$, as follows. Take any $\Delta^{d+2-m} \subset X$ and restrict $A$ to it. We define
\[\int_{\Delta^{d+2-m}}\omega_m(A,f_1,\ldots,f_m) := \int_{\Delta^{d+2-m} \times \Delta^{m}} \omega(A^{\vec f}).\]
Note that for $m = d+2$, $X = \star$, the $d+2$nd descendant is equal to the inhomogeneous cocycle $\omega$ evaluated on $f_1,\ldots,f_{d+2} \in G$.

These descendants are extremely useful. For example, the 1st descendant satisfies
\[\omega(A^f) - \omega(A) + d\omega_1(A,f) = 0.\]
This can be proven by considering $A^{(f)}$ on $\Delta^{d+2} \times [0,1]$. Since $d\omega = 0$, we have
\[0 = \int_{\Delta^{d+2} \times [0,1]} d\omega(A^{(f)}) = \int_{\partial(\Delta^{d+2} \times [0,1])} \omega(A^{(f)}) \\
= \int_{\Delta^{d+2} \times 1} \omega(A^f) - \int_{\Delta^{d+2} \times 0} \omega(A) + \int_{(\partial \Delta^{d+2}) \times [0,1]} \omega(A^{(f)} \\
= \int_{\Delta^{d+2} \times 1} \omega(A^f) - \int_{\Delta^{d+2} \times 0} \omega(A) + \int_{\partial \Delta^{d+2}} \omega_1(A,f).\]
Likewise, the 2nd descendant satisfies
\[\omega_1(A,f_2 f_1) - \omega_1(A,f_1) - \omega_1(A^{f_1},f_2) + d\omega_2(A,f_1,f_2) = 0\]
which can be proven the same way. In general, $\omega_m$ measures the boundary term needed to make $\omega_{m-1}$ an $m-1$-cocycle for the group law on $C^0(X,G)$.

To warm up with a physics application, consider a $d+1$-manifold $X$ with Hilbert space $\IC[G]$ associated to each vertex. We can define on it the wavefunction
\[\ket{\omega_1} = \sum_{f \in C^0(X,G)} e^{i \int_X \omega_1(0,f)} \ket{f},\]
where 0-cochains $f \in C^0(X,G)$ label the product state group basis. Suppose we act on this Hilbert space by left multiplication by some $g \in G$. From the 2nd descendant we have the identity
\[\omega_1(0,gf)-\omega_1(0,f) + d\omega_2(0,f,g) = 0\]
(note $\omega_1(0,g) = 0$, which can easily be proven from the definition). This implies that $\ket{\omega_1}$ is symmetric up to a boundary term given by the 2nd descendant. This boundary term will give us our FDQC $G$-representation.

Let $Y^d$ be a triangulated $d$-manifold with a Hilbert space $\IC[G]$ assigned to each vertex. For example, we could take $Y^d = \IR^d$ with a triangulation whose vertices live on the integer lattice $\IZ^d$. Then we would be in the usual setting of QCA $G$-representations studied in this work. However, the construction generally works as long as $Y^d$ has no boundary. As before, we work in a basis labeled by 0-cochains $f \in C^0(Y,G)$. We define the following operator in this basis
\[U_g \ket{f} = e^{i \int_Y \omega_2(0,f,g)} \ket{g \cdot f}.\]
This can be expressed as a finite depth quantum circuit, with one layer given by on-site left-multiplication $L_g$ and the other layers defined by the mutually commuting local phase operators
\[\ket{f} \mapsto e^{i \int_{\Delta^d} \omega_2(0,f,g)} \ket{f}\]
on each $d$-simplex $\Delta^d$. The circuit is finite depth because each vertex only belongs to a finite set of $d$-simplices, and $\int_{\Delta^d} \omega_2(0,f,g)$ only depends on $f$ restricted to $\Delta^d$.

The 3rd descendant $\omega_3(A,f,g_1,g_2)$ satisfies
\[d\omega_3(A,f,g_1,g_2) = \omega_2(A,f,g_1 g_2) - \omega_2(A,f,g_1) - \omega_2(A,g_1\cdot f, g_2).\]
The existence of this descendant implies that $U_g$ satisfies the $G$ group law
\[U_{g_1} U_{g_2} = U_{g_1 g_2}\]
up to boundary terms. Thus these define an FDQC $G$-representation if $Y$ has no boundary.

We can define a very convenient set of blends by taking our symmetry to act only on a triangulated submanifold $W \subset Y$ with the same definition as above, applying left multiplication to vertices in $W$ and integrating over $d$-simplices in $W$. Then we will have
\[(U^W_{gh})^{-1} U^W_g U^W_h \ket{f} = e^{i \int_{\partial W} \omega_3(0,f,g,h)}\ket{f}.\]
We can then define a blend-of-blend to be
\[V^\Gamma_{g,h} \ket{f} = e^{i \int_\Gamma \omega_3(0,f,g,h)}\ket{f},\]
where $\Gamma$ is any codimension 1 triangulated submanifold of $Y$. Proceeding in this way, we get a series of blends which we can always continue using the descendants. Note that they are all diagonal in the $C^0(Y,G)$ basis. At the last stage, we find the local operator
\[N^x_{g_1,\ldots,g_{d+1}}\ket{f} = e^{i \int_{x}\omega_{d+2}(0,f,g_1,\ldots,g_{d+1})}\ket{f} = e^{i \omega(f(x),g_1,\ldots,g_{d+1})}\ket{f}.\]
We then ask if this operator satisfies a (twisted) $d+1$-cocycle equation for $g_1,\ldots,g_{d+1}$. The Else-Nayak index \cite{Else_2014} is
\[(N^x_{g_1,\ldots,g_{d+1}})^\pm (N^x_{g_1,\ldots,g_d,g_{d+1} g_{d+2}})^{\mp} \cdots \\ \cdots (N^x_{g_1 g_2, g_3,\ldots,g_{d+2}})^{- 1} {\rm Ad}_{U_{g_1}}(N^x_{g_2,\ldots,g_{d+2}}),\]
where $\pm = (-1)^{d+1}$. This yields a pure phase operator whose exponent is
\[\pm \omega(f(x),g_1,\ldots,g_{d+1}) \mp \omega(f(x),g_1,\ldots,g_{d+1} g_{d+2}) + \cdots \\ \ + \omega(f(x)g_1, g_2,\ldots,g_{d+2}).\]
This is equal to
\[(d\omega)(f(x),g_1,\ldots,g_{d+2}) + \omega(g_1,\ldots,g_{d+2}).\]
The first term is zero because $\omega$ is a $d+2$-cocycle. Thus, the Else-Nayak index we obtain from these blends is precisely the cocycle we began with!

\section{Dependence of Anomaly Indices on Choices of Lifts}\label{appdependenceonlifts}

In this appendix we consider stable anomalies for simplicity. We will show that the anomaly indices, including the Else-Nayak index, in general depend on the choices of lifts. We will prove the following proposition as well as provide some physical interpretation and an example.

\vspace{3mm}
\begin{mdframed}[backgroundcolor=blue!5]
\begin{proposition}\label{propanomalyindexdependence}
    Suppose we have two lifts $\alpha_k$ and $\tilde \alpha_k$
\begin{equation}
\begin{tikzcd}
  & \CQ^{d,k} \arrow[d, "p^k"] \arrow[r,"c^k"] & B^{k+1}\pi_{k+1} \CQ^d \\
BG \arrow[r,"\alpha"] \arrow[ur,"{\alpha^k,\tilde \alpha^k}"] & \CQ^{d} &
\end{tikzcd} 
\end{equation}
then the $k+1$st anomaly indices we compute between them differ by
    \[ [c^k \circ \tilde \alpha^k] - [c^k \circ \alpha^k] = [c^k \circ i^k \circ \beta^k],\]
    where $\beta^k:BG \to \CQ_k^d$ is a map into the fiber of the Whitehead map
\begin{equation}
    \begin{tikzcd}
        \CQ^d_k \arrow[r,"i^k"] & \CQ^{d,k}  \arrow[d,"p^k"] \\
        & \CQ^d 
    \end{tikzcd}
\end{equation}
and measures the ``difference'' between the lifts, satisfying
\[ [\tilde \alpha^k] - [\alpha^k] = [i^k \circ \beta^k]\]
in the abelian group of homotopy classes of maps $BG \to \CQ^{d,k}$.

These fibers $\CQ^d_k$ form the Postnikov tower for $\Omega_\star \CQ^d$:
\[\Omega_\star\CQ^d=\CQ^d_{d+2} \to \cdots \to\CQ_k^d \to \CQ_{k-1}^d \to \cdots \to \CQ_1^d=\pi_1\CQ^d \to \CQ_0^d = \star\]
and
\[ [c^k \circ i^k] \in H^{k+1}(\CQ^d_k,\pi_{k+1} \CQ^d)\]
is the Postnikov invariant.

For lifts $\alpha^k$, $\tilde \alpha^k$ which agree at the $k-1$ level:
\begin{equation}
\begin{tikzcd}
 B^{k-1} \pi_k \CQ^d \arrow[r,"j^k"] & \CQ^{d,k} \arrow[d] \arrow[r,"c^k"] & B^{k+1}\pi_{k+1} \CQ^{d,k-1} \\
BG \arrow[r,"\alpha"] \arrow[ur,"{\alpha^k,\tilde \alpha^k}"] & \CQ^{d,k-1} &
\end{tikzcd} 
\end{equation}
their difference is measured by $\gamma^k:BG \to B^{k-1} \pi_k\CQ^d$, and their $k+1$st anomalies indices satisfy
\[ [c^k \circ \tilde \alpha^k] -[c^k \circ \alpha^k] = [c^k \circ j^k \circ \gamma^k],\]
where
\[ [c^k \circ j^k] \in H^{k+1}(B^{k-1} \pi_{k} \CQ,\pi_{k+1} \CQ)\]
is a ``mini Postnikov invariant'' classifying the ``mini tower'' $\CQ^{d,k} \to \CQ^{d,k-1}$.

In particular, the first and second anomaly indices are independent of choices, since $\alpha^0$ is determined for us and
\[ [c^1 \circ j^1] \in H^2(\pi_1 \CQ, \pi_2 \CQ) = 0\]
since $\pi_1 \CQ$ is a discrete space.
\end{proposition}
\end{mdframed}
\vspace{3mm}

\begin{proof}
    Let us recall the Whitehead tower of $\CQ^d$
\[B^{d+2}U(1)_{disc} = \CQ^{d,d+2}\to \cdots \to \CQ^{d,2} \xrightarrow{p_2} \CQ^{d,1} \xrightarrow{p_1} \CQ^{d,0}=\CQ^d.\]
Suppose that we have a stable $d$-QCA $G$-representation with corresponding map
\[\alpha:BG \to \CQ^d,\]
and we are trying to determine its anomaly by taking lifts up the Whitehead tower. Suppose that we have constructed a lift up to step $k$, giving the following homotopy-commutative diagram (The square is a homotopy pullback square): 
\begin{equation}
\begin{tikzcd}
B^k \pi_{k+1} \CQ^d \arrow[r] & \CQ^{d,k+1} \arrow[d] \arrow[r]  & \star \arrow[d] \\
  & \CQ^{d,k} \arrow[d, "p^k"] \arrow[r, "c^k"] & B^{k+1} \pi_{k+1}\CQ^d\\
BG \arrow[r,"\alpha"] \arrow[ur,"\alpha^k"] & \CQ^{d} &
\end{tikzcd}
\end{equation}
The $k+1$st anomaly index is the homotopy class of the map
\[c^k \circ \alpha^k: BG \to B^{k+1} \pi_{k+1} \CQ^d.\]
We want to understand the dependence of this anomaly index on the choice of lift $\alpha^k$ for fixed $\alpha$.

Suppose we choose a different lift $\tilde \alpha^k$. Since $p^k \circ \alpha^k = p^k \circ \tilde \alpha^k$ up to homotopy, we can compare them fiberwise, obtaining a (homotopy class of) map
\[\beta^k:BG \to \CQ^d_k\]
where $\CQ^d_k$ is the fiber of $p^k$. In fact, since we are in the stable case, $\CQ^d$ is an infinite loop space, and so are each of the spaces $\CQ^{d,k}$ in the Whitehead tower. Maps $BG \to \CQ^{d,k}$ thus form an abelian group up to homotopy, and we in fact have
\[ [\tilde \alpha^k] - [\alpha^k] = [i^k \circ \beta^k],\]
where brackets denote homotopy class and $i^k:\CQ^d_k \to \CQ^{d,k}$ is the inclusion of the fiber. To summarize, the situation is now
\begin{equation}
\begin{tikzcd}
B^k \pi_{k+1} \CQ^d \arrow[r] & \CQ^{d,k+1} \arrow[d] \arrow[r]  & \star \arrow[d] \\
 \CQ^d_k \arrow[r,"i^k"] & \CQ^{d,k} \arrow[d, "p^k"] \arrow[r, "c^k"] & B^{k+1} \pi_{k+1}\CQ^d\\
BG \arrow[u,"\beta^k"] & \CQ^{d} &
\end{tikzcd}
\end{equation}
In terms of the anomaly indices, we have
\[ [c^k \circ \tilde \alpha^k] - [c^k \circ \alpha_k] = [c^k \circ i^k \circ \beta^k] \in H^{k+1}(BG, \pi_{k+1} \CQ^d).\]
Thus we see that the anomaly indices can depend on the lifts if
\[ [c^k \circ i^k] \in H^{k+1}(\CQ_k^d,\pi_{k+1} \CQ^d)\]
is nonzero.

The fibers $\CQ_k^d$ appearing in the Whitehead tower form their own tower called the ``Postnikov tower'', in this case of $\Omega_\star \CQ^d$ \cite{hatcher2005algebraic,may1999concise}:
\[\Omega_\star\CQ^d=\CQ^d_{d+2} \to \cdots \to\CQ_k^d \to \CQ_{k-1}^d \to \cdots \to \CQ_1^d=\pi_1\CQ^d \to \CQ_0^d = \star\]
These form another sequence of fibrations (compare the Whitehead tower \eqref{eqnwhiteheadtower}):
\begin{equation}
\begin{tikzcd}
B^{k} \pi_{k+1}(\CQ^d) \arrow[r] & \CQ^{d}_{k+1} \arrow[d, "m^k"] \arrow[r] & \star \arrow[d] \\
& \CQ^{d}_k \arrow[r, "c^k \circ i^k"] & B^{k+1} \pi_{k+1} \CQ^d
\end{tikzcd}    
\end{equation}
Note that the classifying map $c^k \circ i^k$, known as the Postnikov invariant, is the same combination appearing above.

Thus we find that in general, \textbf{nonzero Postnikov invariants can lead to dependence of the anomaly indices on choices of lifts}.

We can illustrate this more concretely in the case that $\alpha^k$ and $\tilde \alpha^k$ coincide all the way up to the $k-1$st level, meaning we have
\begin{equation}
\begin{tikzcd}
B^k \pi_{k+1} \CQ^d \arrow[r] & \CQ^{d,k+1} \arrow[r] \arrow[d] & \star \arrow[d] \\
  B^{k-1} \pi_k \CQ^d \arrow[r,"j^k"] & \CQ^{d,k} \arrow[d, "p^k"] \arrow[r,"c^k"] & B^{k+1} \pi_{k+1} \CQ^d \\
BG \arrow[r,"\alpha^{k-1}"] \arrow[ur,"\tilde\alpha^k"] & \CQ^{d,k-1} & 
\end{tikzcd}
\end{equation}
Now the difference between $\alpha^k$ and $\tilde \alpha^k$ is measured simply by
\[\gamma^k:BG \to B^{k-1} \pi_k \CQ^d \\
[\tilde \alpha^k] - [\alpha_k] = [j^k \circ \gamma^k].\]
This gives a difference in the $k+1$st anomaly index
\[ [c^k \circ j^k \circ \gamma^k],\]
where the ``mini Postnikov class''
\[ [c^k \circ j^k] \in H^{k+1}(B^{k-1} \pi_k \CQ^d,\pi_{k+1} \CQ^d)\]
appears, which classifies this two-piece subtower of the whole Postnikov tower of $\Omega_\star \CQ^d$. Note these are what appear in the $E^2$ page of the Atiyah-Hirzebruch spectral sequence.
\end{proof}

Although the dependence of the anomaly indices on the choices of lifts may seem worrisome, it is actually a necessary feature for the spectrum of QCA to have. Recall that there is a map
\[B^{d+2}U(1)_{disc} \to \CQ^d.\]
This gives a map (for finite $G$ the topology on $U(1)$ doesn't matter)
\[i:H^{d+2}(BG,U(1)) \to \CQ^d(BG)\]
which can be understood as the inclusion of ``group cohomology anomalies'' in the group of stable blend anomalies. We can define this map in lattice terms by taking the FDQC $G$ symmetries that come from group cohomology classes (see Appendix \ref{appgroupcohfdqcreps}), and measuring its stable blend anomaly.

We expect that the map $i$ is neither injective nor surjective, and moreover that $H^{d+2}(BG,U(1))$ can mix with a non-trivial extension of the quotient
\[\CQ^d(BG)/i(H^{d+2}(BG,U(1))).\]
Such phenomena are only possible if the Postnikov classes are non-vanishing (see eg. \cite{thorngren2019anomaliesbosonization} and references therein).

A similar phenomenon is known in the study of the spectrum of invertible orientable TQFTs in $d+2$-dimensions \cite{Freed_2021}, which we denote $\Omega^{d+2}_{SO}$. There is likewise a map
\[H^{d+2}(BG,U(1)) \to \Omega^{d+2}_{SO}(BG)\]
which is known to have these properties \cite{kapustin2014symmetryprotectedtopologicalphases}. This map can be considered as taking the orientable invertible TQFT defined by the Dijkgraaf-Witten path integral \cite{dijkgraaf1990topological}.

We can also consider systems of fermions, and a spin TQFT spectrum $\Omega^{d+2}_{Spin}$, giving also a map
\[H^{d+2}(BG,U(1)) \to \Omega^{d+2}_{Spin}(BG)\]
corresponding again to Dijkgraaf-Witten theory but now considered as a spin TQFT. This has some low dimensional examples where group cohomology SPTs become trivial. We expect that the corresponding boundary theory has an anomalous symmetry with an ill-defined Else-Nayak invariant, reflecting that the Else-Nayak invariant is not an invariant of the bulk SPT phase.

The simplest example is for $G = \IZ_2$, and corresponds to
\[\frac12 A^5 \in H^5(B\IZ_2,U(1)).\]
On spin 5-manifolds manifolds, this class is always trivial, since
\[A^5 = Sq^2 A^3.\]
We calculated a Postnikov class for $\CQ^3(B\IZ_2)$ in \hyperref[secanomalousferms1d]{Section \ref{secanomalousferms1d}} and found the same $Sq^2$ operation appear. So the Atiyah-Hirzebruch spectral sequence has a differential on the $E^2$ page eating $A^5$, in other words $A^5$ maps to zero in $\CQ^3(B\IZ_2)$. Thus, we should expect that the Else-Nayak invariant is ambiguous by this $A^5$ term when we compute it on a 3d lattice.

\end{oldappendices}

 \bibliography{main}

@article{fredenhagen1982locality,
author = {Buchholz, Detlev and Fredenhagen, Klaus},
year = {1982},
month = {03},
pages = {1-54},
title = {Locality and the structure of particle states},
volume = {84},
journal = {Communications in Mathematical Physics},
doi = {10.1007/BF01208370}
}

@article{Freedman_2020,
   title={Classification of Quantum Cellular Automata},
   volume={376},
   ISSN={1432-0916},
   url={http://dx.doi.org/10.1007/s00220-020-03735-y},
   DOI={10.1007/s00220-020-03735-y},
   number={2},
   journal={Communications in Mathematical Physics},
   publisher={Springer Science and Business Media LLC},
   author={Freedman, Michael and Hastings, Matthew B.},
   year={2020},
   month=apr, pages={1171–1222} }

@misc{schumacher2004reversiblequantumcellularautomata,
      title={Reversible quantum cellular automata}, 
      author={B. Schumacher and R. F. Werner},
      year={2004},
      eprint={quant-ph/0405174},
      archivePrefix={arXiv},
      primaryClass={quant-ph},
      url={https://arxiv.org/abs/quant-ph/0405174}, 
}

@article{haah2023invertible,
  title={Invertible subalgebras},
  author={Haah, Jeongwan},
  journal={Communications in Mathematical Physics},
  volume={403},
  number={2},
  pages={661--698},
  year={2023},
  publisher={Springer}
}

@misc{kitaevIPAMtalk,
  author       = {Alexei Kitaev},
  title        = {Homotopy‐theoretic approach to SPT phases in action: $Z_{16}$ classification of three‐dimensional superconductors},
  howpublished = {Talk at IPAM, UCLA},
  year         = {2015},
  note         = {Abstract available at \url{https://www.ipam.ucla.edu/abstract/?tid=12389}}
}

@article{qi2011topological,
  title={Topological insulators and superconductors},
  author={Qi, Xiao-Liang and Zhang, Shou-Cheng},
  journal={Reviews of modern physics},
  volume={83},
  number={4},
  pages={1057--1110},
  year={2011},
  publisher={APS}
}

@article{hasan2010colloquium,
  title={Colloquium: topological insulators},
  author={Hasan, M Zahid and Kane, Charles L},
  journal={Reviews of modern physics},
  volume={82},
  number={4},
  pages={3045--3067},
  year={2010},
  publisher={APS}
}

@article{lieb1961two,
  title={Two soluble models of an antiferromagnetic chain},
  author={Lieb, Elliott and Schultz, Theodore and Mattis, Daniel},
  journal={Annals of Physics},
  volume={16},
  number={3},
  pages={407--466},
  year={1961},
  publisher={Elsevier}
}

@article{else2025t,
  title={’t Hooft Anomalies in Metals},
  author={Else, Dominic V},
  journal={Annual Review of Condensed Matter Physics},
  volume={17},
  year={2025},
  publisher={Annual Reviews}
}

@article{cho2017anomaly,
  title={Anomaly manifestation of Lieb-Schultz-Mattis theorem and topological phases},
  author={Cho, Gil Young and Hsieh, Chang-Tse and Ryu, Shinsei},
  journal={Physical Review B},
  volume={96},
  number={19},
  pages={195105},
  year={2017},
  publisher={APS}
}

@article{seiberg2024majorana,
  title={Majorana chain and Ising model-(non-invertible) translations, anomalies, and emanant symmetries},
  author={Seiberg, Nathan and Shao, Shu-Heng},
  journal={SciPost Physics},
  volume={16},
  number={3},
  pages={064},
  year={2024}
}

@article{jones2021one,
  title={One-dimensional lattice models for the boundary of two-dimensional Majorana fermion symmetry-protected topological phases: Kramers-Wannier duality as an exact Z 2 symmetry},
  author={Jones, Robert A and Metlitski, Max A},
  journal={Physical Review B},
  volume={104},
  number={24},
  pages={245130},
  year={2021},
  publisher={APS}
}

@misc{malkiewich2023spectra,
  title={Spectra and stable homotopy theory (draft version, first 6 chapters)},
  author={Malkiewich, Cary},
  year={2023}
}

@article{Metlitski_2018,
   title={Intrinsic and emergent anomalies at deconfined critical points},
   volume={98},
   ISSN={2469-9969},
   url={http://dx.doi.org/10.1103/PhysRevB.98.085140},
   DOI={10.1103/physrevb.98.085140},
   number={8},
   journal={Physical Review B},
   publisher={American Physical Society (APS)},
   author={Metlitski, Max A. and Thorngren, Ryan},
   year={2018},
   month=aug }

@article{Cheng_2023,
   title={Lieb-Schultz-Mattis, Luttinger, and ’t Hooft - anomaly matching in lattice systems},
   volume={15},
   ISSN={2542-4653},
   url={http://dx.doi.org/10.21468/SciPostPhys.15.2.051},
   DOI={10.21468/scipostphys.15.2.051},
   number={2},
   journal={SciPost Physics},
   publisher={Stichting SciPost},
   author={Cheng, Meng and Seiberg, Nathan},
   year={2023},
   month=aug }

@article{else2020topological,
  title={Topological theory of Lieb-Schultz-Mattis theorems in quantum spin systems},
  author={Else, Dominic V and Thorngren, Ryan},
  journal={Physical Review B},
  volume={101},
  number={22},
  pages={224437},
  year={2020},
  publisher={APS}
}

@article{Kitaev_2001,
   title={Unpaired Majorana fermions in quantum wires},
   volume={44},
   ISSN={1468-4780},
   url={http://dx.doi.org/10.1070/1063-7869/44/10S/S29},
   DOI={10.1070/1063-7869/44/10s/s29},
   number={10S},
   journal={Physics-Uspekhi},
   publisher={Uspekhi Fizicheskikh Nauk (UFN) Journal},
   author={Kitaev, A Yu},
   year={2001},
   month=oct, pages={131–136} }

@article{kapustin2025higher,
  title={Higher symmetries and anomalies in quantum lattice systems},
  author={Kapustin, Anton and Xu, Shixiong},
  journal={arXiv preprint arXiv:2505.04719},
  year={2025}
}

@article{street1991parity,
  title={Parity complexes},
  author={Street, Ross},
  journal={Cahiers de topologie et g{\'e}om{\'e}trie diff{\'e}rentielle cat{\'e}goriques},
  volume={32},
  number={4},
  pages={315--343},
  year={1991}
}

@article{Haah_2022,
   title={Nontrivial Quantum Cellular Automata in Higher Dimensions},
   volume={398},
   ISSN={1432-0916},
   url={http://dx.doi.org/10.1007/s00220-022-04528-1},
   DOI={10.1007/s00220-022-04528-1},
   number={1},
   journal={Communications in Mathematical Physics},
   publisher={Springer Science and Business Media LLC},
   author={Haah, Jeongwan and Fidkowski, Lukasz and Hastings, Matthew B.},
   year={2022},
   month=nov, pages={469–540} }

@article{Gross_2012,
   title={Index Theory of One Dimensional Quantum Walks and Cellular Automata},
   volume={310},
   ISSN={1432-0916},
   url={http://dx.doi.org/10.1007/s00220-012-1423-1},
   DOI={10.1007/s00220-012-1423-1},
   number={2},
   journal={Communications in Mathematical Physics},
   publisher={Springer Science and Business Media LLC},
   author={Gross, D. and Nesme, V. and Vogts, H. and Werner, R. F.},
   year={2012},
   month=jan, pages={419–454} }

@article{kogut1979introduction,
  title={An introduction to lattice gauge theory and spin systems},
  author={Kogut, John B},
  journal={Reviews of Modern Physics},
  volume={51},
  number={4},
  pages={659},
  year={1979},
  publisher={APS}
}

@article{Tarantino_2016,
   title={Discrete spin structures and commuting projector models for two-dimensional fermionic symmetry-protected topological phases},
   volume={94},
   ISSN={2469-9969},
   url={http://dx.doi.org/10.1103/PhysRevB.94.115115},
   DOI={10.1103/physrevb.94.115115},
   number={11},
   journal={Physical Review B},
   publisher={American Physical Society (APS)},
   author={Tarantino, Nicolas and Fidkowski, Lukasz},
   year={2016},
   month=sep }

@misc{sopenko2025reflectionpositivityrefinedindex,
      title={Reflection positivity and a refined index for 2d invertible phases}, 
      author={Nikita Sopenko},
      year={2025},
      eprint={2509.01711},
      archivePrefix={arXiv},
      primaryClass={math-ph},
      url={https://arxiv.org/abs/2509.01711}, 
}

@misc{baez2004higherdimensionalalgebrav2groups,
      title={Higher-Dimensional Algebra V: 2-Groups}, 
      author={John C. Baez and Aaron D. Lauda},
      year={2004},
      eprint={math/0307200},
      archivePrefix={arXiv},
      primaryClass={math.QA},
      url={https://arxiv.org/abs/math/0307200}, 
}

@incollection{hooft1980naturalness,
  title={Naturalness, chiral symmetry, and spontaneous chiral symmetry breaking},
  author={Hooft, G’t},
  booktitle={Recent developments in gauge theories},
  pages={135--157},
  year={1980},
  publisher={Springer}
}

@misc{friedman2023elementaryillustratedintroductionsimplicial,
      title={An elementary illustrated introduction to simplicial sets}, 
      author={Greg Friedman},
      year={2023},
      eprint={0809.4221},
      archivePrefix={arXiv},
      primaryClass={math.AT},
      url={https://arxiv.org/abs/0809.4221}, 
}

@article{Ara_2013,
   title={On the homotopy theory of Grothendieck's infinity-groupoids},
   volume={217},
   ISSN={0022-4049},
   url={http://dx.doi.org/10.1016/j.jpaa.2012.10.010},
   DOI={10.1016/j.jpaa.2012.10.010},
   number={7},
   journal={Journal of Pure and Applied Algebra},
   publisher={Elsevier BV},
   author={Ara, Dimitri},
   year={2013},
   month=jul, pages={1237–1278} }

@misc{followup,
title={Quantum Cellular Automata and Higher Categories},
author={Alexander M. Czajka and Roman Geiko and Ryan Thorngren},
journal={to appear}}

@article{haegeman2015gauging,
  title={Gauging quantum states: from global to local symmetries in many-body systems},
  author={Haegeman, Jutho and Van Acoleyen, Karel and Schuch, Norbert and Cirac, J Ignacio and Verstraete, Frank},
  journal={Physical Review X},
  volume={5},
  number={1},
  pages={011024},
  year={2015},
  publisher={APS}
}

@incollection{chen2022exactly,
  title={Exactly Solvable Topological and Fracton Models as Gauge Theories},
  author={Chen, Xie},
  booktitle={A Festschrift in Honor of the CN Yang Centenary: Scientific Papers},
  pages={151--166},
  year={2022},
  series={World Scientific}
}

@article{greenlees1995equivariant,
  title={Equivariant stable homotopy theory},
  author={Greenlees, John PC and May, J Peter},
  journal={Handbook of algebraic topology},
  volume={277},
  pages={323},
  year={1995},
  publisher={North-Holland}
}

@book{whitehead2012elements,
  title={Elements of homotopy theory},
  author={Whitehead, George W},
  volume={61},
  year={2012},
  series={Springer Science \& Business Media}
}

@book{brown2012cohomology,
  title={Cohomology of Groups},
  author={Brown, K.S.},
  isbn={9781468493290},
  series={Graduate Texts in Mathematics},
  url={https://books.google.com/books?id=6aUVoAEACAAJ},
  year={2012}
}

@book{hatcher2005algebraic,
  title={Algebraic topology},
  author={Hatcher, Allen},
  year={2005},
  publisher={}
}

@phdthesis{thorngren2018combinatorial,
  title={Combinatorial topology and applications to quantum field theory},
  author={Thorngren, Ryan George},
  year={2018},
  school={University of California, Berkeley}
}

@book{bratteli2012operator,
  title={Operator Algebras and Quantum Statistical Mechanics 1},
  author={Ola Bratteli, Derek W. Robinson},
  series={Theoretical and Mathematical Physics},
  year={2010}
}

@book{may1999concise,
  title={A Concise Course in Algebraic Topology},
  author={May, J.P.},
  series={Chicago Lectures in Mathematics},
  year={1999}
}

@misc{kubota2025stablehomotopytheoryinvertible,
      title={Stable homotopy theory of invertible gapped quantum spin systems I: Kitaev's $\Omega$-spectrum}, 
      author={Yosuke Kubota},
      year={2025},
      eprint={2503.12618},
      archivePrefix={arXiv},
      primaryClass={math-ph},
      url={https://arxiv.org/abs/2503.12618}, 
}

@article{Zhang_2022,
   title={Vanishing Hall conductance for commuting Hamiltonians},
   volume={105},
   ISSN={2469-9969},
   url={http://dx.doi.org/10.1103/PhysRevB.105.L081103},
   DOI={10.1103/physrevb.105.l081103},
   number={8},
   journal={Physical Review B},
   publisher={American Physical Society (APS)},
   author={Zhang, Carolyn and Levin, Michael and Bachmann, Sven},
   year={2022},
   month=feb }

@misc{zhang2022topologicalinvariantssptentanglers,
      title={Topological invariants for SPT entanglers}, 
      author={Carolyn Zhang},
      year={2022},
      eprint={2210.02485},
      archivePrefix={arXiv},
      primaryClass={cond-mat.str-el},
      url={https://arxiv.org/abs/2210.02485}, 
}

@book{may2012more,
  title={More concise algebraic topology: localization, completion, and model categories},
  author={May, J Peter and Ponto, Kate},
  year={2012},
  series={University of Chicago Press}
}

@misc{kawagoe2025anomalydiagnosissymmetryrestriction,
      title={Anomaly diagnosis via symmetry restriction in two-dimensional lattice systems}, 
      author={Kyle Kawagoe and Wilbur Shirley},
      year={2025},
      eprint={2507.07430},
      archivePrefix={arXiv},
      primaryClass={cond-mat.str-el},
      url={https://arxiv.org/abs/2507.07430}, 
}

@misc{shirley2025anomalyfreesymmetriesobstructionsgauging,
      title={Anomaly-free symmetries with obstructions to gauging and onsiteability}, 
      author={Wilbur Shirley and Carolyn Zhang and Wenjie Ji and Michael Levin},
      year={2025},
      eprint={2507.21267},
      archivePrefix={arXiv},
      primaryClass={cond-mat.str-el},
      url={https://arxiv.org/abs/2507.21267}, 
}

@article{Beaudry_2024,
   title={Homotopical foundations of parametrized quantum spin systems},
   volume={36},
   ISSN={1793-6659},
   url={http://dx.doi.org/10.1142/S0129055X24600031},
   DOI={10.1142/s0129055x24600031},
   number={09},
   journal={Reviews in Mathematical Physics},
   publisher={World Scientific Pub Co Pte Ltd},
   author={Beaudry, Agnès and Hermele, Michael and Moreno, Juan and Pflaum, Markus J. and Qi, Marvin and Spiegel, Daniel D.},
   year={2024},
   month=mar }

@article{Freed_2021,
   title={Reflection positivity and invertible topological phases},
   volume={25},
   ISSN={1465-3060},
   url={http://dx.doi.org/10.2140/gt.2021.25.1165},
   DOI={10.2140/gt.2021.25.1165},
   number={3},
   journal={Geometry; Topology},
   publisher={Mathematical Sciences Publishers},
   author={Freed, Daniel S and Hopkins, Michael J},
   year={2021},
   month=may, pages={1165–1330} }

@misc{thorngren2019anomaliesbosonization,
      title={Anomalies and Bosonization}, 
      author={Ryan Thorngren},
      year={2019},
      eprint={1810.04414},
      archivePrefix={arXiv},
      primaryClass={cond-mat.str-el},
      url={https://arxiv.org/abs/1810.04414}, 
}

@article{tantivasadakarn2018full,
  title={Full commuting projector Hamiltonians of interacting symmetry-protected topological phases of fermions},
  author={Tantivasadakarn, Nathanan and Vishwanath, Ashvin},
  journal={Physical Review B},
  volume={98},
  number={16},
  pages={165104},
  year={2018},
  publisher={APS}
}

@article{Ellison_2019,
   title={Disentangling Interacting Symmetry-Protected Phases of Fermions in Two Dimensions},
   volume={9},
   ISSN={2160-3308},
   url={http://dx.doi.org/10.1103/PhysRevX.9.011016},
   DOI={10.1103/physrevx.9.011016},
   number={1},
   journal={Physical Review X},
   publisher={American Physical Society (APS)},
   author={Ellison, Tyler D. and Fidkowski, Lukasz},
   year={2019},
   month=jan }

@article{Gaiotto_2019,
   title={Symmetry protected topological phases and generalized cohomology},
   volume={2019},
   ISSN={1029-8479},
   url={http://dx.doi.org/10.1007/JHEP05(2019)007},
   DOI={10.1007/jhep05(2019)007},
   number={5},
   journal={Journal of High Energy Physics},
   publisher={Springer Science and Business Media LLC},
   author={Gaiotto, Davide and Johnson-Freyd, Theo},
   year={2019},
   month=may }

@article{Xiong_2018,
   title={Minimalist approach to the classification of symmetry protected topological phases},
   volume={51},
   ISSN={1751-8121},
   url={http://dx.doi.org/10.1088/1751-8121/aae0b1},
   DOI={10.1088/1751-8121/aae0b1},
   number={44},
   journal={Journal of Physics A: Mathematical and Theoretical},
   publisher={IOP Publishing},
   author={Xiong, Charles Zhaoxi},
   year={2018},
   month=oct, pages={445001} }

@misc{tu2025anomaliesglobalsymmetrieslattice,
      title={Anomalies of global symmetries on the lattice}, 
      author={Yi-Ting Tu and David M. Long and Dominic V. Else},
      year={2025},
      eprint={2507.21209},
      archivePrefix={arXiv},
      primaryClass={cond-mat.str-el},
      url={https://arxiv.org/abs/2507.21209}, 
}

@article{Bultinck_2019,
   title={UV perspective on mixed anomalies at critical points between bosonic symmetry-protected phases},
   volume={100},
   ISSN={2469-9969},
   url={http://dx.doi.org/10.1103/PhysRevB.100.165132},
   DOI={10.1103/physrevb.100.165132},
   number={16},
   journal={Physical Review B},
   publisher={American Physical Society (APS)},
   author={Bultinck, Nick},
   year={2019},
   month=oct }

@article{Else_2014,
   title={Classifying symmetry-protected topological phases through the anomalous action of the symmetry on the edge},
   volume={90},
   ISSN={1550-235X},
   url={http://dx.doi.org/10.1103/PhysRevB.90.235137},
   DOI={10.1103/physrevb.90.235137},
   number={23},
   journal={Physical Review B},
   publisher={American Physical Society (APS)},
   author={Else, Dominic V. and Nayak, Chetan},
   year={2014},
   month=dec }

@misc{kapustin2014symmetryprotectedtopologicalphases,
      title={Symmetry Protected Topological Phases, Anomalies, and Cobordisms: Beyond Group Cohomology}, 
      author={Anton Kapustin},
      year={2014},
      eprint={1403.1467},
      archivePrefix={arXiv},
      primaryClass={cond-mat.str-el},
      url={https://arxiv.org/abs/1403.1467}, 
}

@article{Chen_2013,
   title={Symmetry protected topological orders and the group cohomology of their symmetry group},
   volume={87},
   ISSN={1550-235X},
   url={http://dx.doi.org/10.1103/PhysRevB.87.155114},
   DOI={10.1103/physrevb.87.155114},
   number={15},
   journal={Physical Review B},
   publisher={American Physical Society (APS)},
   author={Chen, Xie and Gu, Zheng-Cheng and Liu, Zheng-Xin and Wen, Xiao-Gang},
   year={2013},
   month=apr }

@article{pollmann2012symmetry,
  title={Symmetry protection of topological phases in one-dimensional quantum spin systems},
  author={Pollmann, Frank and Berg, Erez and Turner, Ari M and Oshikawa, Masaki},
  journal={Physical Review B—Condensed Matter and Materials Physics},
  volume={85},
  number={7},
  pages={075125},
  year={2012},
  publisher={APS}
}

@article{kapustin2014anomalous,
  title={Anomalous discrete symmetries in three dimensions and group cohomology},
  author={Kapustin, Anton and Thorngren, Ryan},
  journal={Physical review letters},
  volume={112},
  number={23},
  pages={231602},
  year={2014},
  publisher={APS}
}

@misc{kapustin2015highersymmetrygappedphases,
      title={Higher symmetry and gapped phases of gauge theories}, 
      author={Anton Kapustin and Ryan Thorngren},
      year={2015},
      eprint={1309.4721},
      archivePrefix={arXiv},
      primaryClass={hep-th},
      url={https://arxiv.org/abs/1309.4721}, 
}

@article{dijkgraaf1990topological,
  title={Topological gauge theories and group cohomology},
  author={Dijkgraaf, Robbert and Witten, Edward},
  journal={Communications in Mathematical Physics},
  volume={129},
  number={2},
  pages={393--429},
  year={1990},
  publisher={Springer}
}

@misc{kitaev2013sre,
  author       = {Kitaev, Alexei},
  title        = {On the classification of {Short-Range Entangled} states},
  howpublished = {Talk at the Simons Center for Geometry and Physics (SCGP)},
  address      = {Stony Brook, NY},
  month        = jun,
  year         = {2013},
  url          = {http://scgp.stonybrook.edu/archives/7874},
  note         = {Accessed 2025-10-10}
}

@misc{arrighi2009unitaritypluscausalityimplies,
      title={Unitarity plus causality implies localizability}, 
      author={Pablo Arrighi and Vincent Nesme and Reinhard Werner},
      year={2009},
      eprint={0711.3975},
      archivePrefix={arXiv},
      primaryClass={quant-ph},
      url={https://arxiv.org/abs/0711.3975}, 
}

@article{schuch2011classifying,
  title={Classifying quantum phases using matrix product states and projected entangled pair states},
  author={Schuch, Norbert and P{\'e}rez-Garc{\'\i}a, David and Cirac, Ignacio},
  journal={Physical Review B—Condensed Matter and Materials Physics},
  volume={84},
  number={16},
  pages={165139},
  year={2011},
  publisher={APS}
}

@article{fidkowski2011topological,
  title={Topological phases of fermions in one dimension},
  author={Fidkowski, Lukasz and Kitaev, Alexei},
  journal={Physical Review B—Condensed Matter and Materials Physics},
  volume={83},
  number={7},
  pages={075103},
  year={2011},
  publisher={APS}
}

@article{Fidkowski_2019,
   title={Interacting invariants for Floquet phases of fermions in two dimensions},
   volume={99},
   ISSN={2469-9969},
   url={http://dx.doi.org/10.1103/PhysRevB.99.085115},
   DOI={10.1103/physrevb.99.085115},
   number={8},
   journal={Physical Review B},
   publisher={American Physical Society (APS)},
   author={Fidkowski, Lukasz and Po, Hoi Chun and Potter, Andrew C. and Vishwanath, Ashvin},
   year={2019},
   month=feb }

@article{Gu_2014,
   title={Symmetry-protected topological orders for interacting fermions: Fermionic topological nonlinear $\sigma$-models and a special group supercohomology theory},
   volume={90},
   ISSN={1550-235X},
   url={http://dx.doi.org/10.1103/PhysRevB.90.115141},
   DOI={10.1103/physrevb.90.115141},
   number={11},
   journal={Physical Review B},
   publisher={American Physical Society (APS)},
   author={Gu, Zheng-Cheng and Wen, Xiao-Gang},
   year={2014},
   month=sep }

@book{goerss1999simplicial,
  title={Simplicial Homotopy Theory},
  author={Goerss, P.G. and Jardine, J.F.},
  isbn={9783764360641},
  lccn={99029941},
  series={Progress in mathematics (Boston, Mass.) v. 174},
  url={https://books.google.com/books?id=xFwXQCtNcUoC},
  year={1999},
  publisher={Springer}
}

@misc{jones2025localtopologicalorderboundary,
      title={Local topological order and boundary algebras}, 
      author={Corey Jones and Pieter Naaijkens and David Penneys and Daniel Wallick},
      year={2025},
      eprint={2307.12552},
      archivePrefix={arXiv},
      primaryClass={math-ph},
      url={https://arxiv.org/abs/2307.12552}, 
}

\end{document}